\documentclass[aps, prd, twocolumn, amsmath, floats,floatfix, superscriptaddress, nofootinbib]{revtex4-2}

\usepackage{graphicx,amssymb,amsmath,amsthm,amsfonts,epsfig}
\usepackage[linktocpage]{hyperref}
\usepackage[usenames]{color}
\usepackage{epstopdf}

\usepackage{bm}
\usepackage{dcolumn}
\usepackage[utf8]{inputenc}
\usepackage{latexsym}
\usepackage{rotating}
\usepackage{soul}

\usepackage{longtable}
\usepackage{enumerate}
\usepackage{mathtools}
\usepackage{url}

\def\p{\partial}
\def\na{\nabla}

\newcommand{\ben}{\begin{enumerate}}
\newcommand{\een}{\end{enumerate}}

\def\be{\begin{equation}}
\def\ee{\end{equation}}
\def\bea{\begin{eqnarray}}
\def\eea{\end{eqnarray}}
\newcommand{\beq}{\begin{eqnarray}}
\newcommand{\eeq}{\end{eqnarray}} 
\newcommand{\ba}{\begin{align}}
\newcommand{\ea}{\end{align}}

\begin{document}

\title{Electromagnetic emission from axionic boson star collisions}

\author{Nicolas~Sanchis-Gual}
 \affiliation{Departamento de
  Astronom\'{\i}a y Astrof\'{\i}sica, Universitat de Val\`encia,
  Dr.\ Moliner 50, 46100, Burjassot (Val\`encia), Spain}
\affiliation{Departamento  de  Matem\'{a}tica  da  Universidade  de  Aveiro and 
  Centre for Research and Development in Mathematics and Applications (CIDMA),
  Campus de Santiago, 3810-183 Aveiro, Portugal}

\author{Miguel~Zilh\~ao}
  \affiliation{Departamento  de  Matem\'{a}tica  da  Universidade  de  Aveiro and 
  Centre for Research and Development in Mathematics and Applications (CIDMA),
  Campus de Santiago, 3810-183 Aveiro, Portugal}
\affiliation{Centro de Astrof\'\i sica e Gravita\c c\~ao -- CENTRA,
  Departamento de F\'\i sica, Instituto Superior T\'ecnico -- IST,
  Universidade de Lisboa -- UL, Avenida Rovisco Pais 1, 1049-001 Lisboa, Portugal}

\author{Vitor~Cardoso}
\affiliation{Centro de Astrof\'\i sica e Gravita\c c\~ao -- CENTRA,
  Departamento de F\'\i sica, Instituto Superior T\'ecnico -- IST,
  Universidade de Lisboa -- UL, Avenida Rovisco Pais 1, 1049-001 Lisboa, Portugal}
\affiliation{Niels Bohr International Academy, Niels Bohr Institute, Blegdamsvej 17, 2100 Copenhagen, Denmark}

\begin{abstract}
  We explore the dynamics of boson stars in the presence of axionic couplings
  through nonlinear evolutions of Einstein's field equations. 
	We show that, for large axionic couplings, isolated boson stars
	become unstable, and decay via a large burst of electromagnetic radiation, becoming less massive and more dilute.
	Our full nonlinear results are in good agreement with flat-space estimates for the critical couplings.
  We then consider head-on collisions of sub-critical boson stars and study	the
  electromagnetic and gravitational signal. Boson stars cluster around the critical point via interactions, and we argue that
	mergers will generically be also sources of electromagnetic radiation, in addition to gravitational waves,
	which can be used to place constraints on the axionic coupling if such multimessenger signals are detected.
\end{abstract}

\maketitle

\section{Introduction}

First introduced as a possible solution to the strong CP problem in
QCD~\cite{Peccei:1977hh}, axion (or axion-like) particles are now regarded as
possible candidates for cold dark matter~\cite{Bergstr_m_2009,Fairbairn:2014zta,Chadha-Day:2021szb}. These particles
can arise from simple extensions of the Standard Model~\cite{Freitas:2021cfi}, from string theory compactifications~\cite{Cicoli:2013ana}, and can have implications for cosmological and astronomical observations~\cite{Arvanitaki:2009fg}.
Typically such axion models make use of very weak coupling constants, for compatibility with prior experiments, making their detection very challenging.
Indeed, no evidence has been found so far of interactions between dark matter and Standard Model particles.

For dense axionic environments, however, photon emission can be triggered~\cite{Kephart:1986vc,Kephart:1994uy,Rosa:2017ury} which can lead to possible observational signatures both in the electromagnetic (EM) as well as in the gravitational-wave (GW) spectrum~\cite{Boskovic:2018lkj,Ikeda:2018nhb}. In this context, axion-like particles can trigger the superradiant instability around spinning black holes (BHs)~\cite{Detweiler:1980uk,Cardoso:2005vk,Dolan:2012yt,Yoshino:2013ofa,Brito:2015oca,East:2017ovw}, leading to GW emission that could be observed by current or future LIGO-Virgo-KAGRA detectors~\cite{Arvanitaki:2010sy,Brito:2014wla,Isi:2018pzk,Palomba:2019vxe,Yuan:2021ebu,LIGOScientific:2021jlr} or by third-generation detectors, such as the Einstein Telescope~\cite{Hild:2011}.

Presently, the only compact objects known to source detectable GW signals are BHs and neutron stars~\cite{Barack:2018yly,LIGOScientific:2016aoc,LIGOScientific:2017vwq,LIGOScientific:2018mvr,LIGOScientific:2020ibl,LIGOScientific:2021usb,LIGOScientific:2020iuh}. However, other kinds of
compact objects -- such as bosonic stars -- could also be detected through the
emitted GW signal during a merger~\cite{Bustillo:2020syj}. Bosonic stars are theoretical exotic compact objects composed of fundamental bosonic fields minimally coupled to Einstein's gravity forming macroscopic self-gravitating Bose-Einstein configurations~\cite{Kaup:1968zz,Ruffini:1969qy,Brito:2015pxa}.
In recent years, several studies have investigated the stability and dynamical formation of bosonic stars~\cite{jetzer1990stability,jetzer1992boson,seidel1994formation,DiGiovanni:2018bvo,Sanchis-Gual:2019ljs}, and their GW emission in binaries~\cite{Palenzuela:2006wp,Palenzuela:2007dm,Palenzuela:2017kcg,Sanchis-Gual:2018oui,Bezares:2022obu}.

In this work we explore the dynamics of axionic bosonic stars, described by a complex (pseudo)scalar field minimally coupled to Einstein's gravity and non-minimally coupled to the EM tensor as in~\cite{Boskovic:2018lkj,Ikeda:2018nhb}. In this context, bosonic fields would not form completely dark objects, but could emit EM radiation in some regimes.
We show that there exists a threshold for the value of the axionic coupling
above which certain star configurations are unstable, decaying to a more dilute solution.
We also measure the EM and GW emission during axionic boson star (BS) collisions. If the coupling is above the threshold, the EM field becomes nonlinear and sources the emission of GWs. The scalar field energy is largely reduced, preventing the collapse to a BH. Therefore, axion BSs may only exist below the critical threshold, but mergers of these objects leading to more compact and potentially unstable configurations can trigger EM emission\footnote{In fact, the parametric instability provides a natural mechanism through which isolated boson stars naturally grow to close to the critical threshold mass via accretion, and therefore mergers should always produce electromagnetic counterparts.}.
If such signals are detected together with GWs in a multimessenger event, this information can be used to constrain possible
values of the axionic coupling to the Maxwell sector.

We use geometrical units $G=c=1$ throughout and let Greek indices run from 0 to
3 and Latin indices from 1 to 3, as usual.

\section{Setup}
\label{sec:setup}

We consider the action describing a complex massive (pseudo)scalar field $\Phi$ with axionic coupling to the EM field through the coupling constant $k_{\rm axion}$
\begin{align}
  \label{eq:action}
{\cal L}&=\frac{R}{4}- \frac{1}{4} F^{\mu\nu} F_{\mu\nu} - \frac{1}{2} g^{\mu\nu} \p_{\mu} \bar \Phi\p_{\nu} \Phi
        - \frac{\mu^{2}}{2} \bar \Phi \Phi \nonumber \\
        &{}- \frac{k_{\rm axion}}{2} \left( \Re(\Phi) + \Im(\Phi) \right)
          \,{\star}F^{\mu\nu} F_{\mu\nu}\,.
\end{align}
The mass of the scalar $\Phi$ is given by $m_{\rm S} = \mu \hbar$, $F_{\mu\nu} \equiv
\na_{\mu}A_{\nu} - \na_{\nu} A_{\mu}$ is the Maxwell tensor and $\star F^{\mu\nu} \equiv \frac{1}{2}\epsilon^{\mu\nu\rho\sigma}F_{\rho\sigma}$
is its dual.
%
The corresponding equations of motion are given by
\begin{subequations}
\label{eq:MFEoMgen}
\begin{align}
\label{eq:MFEoMTensor}
  R_{\mu\nu} - \frac{R}{2} g_{\mu\nu} & =
     T_{\mu\nu}^{\rm EM} + T_{\mu\nu}^{\rm S} \,, \\
\label{eq:MFEoMScalar1}
  \left(\nabla^{\mu}\nabla_{\mu} - \mu^{2} \right) \Re(\Phi) & =\frac{k_{\rm axion}}{2}
    \star \! F^{\mu\nu} F_{\mu\nu} \,,\\
\label{eq:MFEoMScalar2}
  \left(\nabla^{\mu}\nabla_{\mu} - \mu^{2} \right) \Im(\Phi) & =\frac{k_{\rm axion}}{2}
    \star \! F^{\mu\nu} F_{\mu\nu} \,,\\
\label{eq:MFEoMVector}
  \nabla_{\nu}F^{\mu\nu} & = - 2 k_{\rm axion} \star \! F^{\mu\nu} \, \nabla_{\nu}
    \left( \Re(\Phi) + \Im(\Phi) \right)\,,
\end{align}
\end{subequations}
where
\begin{align}
\label{eq:TabEM}
  T_{\mu\nu}^{\rm EM} \equiv 2 F_{\mu\rho} F_{\nu}{}^{\rho}
  - \frac{1}{2} g_{\mu\nu} F_{\rho \sigma} F^{\rho \sigma}\,,
\end{align}
and
\begin{align}
\label{eq:TabS}
T_{\mu\nu}^{\rm S} \equiv \partial_{\mu} \bar \Phi \partial_{\nu} \Phi
  + \partial_{\mu} \Phi \partial_{\nu} \bar  \Phi
  - g_{\mu \nu} \left( g^{\sigma \rho} \partial_{\sigma} \bar  \Phi \partial_{\rho} \Phi
  + \mu^2 \bar \Phi \Phi \right)\,.
\end{align}

We perform a 3+1 decomposition on Eqs.~\eqref{eq:MFEoMgen} to formulate
the system as a Cauchy problem~\cite{Gourgoulhon:2007ue}. To this end, we impose the Lorenz gauge on the vector field
\[
  \nabla_{\mu}A^{\mu}=0 \,,
\]
and introduce the 3-metric
\begin{equation}
\label{eq:3metric}
\gamma_{\mu \nu} = g_{\mu \nu} + n_{\mu} n_{\nu} \,,
\end{equation}
where $n^{\mu} = \frac{1}{\alpha} (1, -\beta^i )$ is a timelike unit vector, and write the full spacetime metric in the form
\begin{equation}
  \label{eq:3+1}
ds^2=-\alpha^{2}dt^{2}+\gamma_{ij}(dx^{i}+\beta^{i}dt)(dx^{j}+\beta^{j}dt)\,,
\end{equation}
where the lapse function $\alpha$ and shift vector $\beta^{i}$ encode the coordinate degrees of freedom.
We further decompose the vector field $A_{\mu}$ as
\begin{equation}
A_{\phi}=-n^{\mu}A_{\mu}\,,\qquad \mathcal{A}_{i}=\gamma^{j}_{~i}A_{j}\,,
\end{equation}
and introduce the EM fields
\begin{equation}
  E^{i}=\gamma^{i}_{~j}F^{j\nu}n_{\nu}\,,\qquad
  B^{i}=\gamma^{i}_{~j}\,{\star}F^{j\nu}n_{\nu}\,.
\end{equation}
Note that $B^i$ is not an evolved variable and is introduced here merely as notational convenience. Finally, we introduce the scalar momentum $\Pi$ as
\begin{equation}
\Pi=-n^{\mu}\nabla_{\mu}\Phi\,.
\end{equation}

The evolution equations for the axion field take the form
\begin{subequations}
\begin{align}
\partial_{t} \Re(\Phi)&=-\alpha\Re(\Pi)+\mathcal{L}_{\beta}\Re(\Phi)\,, \\
\partial_{t}\Re(\Pi)&=\alpha(-D^{2}\Re(\Phi)+\mu^{2}\Re(\Phi)+K\Re(\Pi)-2k_{\rm axion}E^{i}B_{i})\nonumber\\
&\quad{} - D^{i}\alpha D_{i}\Re(\Phi)+\mathcal{L}_{\beta}\Re(\Pi)\,,
\end{align}
\end{subequations}
and analogously for the imaginary part of $\Phi$. For the evolution of the EM fields we find
\begin{subequations}
  \begin{align}
\partial_{t}\mathcal{A}_{i}&=-\alpha(E_{i}+D_{i}\mathcal{A}_{\varphi})-A_{\phi}D_{i}\alpha+\mathcal{L}_{\beta}\mathcal{A}_{i}\,,\\
\partial_{t}E^{i}&=\alpha(KE^{i}+D^{i}Z-(D^{2}\mathcal{A}^{i}-D_{k}D^{i}\mathcal{A}^{k}))\nonumber\\
     &\quad{}+2\alpha\, k_{\rm axion}\left(\epsilon^{ijk}E_{k}D_{j}\left(\Re(\Phi) + \Im(\Phi) \right) \right) \nonumber\\
    &\quad{}+2\alpha\, k_{\rm axion}B^{i} \left(\Re(\Pi) + \Im(\Pi)\right) \nonumber\\
                &\quad{}+\epsilon^{ijk}D_{k}\alpha B_{j}+\mathcal{L}_{\beta}E^{i} \,,\\
\partial_{t}A_{\phi}&=\alpha(KA_{\phi}-D_{i}\mathcal{A}^{i}-Z)
             -\mathcal{A}_{j}D^{j}\alpha+\mathcal{L}_{\beta}A_{\phi}\,,\\
\partial_{t}Z&=\alpha(D_{i}E^{i}-\kappa Z)
                   +2\,k_{\rm axion}\alpha B^{i}D_{i} \left( \Re(\Phi) + \Im(\Phi) \right) \nonumber\\
    &\quad{}+\mathcal{L}{_\beta}Z\,,
  \end{align}
\end{subequations}
where $D_i$ denotes the covariant derivative with respect to the 3-metric $\gamma_{ij}$,
$K$ is the trace of the extrinsic curvature,
and the constraint damping variable $Z$ was added to dynamically enforce the Gauss constraint
\[
D_{i}E^{i}+2k_{\rm axion} B^{i}D_{i} \left( \Re(\Phi) + \Im(\Phi) \right)=0\,.
\]

\subsection{Numerical implementation}
\label{sec:implementation}
We use the Baumgarte-Shapiro-Shibata-Nakamura (BSSN) formulation of Einstein's equations~\cite{Shibata:1995we,Baumgarte:1998te} to evolve the system.
For the numerical evolutions we make use of the Einstein~Toolkit
infrastructure~\cite{Loffler:2011ay,Zilhao:2013hia,EinsteinToolkit:2021_05} with
the Carpet package~\cite{Schnetter:2003rb} for mesh refinement
capabilities,
and AHFinderDirect~\cite{Thornburg:2003sf} for finding
apparent horizons.
The spacetime variables in the BSSN formulation are evolved using
McLachlan~\cite{Brown:2008sb,McLachlan:web}, whereas the axion and EM
fields are evolved by adapting the ScalarEvolve and ProcaEvolve codes available
in~\cite{Canuda_zenodo_3565474}, which were first used and described in
Refs.~\cite{Zilhao:2015tya,Cunha:2017wao,Sanchis-Gual:2019ljs}.
The evolution equations are integrated using fourth-order spatial discretization and a Runge-Kutta method.

\subsection{Initial data}
\begin{table}[htb]
\caption{Boson star models: $\omega/\mu$ is the oscillation frequency, $\Phi_0$ is the central value of the scalar field, and $M_{\rm{BS}}\mu$ is the mass of the star.\label{tab:table1}}
\begin{ruledtabular}
\begin{tabular}{cccc}
ID&$\omega/\mu$&$\Phi_0$&$M_{\rm{BS}}\mu$\\
\hline
BSA&0.9140&0.142&0.584\\
BSB&0.9438&0.089&0.512\\
BSC&0.9697&0.046&0.403\\
BSD&0.9812&0.028&0.328
\end{tabular}
\end{ruledtabular}
\end{table}
Scalar BSs were first discussed in~\cite{Kaup:1968zz,Ruffini:1969qy} (see also~\cite{Liebling:2012fv} for a review).
Spherically symmetric BSs can be obtained from the line element
\begin{equation}
\label{eq:ansatzBS}
 ds^2=-N(r)\sigma^2(r) dt^2+\frac{dr^2}{N(r)}+r^2 (d\theta^2+\sin^2\theta d\varphi^2) \,
\end{equation} 
where $N(r)\equiv 1-{2m(r)}/{r}$, $m(r),\sigma(r)$ are radial functions and $r,\theta,\varphi$ are Schwarzschild-type coordinates, and
the matter field is written in the form
\begin{equation}
\label{eq:ansatzBSfield}
\Phi=\phi(r)e^{-i \omega t} \,,
\end{equation}
where $\omega>0$ is the oscillation frequency.
The corresponding equations are numerically solved with appropriate boundary conditions~\cite{Kaup:1968zz,Ruffini:1969qy,Colpi:1986ye,Liebling:2012fv,Macedo:2013jja,Herdeiro:2017fhv,Escorihuela-Tomas:2017uac,Cardoso:2021ehg}.
The solution is then written in isotropic coordinates to perform the numerical evolutions.

We start by constructing isolated nonspinning BSs with the parameters specified in Table~\ref{tab:table1}. We label those four distinct solutions as BSA, BSB, BSC and BSD.
We then superimpose two such solutions at a distance of $\Delta x = D\mu=16.4$ ($y_0=z_0=0$) and let them collide from rest. 
This construction follows that of Refs.~\cite{Palenzuela:2006wp,Palenzuela:2007dm,Palenzuela:2017kcg,Sanchis-Gual:2018oui,Sanchis-Gual:2020mzb} and amounts to setting
%
\begin{align*}
  \Phi(x_{i}) & = \Phi^{(1)}(x_{i}-x_{0}) + \Phi^{(2)}(x_{i}+x_{0}) \,,  \\
  \gamma_{ij}(x_{i}) & = \gamma_{ij}^{(1)}(x_{i}-x_{0}) + \gamma_{ij}^{(2)}(x_{i}+x_{0})-\gamma_{ij}^{\rm flat}(x_{i}) \,, \\
\alpha(x_{i}) & = \alpha^{(1)}(x_{i}-x_{0}) + \alpha^{(2)}(x_{i}+x_{0}) - 1 \,,
\end{align*}
where superindex $(i)$ labels the stars and $\pm x_{0}$ indicates their initial positions.

Note that the maximum mass of a BS is $M\mu\sim 0.633$~\cite{Brito:2015yfh}. When two equal-mass BSs collide with small energy loss to GWs or scalar waves, one expects the final object to have roughly twice the mass of each component. Thus, all configurations BSA, BSB, BSC are expected to give rise to an unstable BS, which eventually collapses to a BH. The only exception is BSD, which is still on the stable branch. These expectations conform to our numerics, as we show below.

Note that, even though this theory includes couplings to the Maxwell sector, a BS solution and a vanishing EM field solves the field equations. We are interested in this initial state. However, we are also interested in understanding the nonlinear stability of such solutions, and for that we assign a small seed electric field, with $A_{i}=A_{\phi}=E^{r}=E^{\theta}=0$ and~\cite{Ikeda:2018nhb}
\begin{equation}
  \label{IDEM}
E^{\varphi}=E_{0}\,e^{-\bigl(\frac{\displaystyle r-r_{0}}{\displaystyle\sigma}\bigl)^2},
\end{equation}
where $E_{0}=0.001/\Phi_0$, $\Phi_0\equiv \Phi(t=0,r=0)$ is the value of the scalar field at the center of the star, and $r_{0}$, $\sigma$ are the position and width of the initial distribution.

\subsection{Wave extraction}
\label{sec:extraction}
To obtain information about waves emitted in the form of GW and EM
radiation we employ the Newman-Penrose (NP) formalism~\cite{Newman:1961qr} as described in~\cite{Zilhao:2015tya}.
We compute the NP scalars $\Psi_{4}$ and $\phi_{2}$, which are expanded -- at a
given extraction radius $R_\mathrm{ex}$ -- into spin-weighted spherical
harmonics of spin weight $s = -2$ and $-1$ respectively
\begin{align}
  \Psi_4(t, \theta, \varphi) & =
      \sum_{l,m} \Psi_4^{lm}(t)\, Y_{lm}^{-2}(\theta,\varphi) \,,
\label{eq:multipole_Psi4} \\
  \phi_2(t, \theta, \varphi) & =
      \sum_{l,m} \phi_{2}^{lm}(t)\, Y_{lm}^{-1}(\theta,\varphi)
\,.\label{eq:multipole_Phi2}
\end{align}

We will show some of the dominant modes of both quantities to illustrate the GW and EM emissions, in particular the $l=m=2$ mode for $\Psi_4$ and the $l=1$, $m=0$ and $l=m=2$ modes for $\phi_2$. Note that for $\Psi_4$ the $l=2$, $m=0$ and $l=2$, $m=-2$ modes, and for $\phi_2$ the $l=2$, $m=-2$ mode are equally dominant, but are not shown here for simplicity.

\section{Results}\label{sec:results}
\subsection{Flat space estimate}
\label{sec:flatspace}
It can be shown through a flat space analysis that a time-dependent background
of axions (oscillating coherently with amplitude $\Phi_0$) is unstable and grows exponentially as $\Phi\sim e^{\lambda_{\star} t}$.
The growth rate is given by~\cite{Ikeda:2018nhb,Sen:2018cjt,Boskovic:2018lkj}\footnote{Note the
additional factor of $\sqrt{2}$ when comparing with
Refs.~\cite{Ikeda:2018nhb,Sen:2018cjt,Boskovic:2018lkj}. This is due to the fact that the
scalar field considered therein is real, whereas ours is complex. The relevant
term in our action~(\ref{eq:action}) takes the form
$\tfrac{k_{\rm axion}}{2} \left( \Re(\Phi) + \Im(\Phi) \right) \sim
\tfrac{k_{\rm axion}}{2} \sqrt{2} \Phi_0 \cos(\omega t + \pi/4) \leqslant
\tfrac{k_{\rm axion}}{2} \sqrt{2}\Phi_0 $, whereas for the real field of
Refs.~\cite{Ikeda:2018nhb,Sen:2018cjt,Boskovic:2018lkj} one has $\tfrac{k_{\rm axion}}{2} \Phi \sim \tfrac{k_{\rm axion}}{2} \Phi_0\cos(\omega
t) \leqslant \tfrac{k_{\rm axion}}{2} \Phi_0$.
}
\begin{equation}
\label{eq:lambdastar}
\lambda_{\star} = \frac{\sqrt{2}}{2} \mu \, k_{\rm axion} \Phi_0 \,.
\end{equation}
Given a localized configuration of axionic field with characteristic size $d$
(such as in a star), this is then expected to be unstable if the photons cannot
leave the field configuration before the instability kicks in,
\begin{equation}
\label{eq:unstability}
\frac{1}{d} \gtrsim \lambda_{\star} \,.
\end{equation}

We can estimate the threshold value for the axionic coupling $k_{\rm axion}^{\star}$ as follows. For Newtonian BSs, the following mass-radius relation holds~\cite{Guzman:2004wj,Annulli:2020lyc}
\begin{equation}
\label{eq:mass-radius}
M_{\rm BS} \mu \simeq \frac{9.1}{R_{\rm BS} \mu} \,.
\end{equation}
Using $d=2R_{\rm BS}$ in Eq.~(\ref{eq:unstability}) we obtain
\begin{equation}
\label{eq:kstar}
k_{\rm axion}^{\star} \simeq \frac{M_{\rm BS} \mu}{9.1 \Phi_0 \sqrt{2} } \,.
\end{equation}

\subsection{Isolated boson stars}
\label{sec:single-BS}
%
\begin{figure}[thb]
\begin{center}
\begin{tabular}{ p{0.32\linewidth}  }
\centering   $k_{\rm{axion}}=0.23$
\end{tabular}\\
\includegraphics[width=0.3\linewidth]{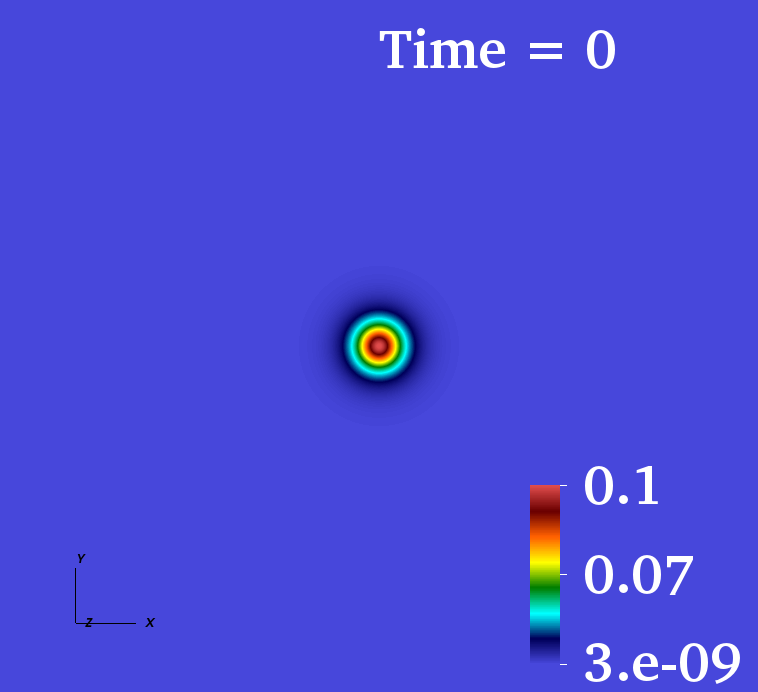}\hspace{-0.005\linewidth}
\includegraphics[width=0.3\linewidth]{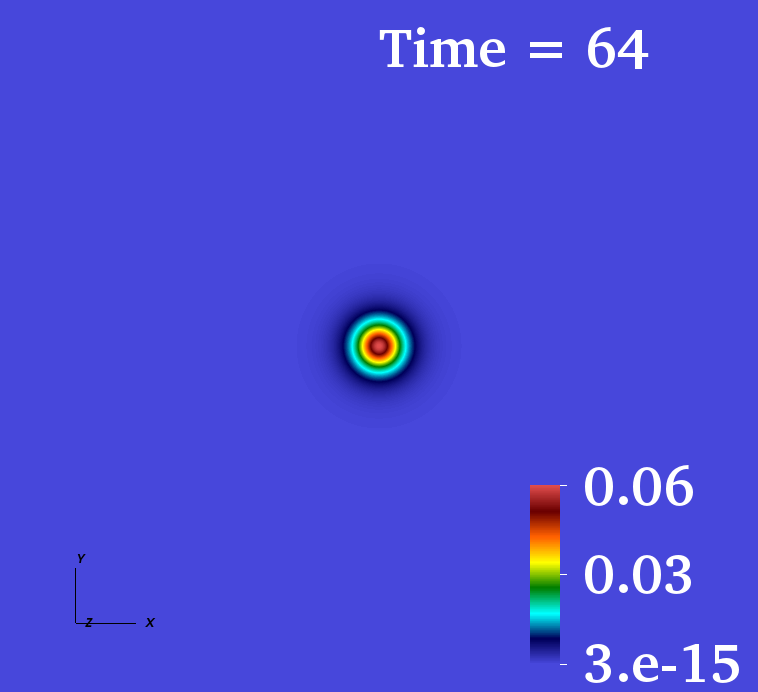}\hspace{-0.005\linewidth}
\includegraphics[width=0.3\linewidth]{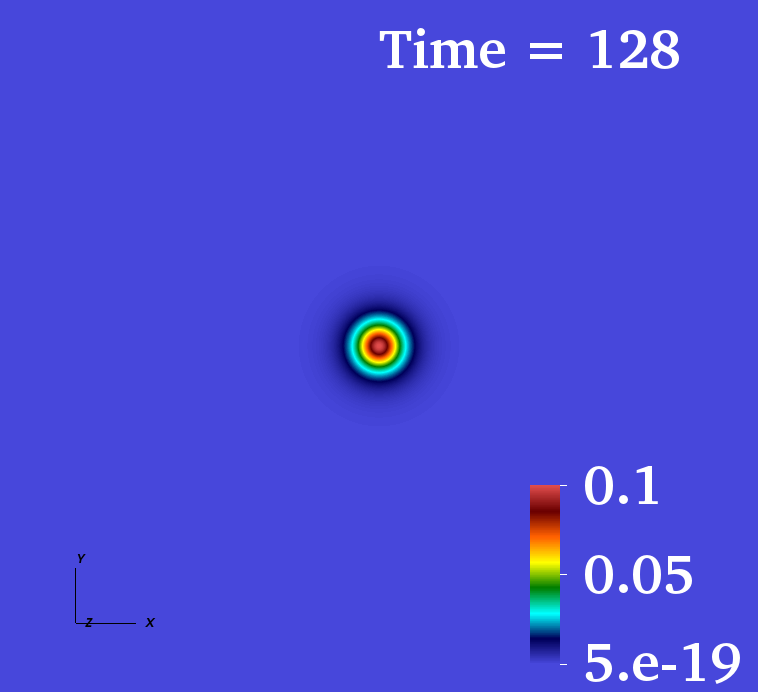}\hspace{-0.005\linewidth}\\
\includegraphics[width=0.3\linewidth]{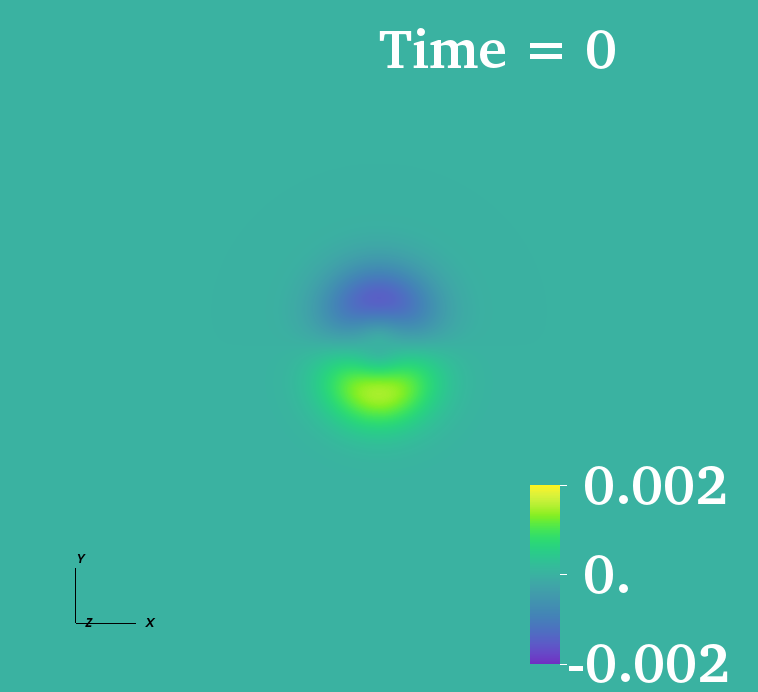}\hspace{-0.005\linewidth}
\includegraphics[width=0.3\linewidth]{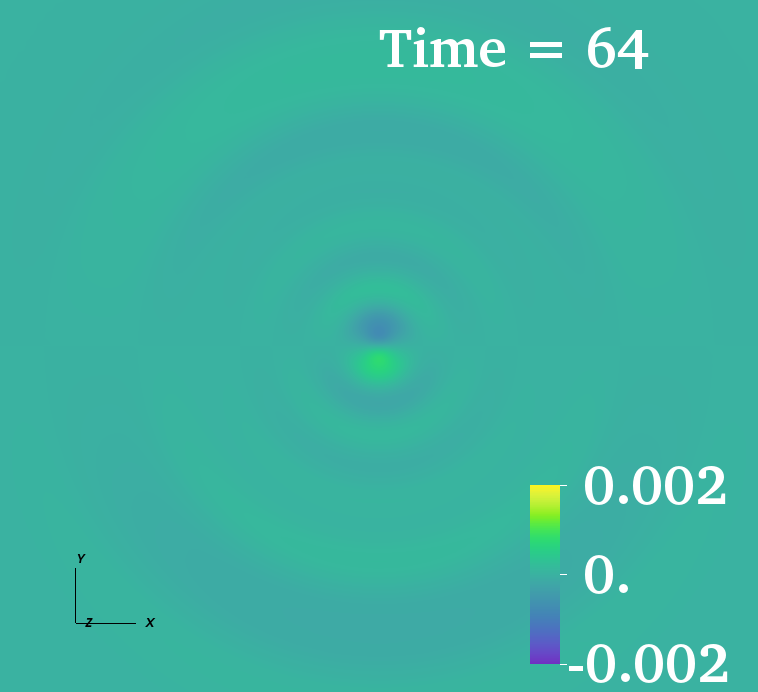}\hspace{-0.005\linewidth}
\includegraphics[width=0.3\linewidth]{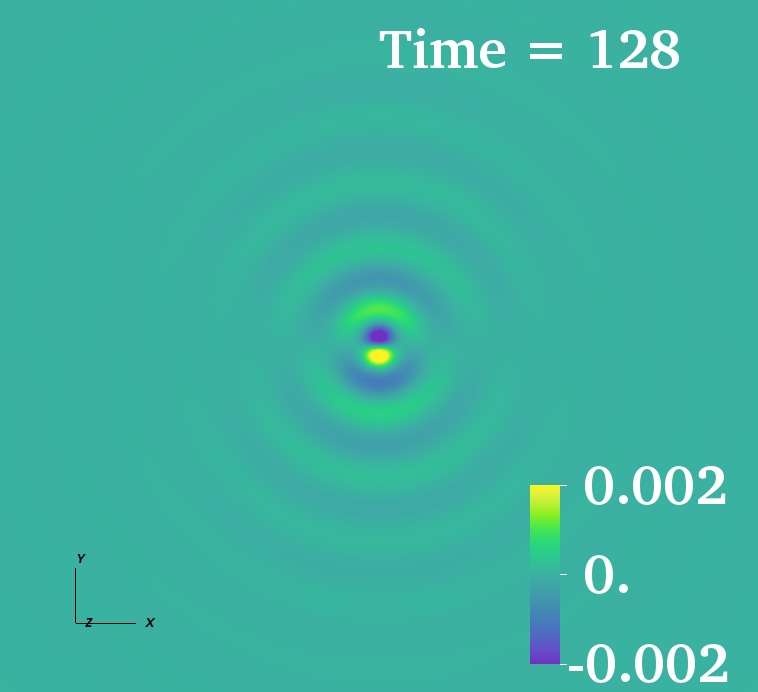}\hspace{-0.005\linewidth}\\
\includegraphics[width=0.3\linewidth]{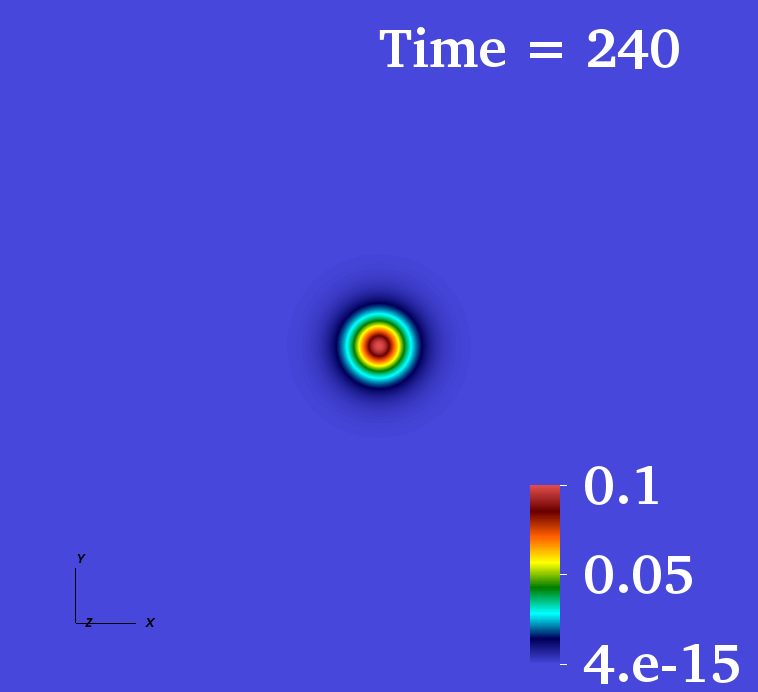}\hspace{-0.005\linewidth}
\includegraphics[width=0.3\linewidth]{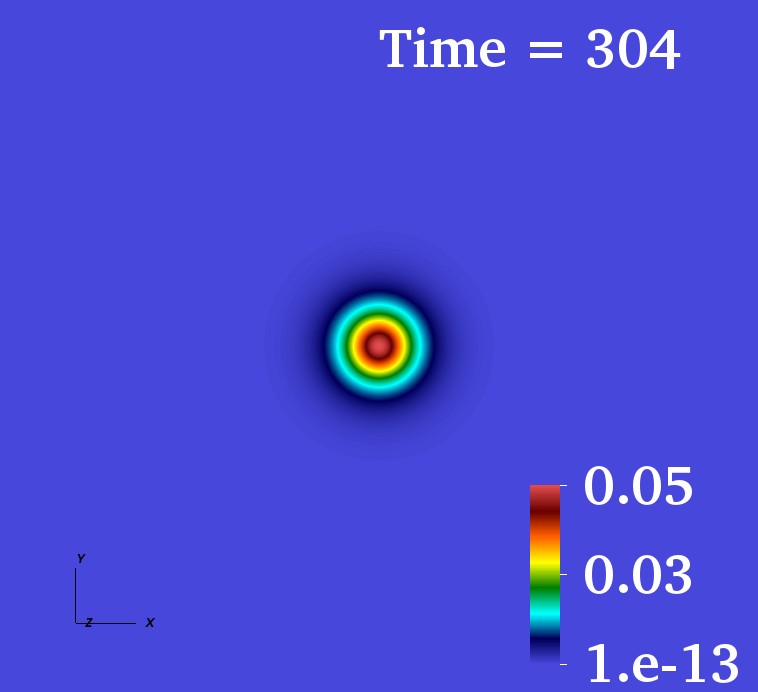}\hspace{-0.005\linewidth}
\includegraphics[width=0.3\linewidth]{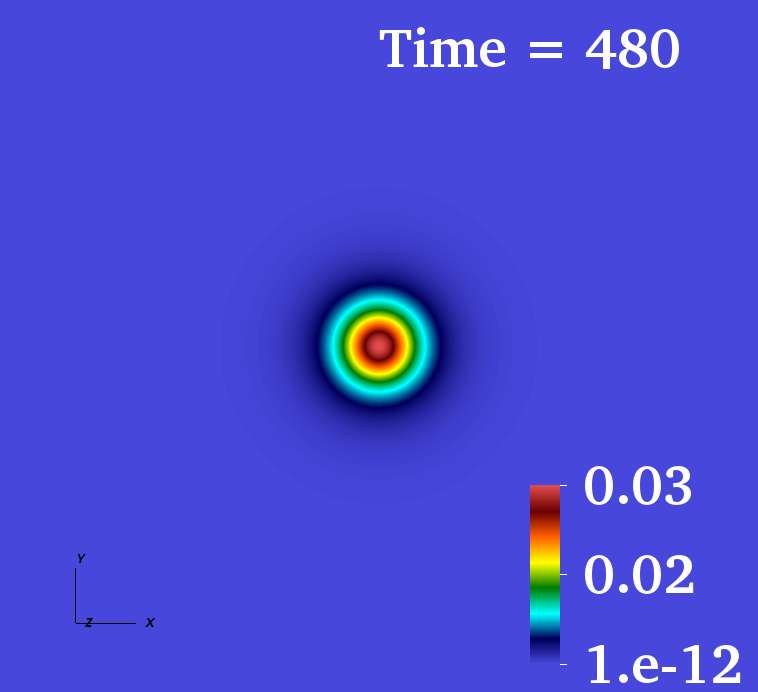}\\
\includegraphics[width=0.3\linewidth]{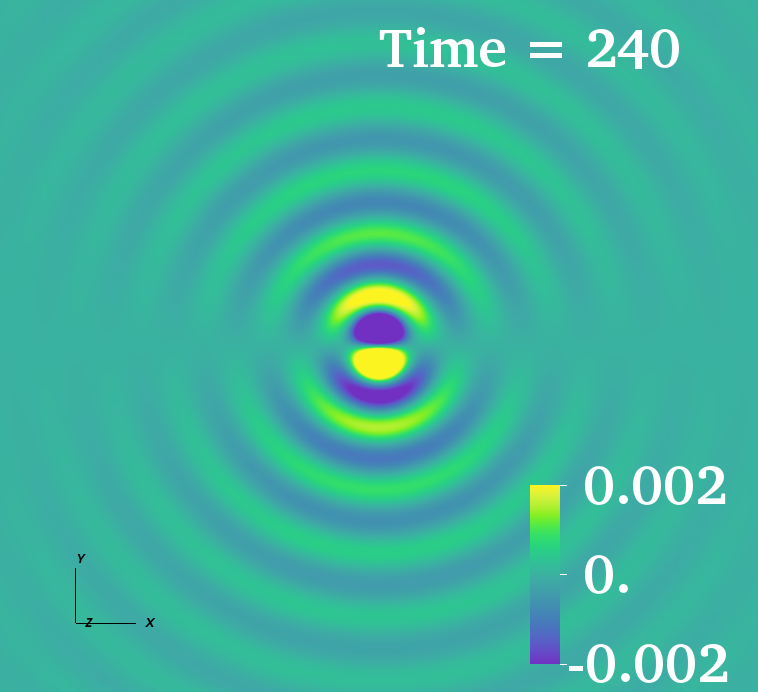}\hspace{-0.005\linewidth}
\includegraphics[width=0.3\linewidth]{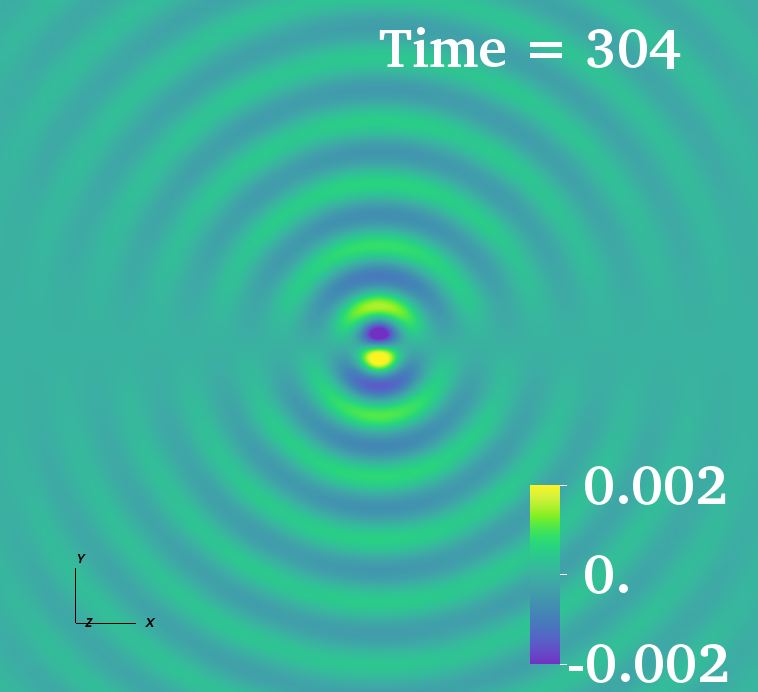}\hspace{-0.005\linewidth}
\includegraphics[width=0.3\linewidth]{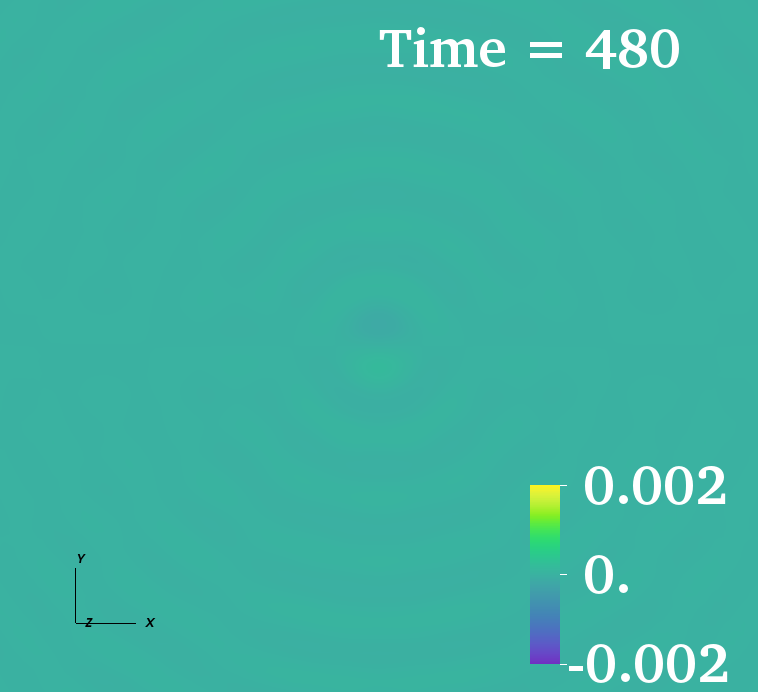}
\caption{Time evolution of the equatorial slice ($z=0$) of $|\Re(\Phi)|$ (first and third rows) and $E^x$ (second and fourth rows) of an isolated BSA configuration (cf. Table~\ref{tab:table1}). The axionic coupling is $k_{\rm{axion}}=0.23$ and time is given in code units where $\mu=1$.\label{fig1}}
\end{center}
\end{figure}
\begin{figure}[thb]
\begin{center}
\begin{tabular}{ p{0.32\linewidth}  }
\centering   $k_{\rm{axion}}=0.43$
\end{tabular}\\
\includegraphics[width=0.3\linewidth]{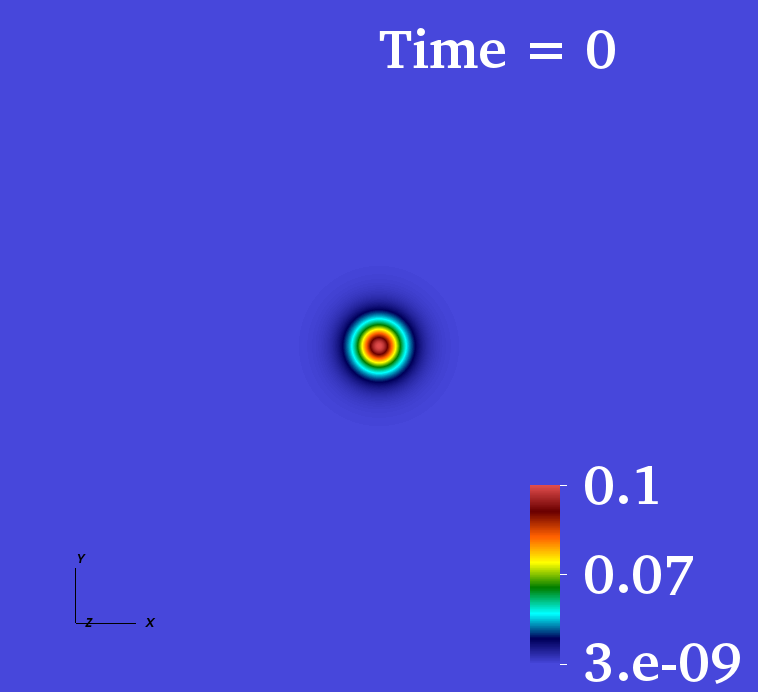}\hspace{-0.005\linewidth}
\includegraphics[width=0.3\linewidth]{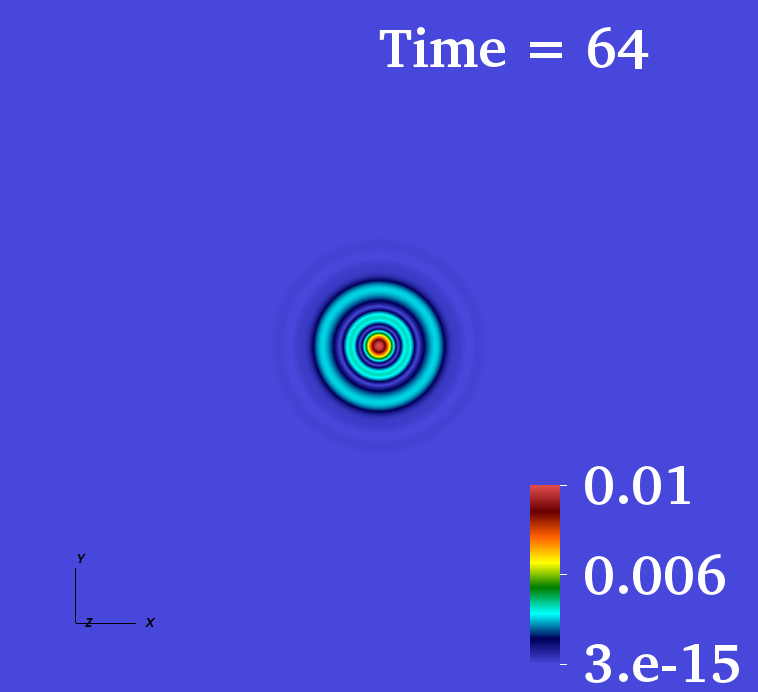}\hspace{-0.005\linewidth}
\includegraphics[width=0.3\linewidth]{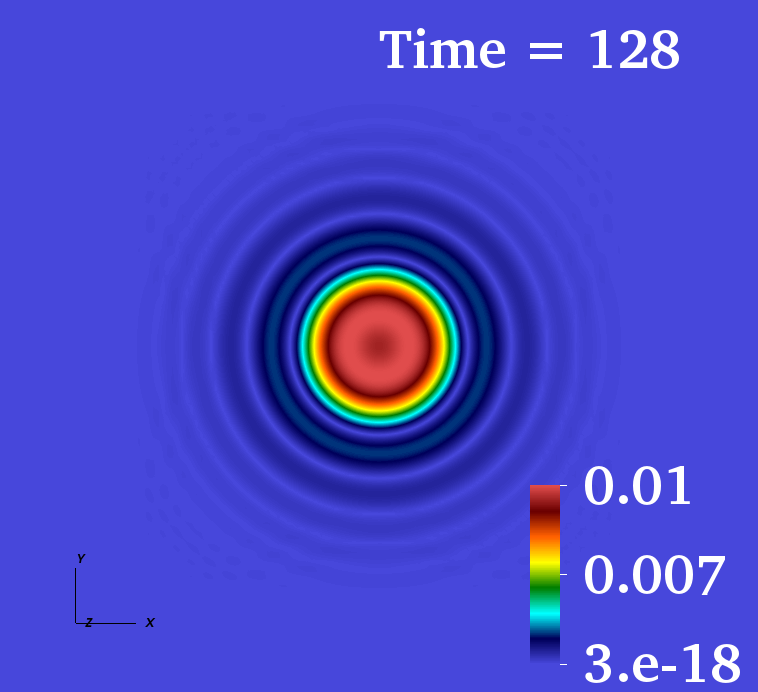}\\
\includegraphics[width=0.3\linewidth]{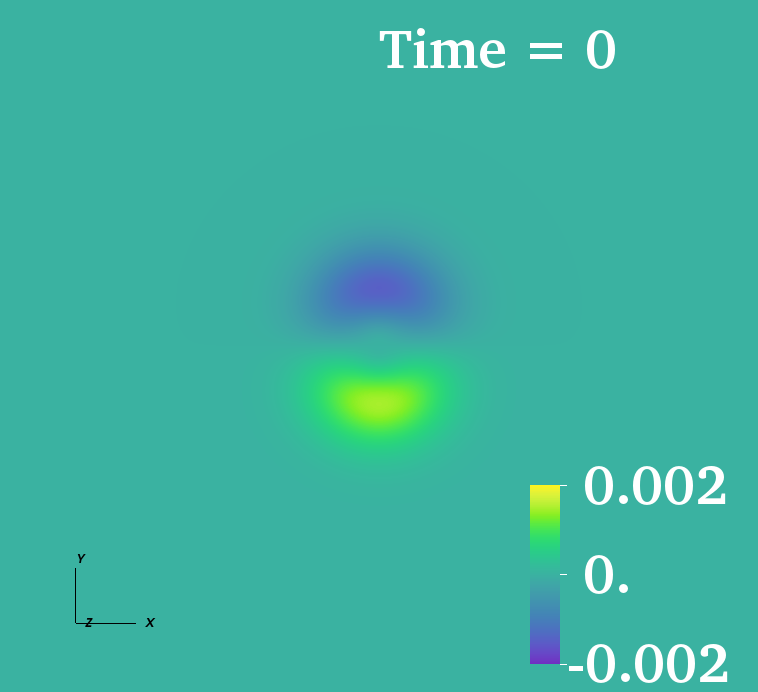}\hspace{-0.005\linewidth}
\includegraphics[width=0.3\linewidth]{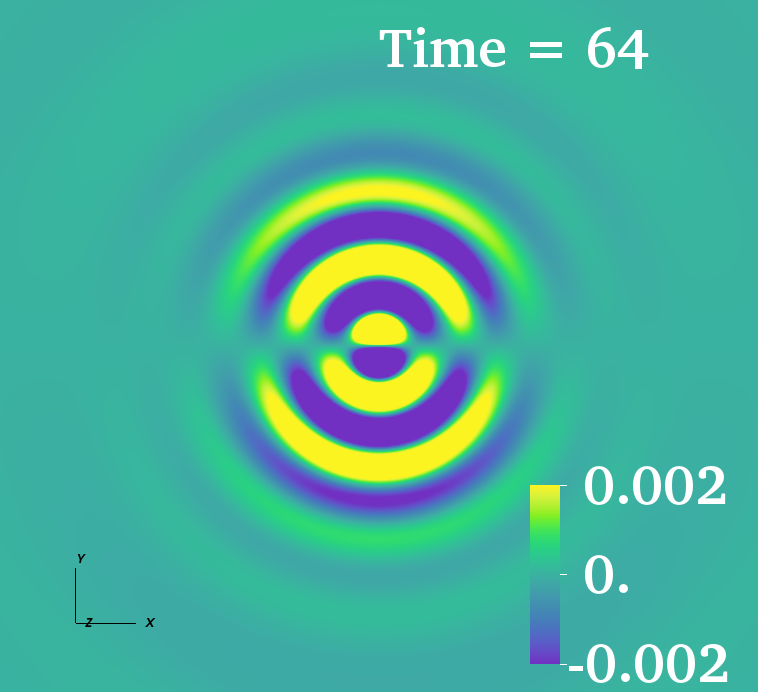}\hspace{-0.005\linewidth}
\includegraphics[width=0.3\linewidth]{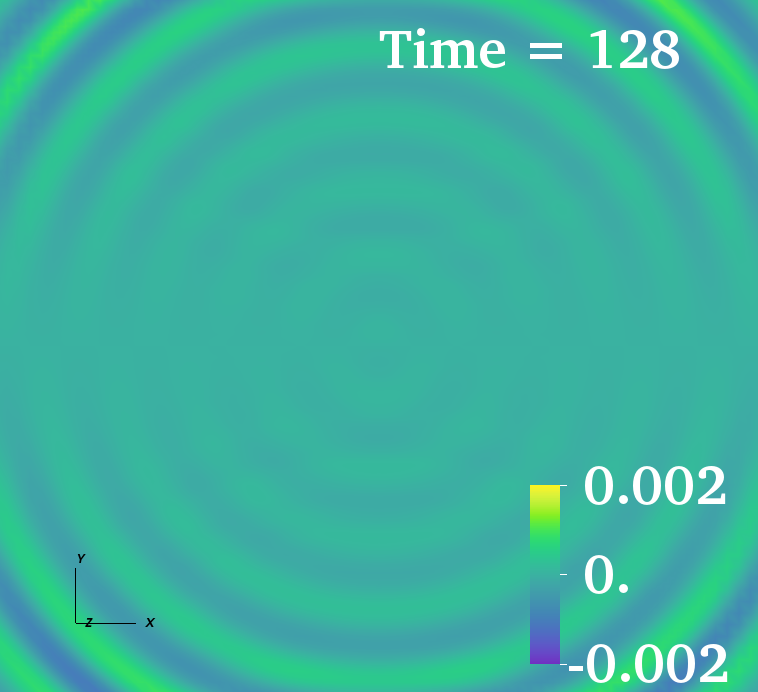}\\
\includegraphics[width=0.3\linewidth]{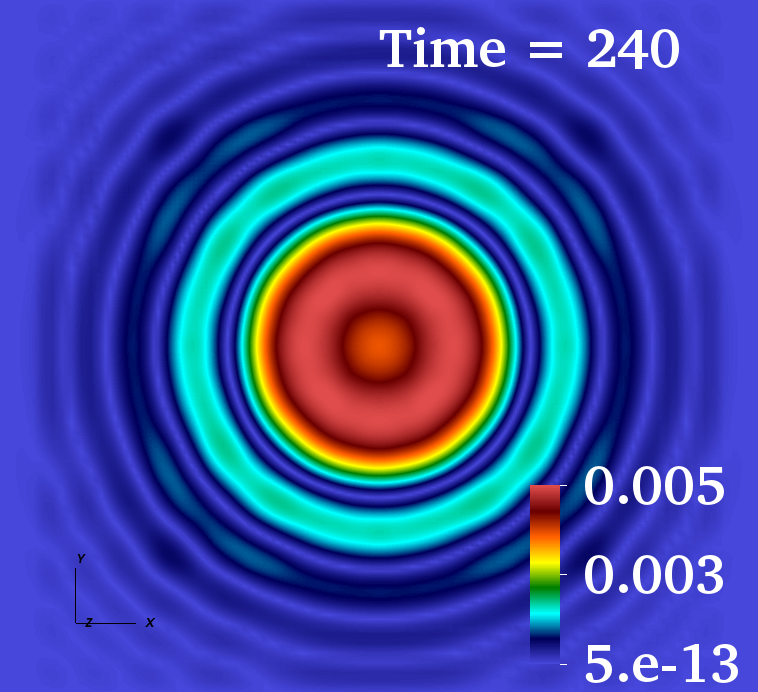}\hspace{-0.005\linewidth}
\includegraphics[width=0.3\linewidth]{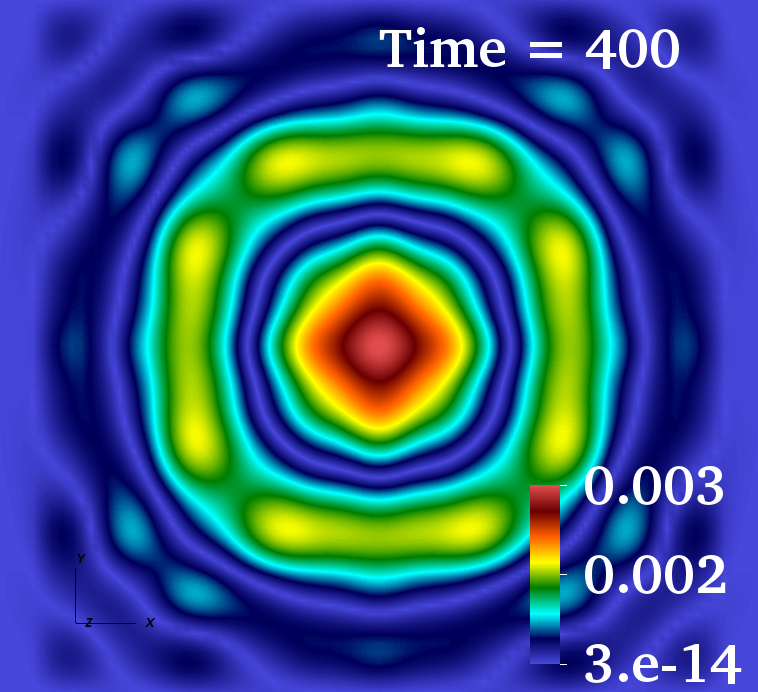}\hspace{-0.005\linewidth}
\includegraphics[width=0.3\linewidth]{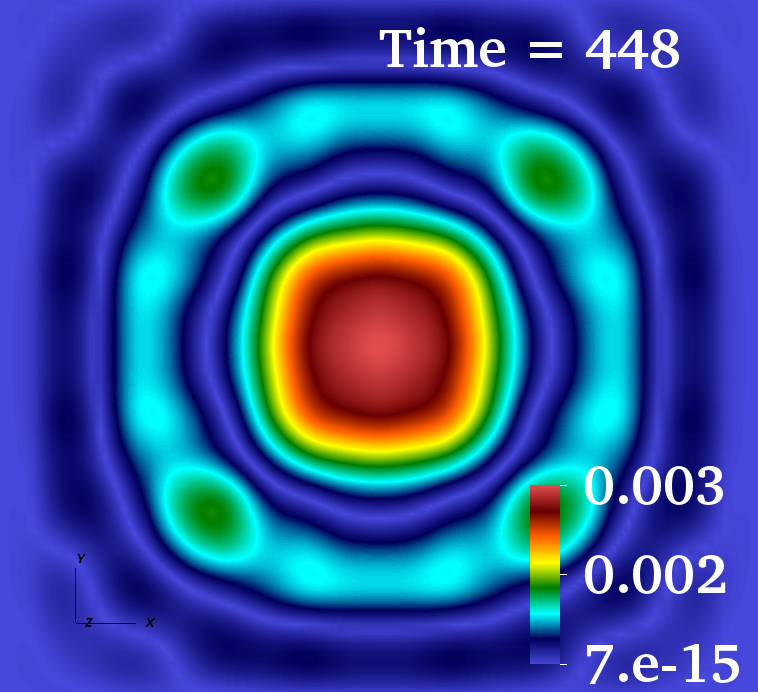}\\
\includegraphics[width=0.3\linewidth]{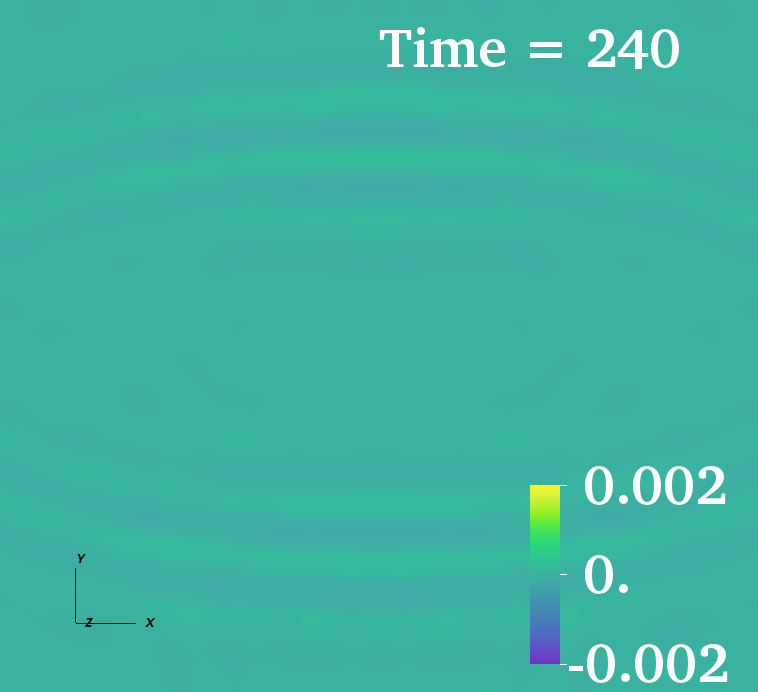}\hspace{-0.005\linewidth}
\includegraphics[width=0.3\linewidth]{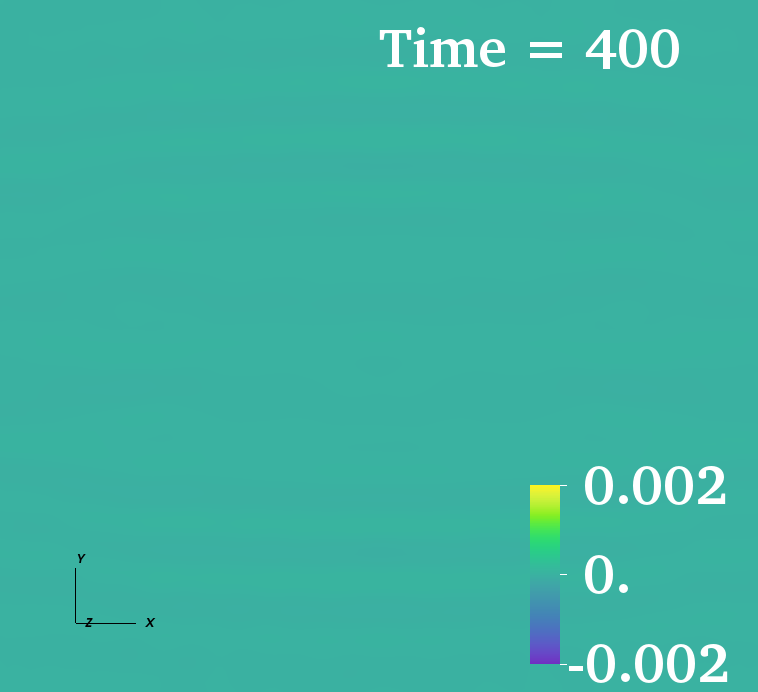}\hspace{-0.005\linewidth}
\includegraphics[width=0.3\linewidth]{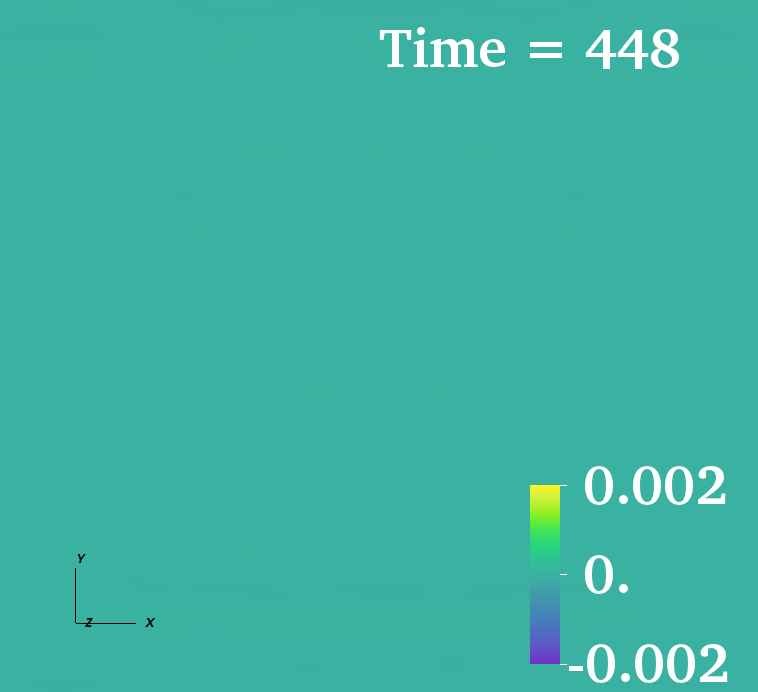}
\caption{Time evolution of the equatorial slice ($z=0$) of $|\Re(\Phi)|$ (first and third rows) and $E^x$ (second and fourth rows) of an isolated BSA configuration (cf. Table~\ref{tab:table1}). The axionic coupling is $k_{\rm{axion}}=0.43$ and time is given in code units where $\mu=1$.\label{fig1b}}
\end{center}
\end{figure}
\begin{figure}[htpb]
\includegraphics[width=0.45\textwidth]{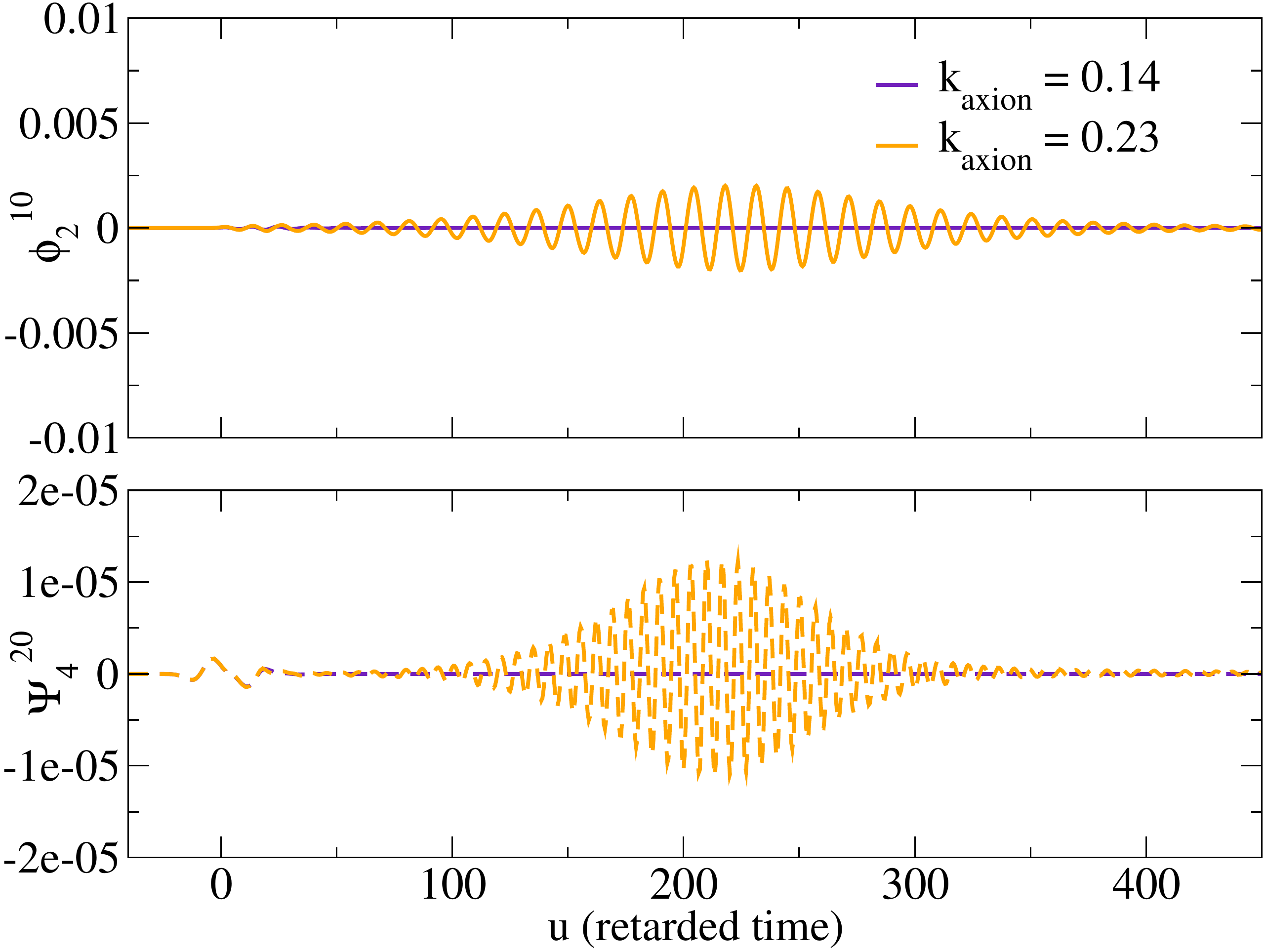} \\
\includegraphics[width=0.45\textwidth]{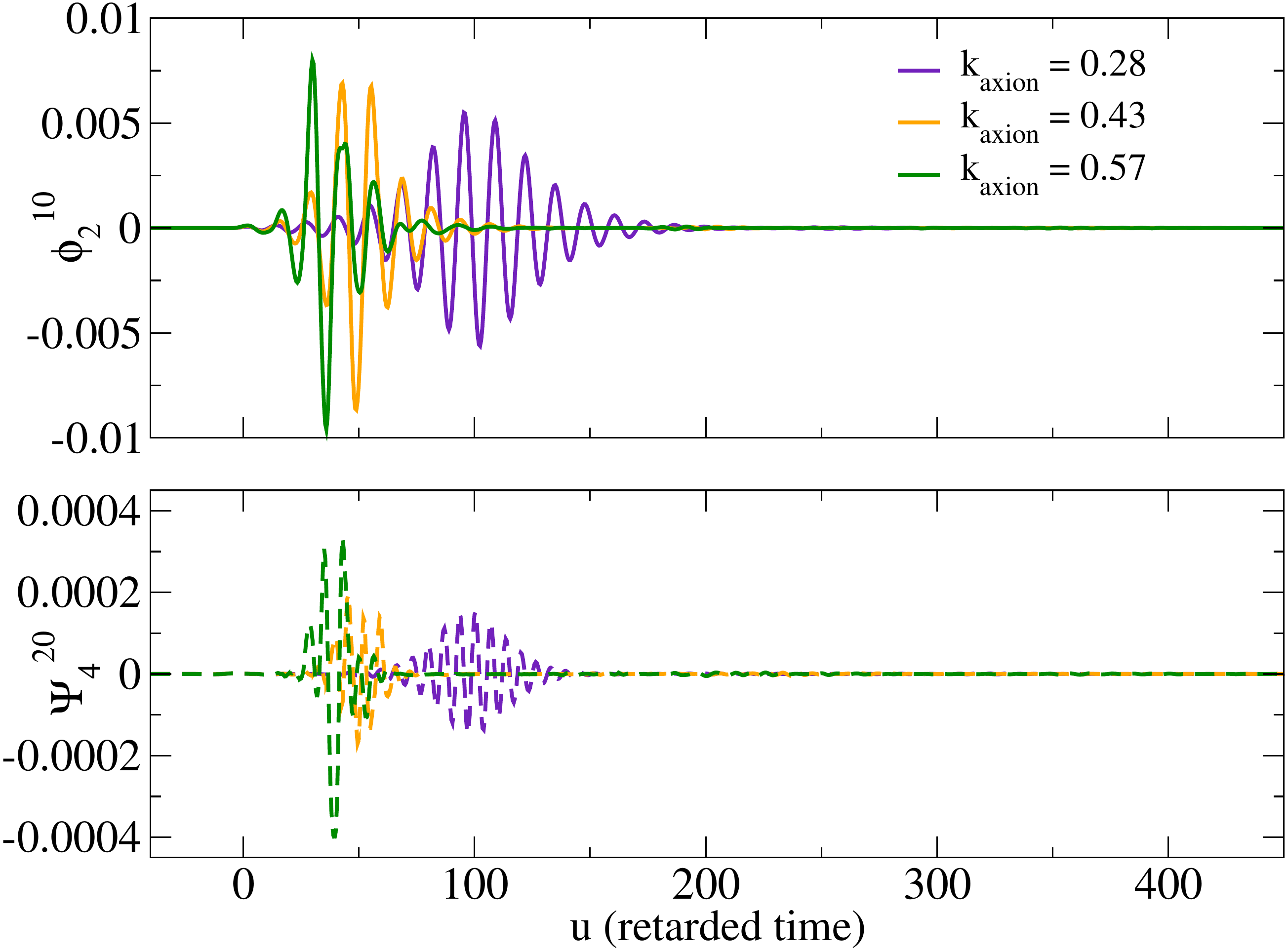} 
\caption{NP scalars $\phi_{2}^{lm}$ and $\Psi_{4}^{lm}$ as a function of the retarded time for the
evolution of BSA (cf. Table~\ref{tab:table1}), for different values of the axionic coupling extracted at $r=40$. \label{fig2}}
\end{figure}

We start by discussing results for the evolution of isolated BSs for different values of the coupling constant $k_{\rm axion}$ and the three different BS models defined in Table~\ref{tab:table1}. 
Our results are summarized in Figs.~\ref{fig1}--\ref{fig2}.
Snapshots of the evolution are shown in Figs.~\ref{fig1} and Figs.~\ref{fig1b}, for BSA ($\omega=0.914\mu$, cf. Table~\ref{tab:table1}) and couplings $k_{\rm axion}=0.23, 0.43$.
For BSA, the analytical estimate from Eq.~(\ref{eq:kstar}) for the coupling threshold gives us
\begin{equation}
\label{eq:kstar-BS1}
k_{\rm axion}^{\star {\rm ana}} \simeq \frac{0.584}{9.1 \times 0.142 \times \sqrt{2}} \simeq 0.32\,.
\end{equation}

As can be seen from these snapshots, both configurations are unstable. Indeed, for $k_{\rm axion}=0.43$ (see Fig.~\ref{fig1b}) soon after the beginning of the evolution, a burst of EM radiation is emitted and the star decays to a very dilute solution. This property can be seen in Table~\ref{tab:table2} where we show the energy radiated in GW and EM waves, and the final mass of the BS. We see that a large part of the initial BS energy is lost through EM emission with a frequency $\omega_{\rm{EM}}\sim\mu/2$, in agreement with the predictions for a parametric instability~\cite{Ikeda:2018nhb,Sen:2018cjt,Boskovic:2018lkj}.

For the coupling $k_{\rm{axion}}=0.23$, the BS also evolves towards a more dilute solution, but the final object is still more compact than the final BS of the previous case with $k_{\rm axion}=0.43$. The EM wave has a smaller amplitude but is emitted for longer time. The final object is a BS with a scalar field amplitude $\Phi_0$ that is below that of the critical value for this given value of the coupling ($k_{\rm{axion}}=0.23$), assuming Eq.~(\ref{eq:kstar}).

In Fig.~\ref{fig2} we show how the amplitude and morphology of the EM and the GW signals, $\phi_{2}^{10}$ and $\Psi_{4}^{20}$ respectively, change with increasing values of $k_{\rm axion}$: the amplitude grows with the coupling, while the width of the wave-packet decreases. The morphology of the GW signal follows that of the EM emission, suggesting that it is the EM field that produces the GWs. The dominant frequency remains almost the same for all values with $\omega_{\rm{EM}}\sim\mu/2$ for the EM waves. The instability is of parametric origin, according to Refs.~\cite{Ikeda:2018nhb,Sen:2018cjt,Boskovic:2018lkj}. On the other hand, the frequency of the GW is found to be $\omega_{\rm{GW}}\sim2\omega_{\rm{EM}}\sim\mu$, in line with the expectation that GWs are mostly driven by the EM field. For high coupling values, significant EM emission is triggered, which in turns triggers GW emission which is not present in the case without coupling. The threshold value of coupling for BSA is found to be around 
\be
k^{\star{\rm num}}_{\rm{axion}}\sim 0.2\,,
\ee
a factor of 1.5 smaller than the flat space estimate [Eq.~(\ref{eq:kstar})]. Below the critical value there is almost no emission of EM waves, mainly the one associated with the initial pulse. The critical value implies that for large couplings it would be impossible to form compact
BSs, since they would decay to a less compact solution emitting EM
waves in the process.

We have repeated the analysis for the three less compact models of BS in Table~\ref{tab:table1} with $M_{\rm{BS}}\mu=0.512$, $M_{\rm{BS}}\mu=0.403$, and $M_{\rm{BS}}\mu=0.328$, finding that the threshold is also within a factor between 1.5 and 2 with respect to the flat space estimate Eq.~(\ref{eq:kstar}). At large coupling, the end state of the evolution of an isolated BS is not universal. As might be anticipated, the star emits larger amounts of EM and GW radiation for larger couplings, since it is unstable. Thus, the end state is less massive and more dilute for larger $k_{\rm axion}$.

\begin{table}[htb]
\caption{Energy emission in the EM and GW channel of an {\it isolated} BSA star, for different couplings. As we remark in the main text, for large couplings GW emission is triggered by a parametric instability in the EM sector, and is therefore subdominant.\label{tab:table2}}
\begin{ruledtabular}
\begin{tabular}{cccc}
$k_{\rm{axion}}$&GW&EM&$M_{\rm{BS}}^{\rm{final}}$\\
\hline
0.14&$3\times10^{-8}$&0.013&0.58\\
0.23&$8\times10^{-5}$&0.082&0.47\\
0.28&$3\times10^{-4}$&0.188&0.30\\
0.43&$4\times10^{-4}$&0.212&0.20\\
0.57&$4\times10^{-4}$&0.301&0.19\\
\end{tabular}
\end{ruledtabular}
\end{table}
%

\subsection{Head-on collisions of boson stars}
\begin{figure*}[thp]
\includegraphics[width=0.321\textwidth]{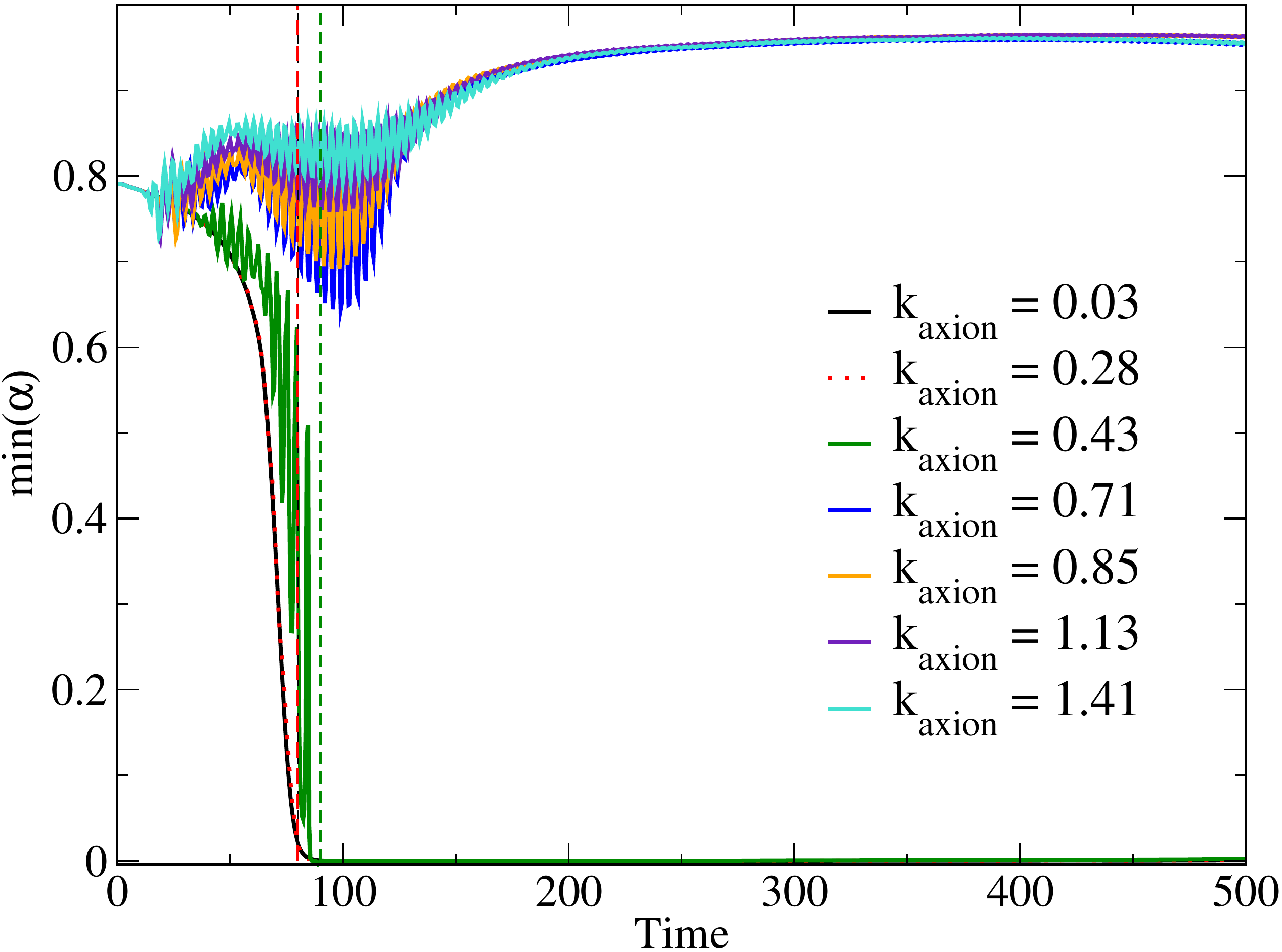} 
\includegraphics[width=0.321\textwidth]{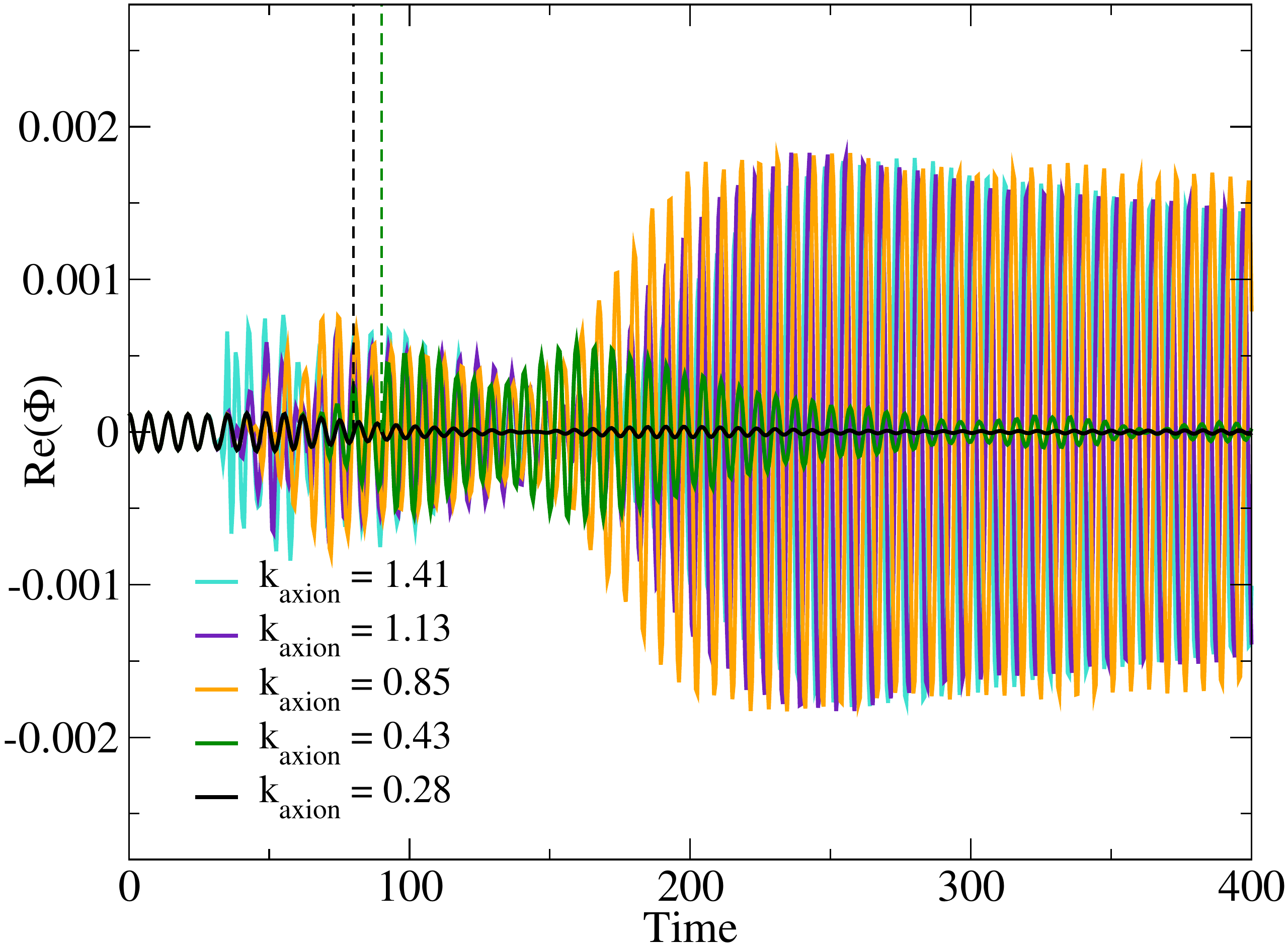} 
\includegraphics[width=0.321\textwidth]{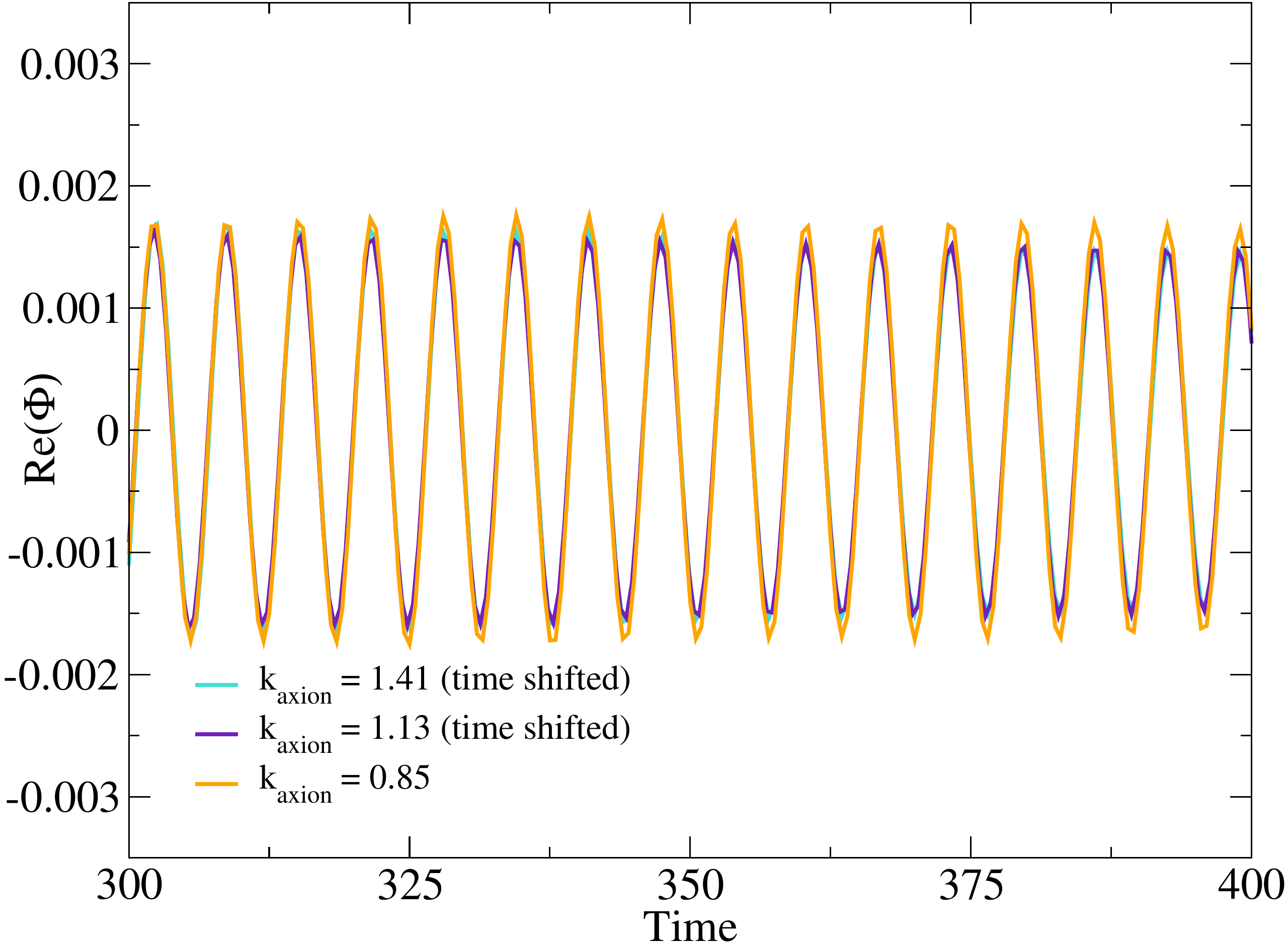} 
\caption{Collisions of two BSA configurations.
{\bf Left panel:} minimum value of the lapse function $\alpha$ for different
  values of the coupling $k_{\rm{axion}}$. Middle panel: Time evolution of the amplitude of the real
  part of the scalar field extracted at $(x, y, z) = (12, 12, 12)$. {\bf Right panel:}
  Detail of the amplitude of the real part of the scalar field for the three largest couplings. Vertical lines mark the appearance of a common apparent horizon, for those configurations that collapse to a BH.
  \label{fig3}}
\end{figure*}

\begin{figure}[thp]
\begin{center}
\begin{tabular}{ p{0.32\linewidth}  }
\centering  $k_{\rm{axion}}=0$
\end{tabular}\\
\includegraphics[width=0.3\linewidth]{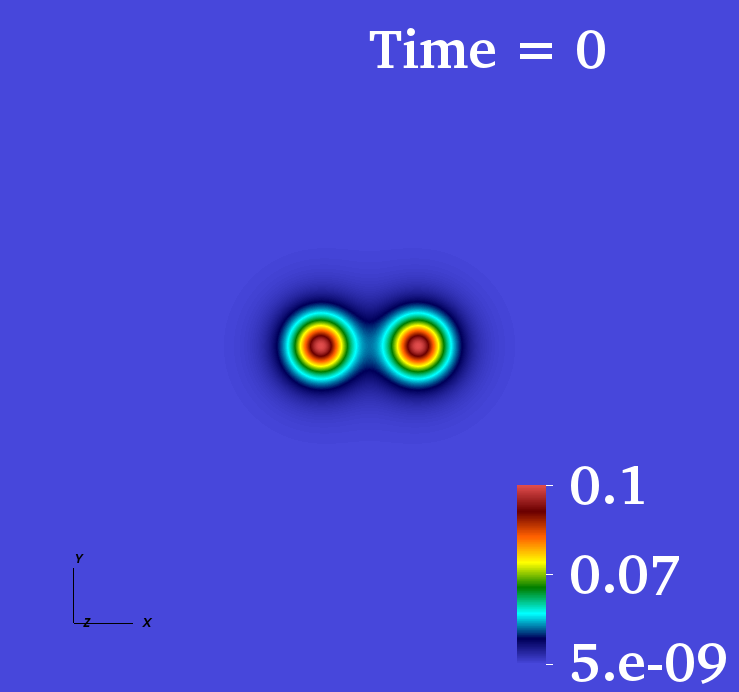}\hspace{-0.005\linewidth}
\includegraphics[width=0.3\linewidth]{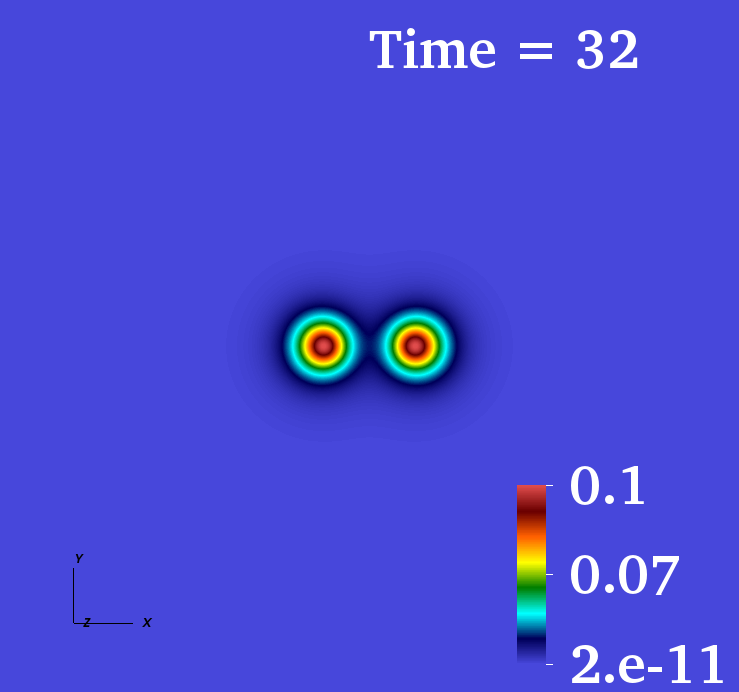}\hspace{-0.005\linewidth}
\includegraphics[width=0.3\linewidth]{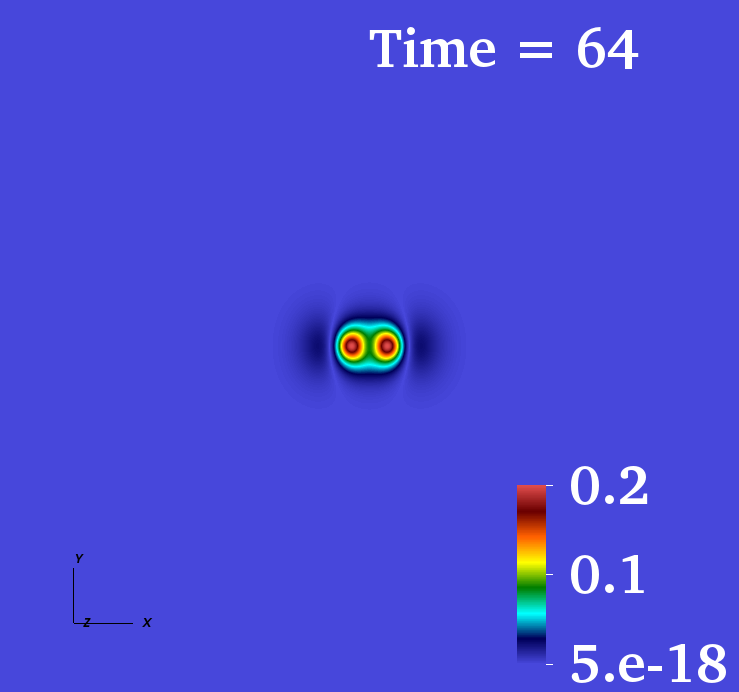}\\
\includegraphics[width=0.3\linewidth]{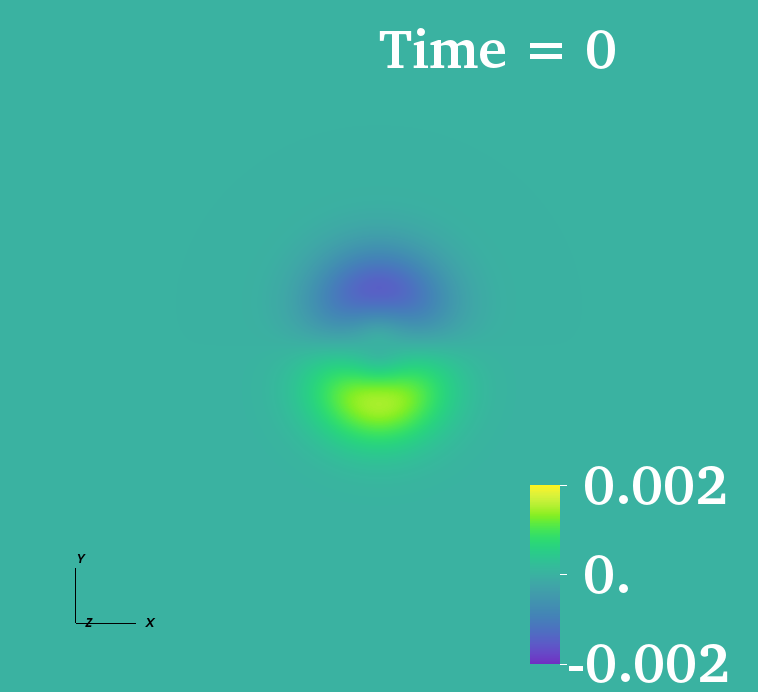}\hspace{-0.005\linewidth}
\includegraphics[width=0.3\linewidth]{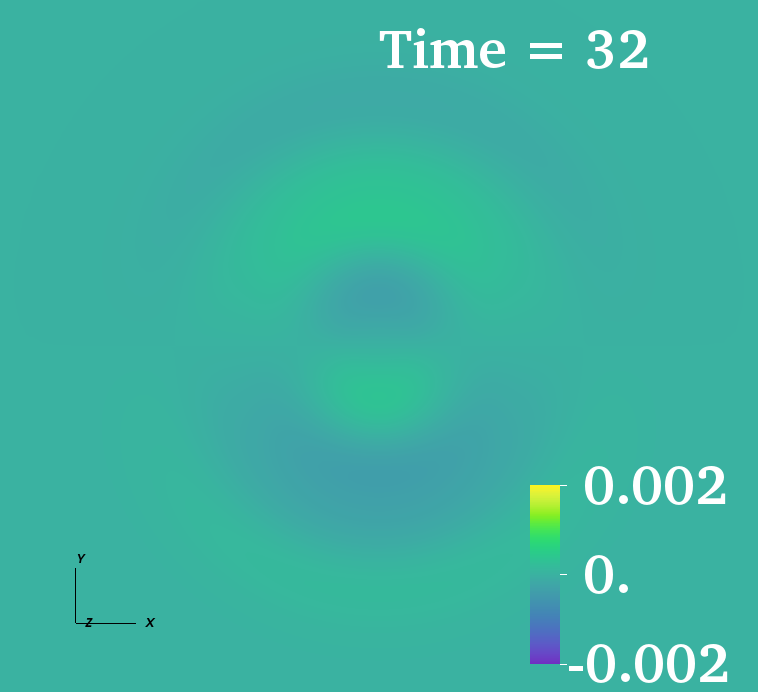}\hspace{-0.005\linewidth}
\includegraphics[width=0.3\linewidth]{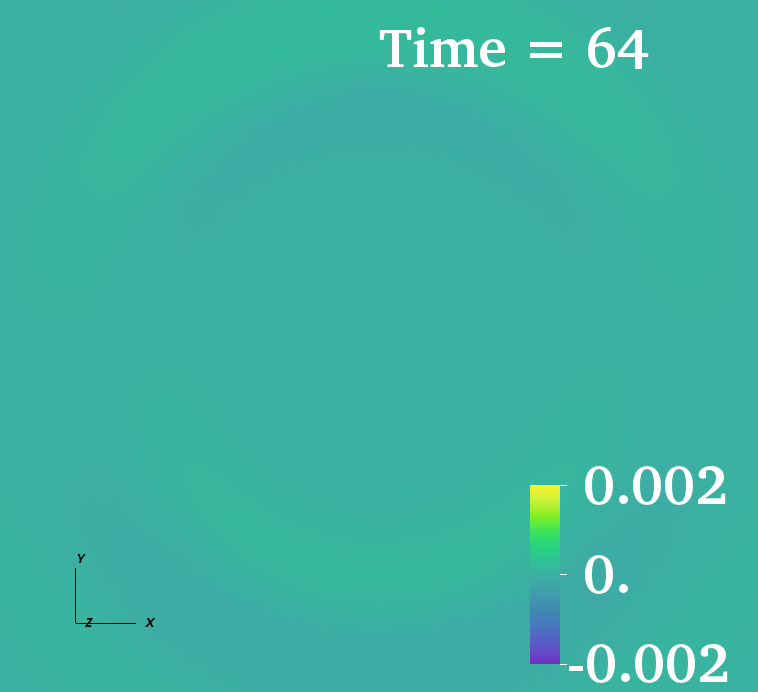}\\
\includegraphics[width=0.3\linewidth]{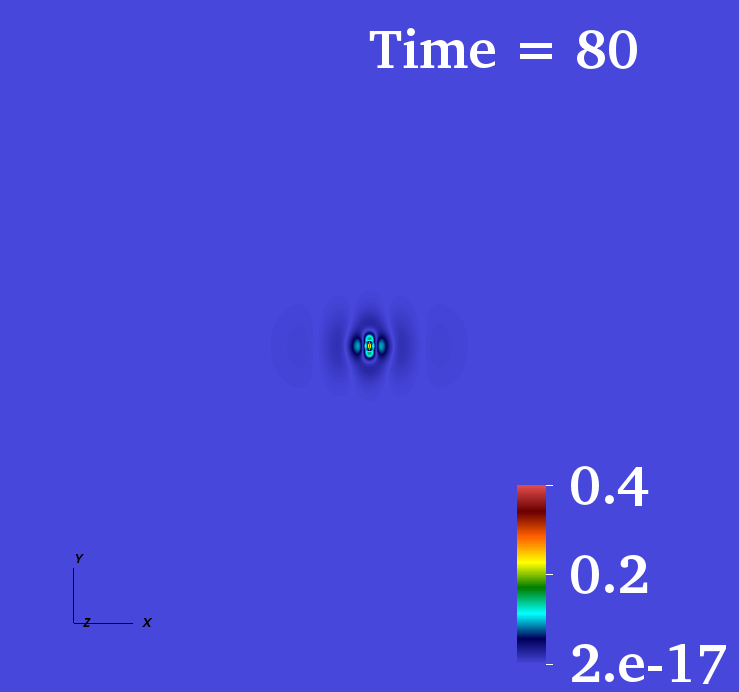}\hspace{-0.005\linewidth}
\includegraphics[width=0.3\linewidth]{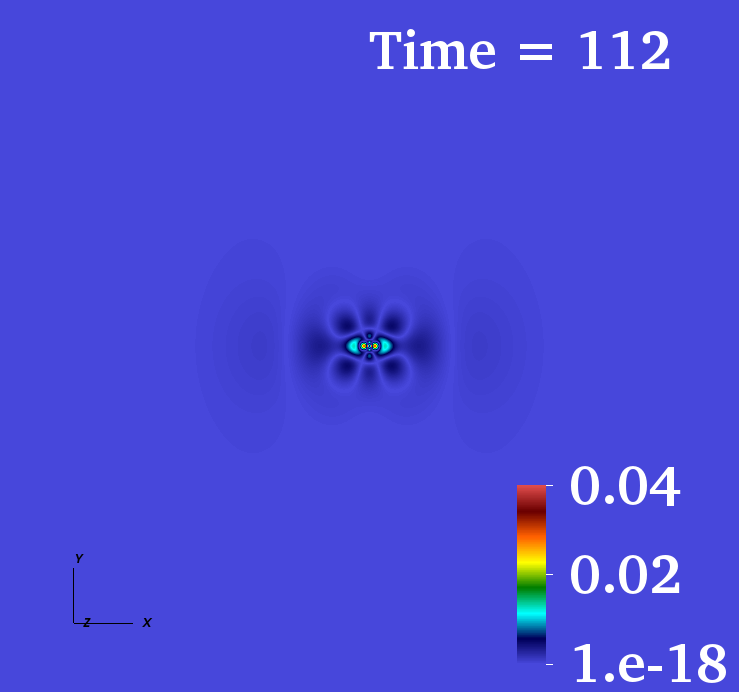}\hspace{-0.005\linewidth}
\includegraphics[width=0.3\linewidth]{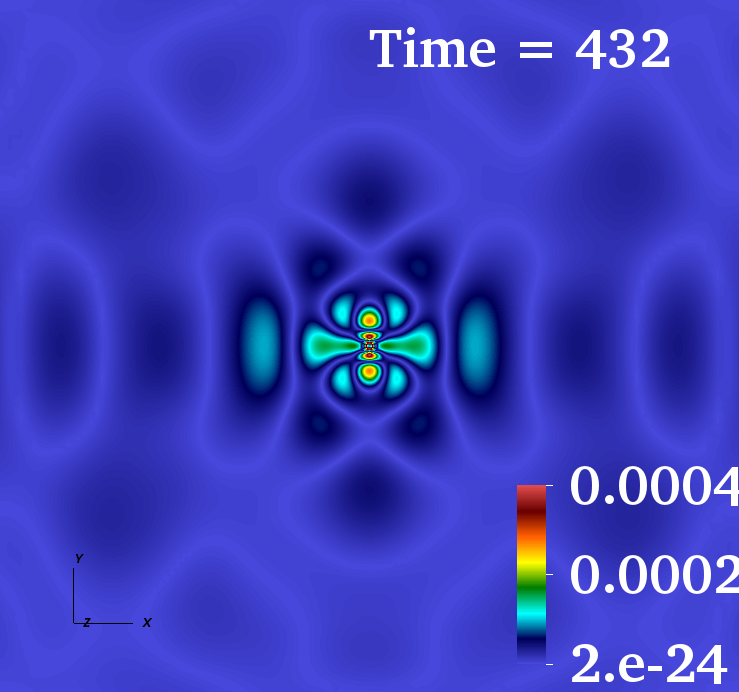}\\
\includegraphics[width=0.3\linewidth]{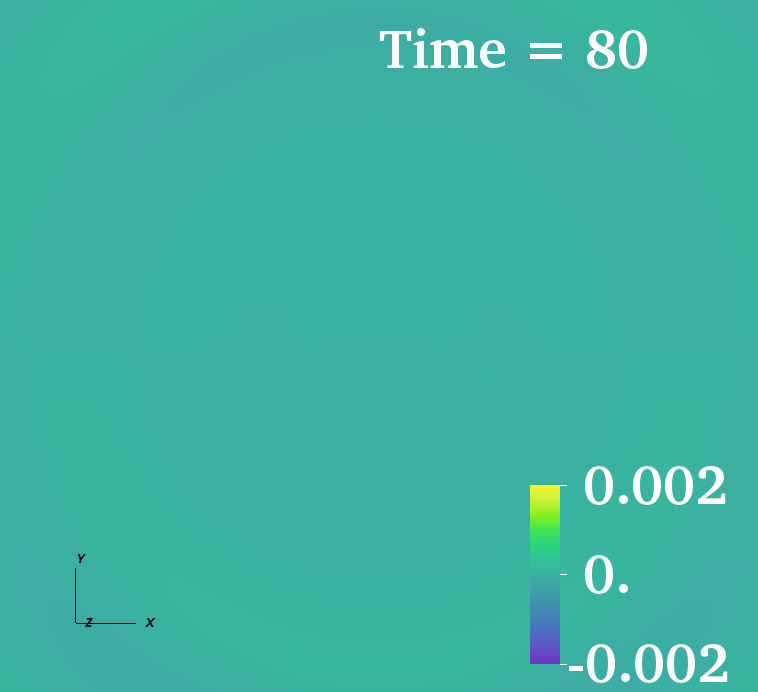}\hspace{-0.005\linewidth}
\includegraphics[width=0.3\linewidth]{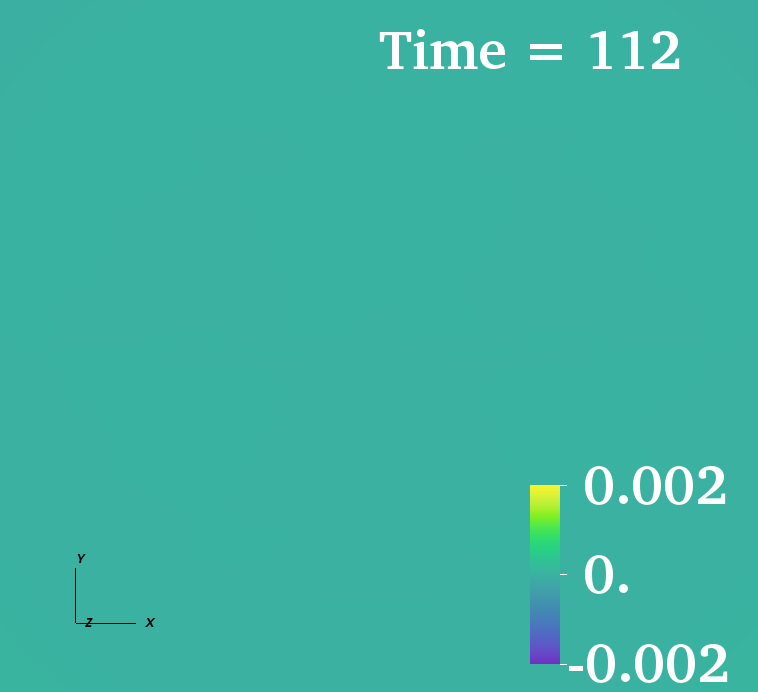}\hspace{-0.005\linewidth}
\includegraphics[width=0.3\linewidth]{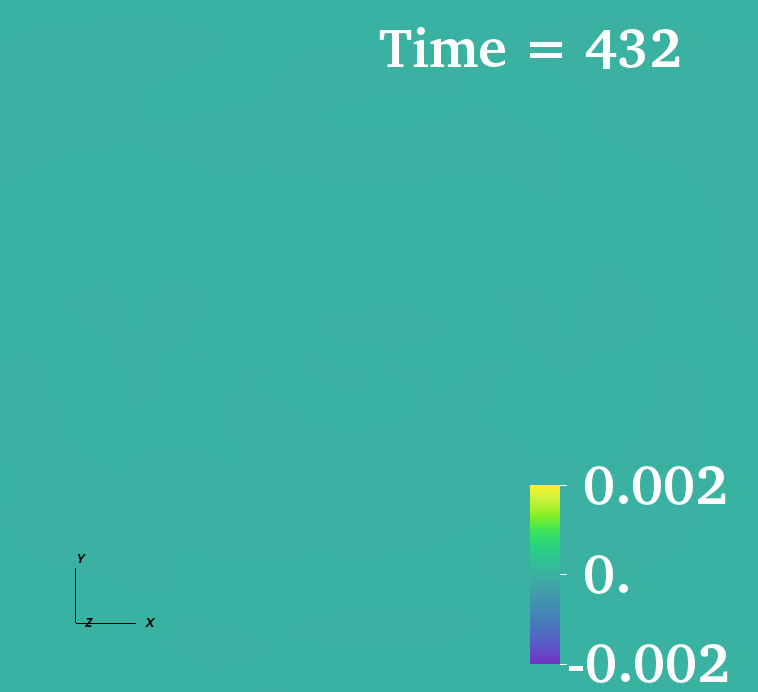}\\
\smallskip
\begin{tabular}{ p{0.32\linewidth}  }
\centering   $k_{\rm{axion}}=0.43$
\end{tabular}\\
\includegraphics[width=0.3\linewidth]{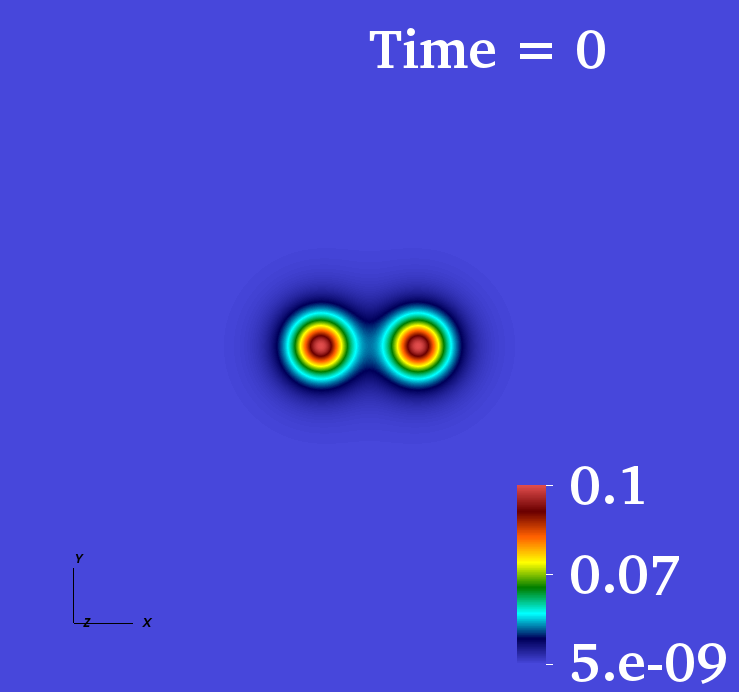}\hspace{-0.005\linewidth}
\includegraphics[width=0.3\linewidth]{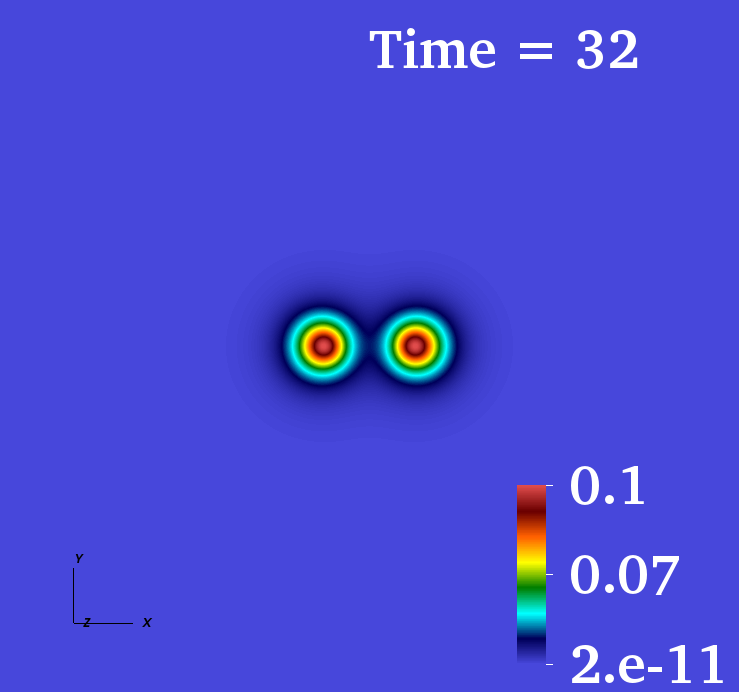}\hspace{-0.005\linewidth}
\includegraphics[width=0.3\linewidth]{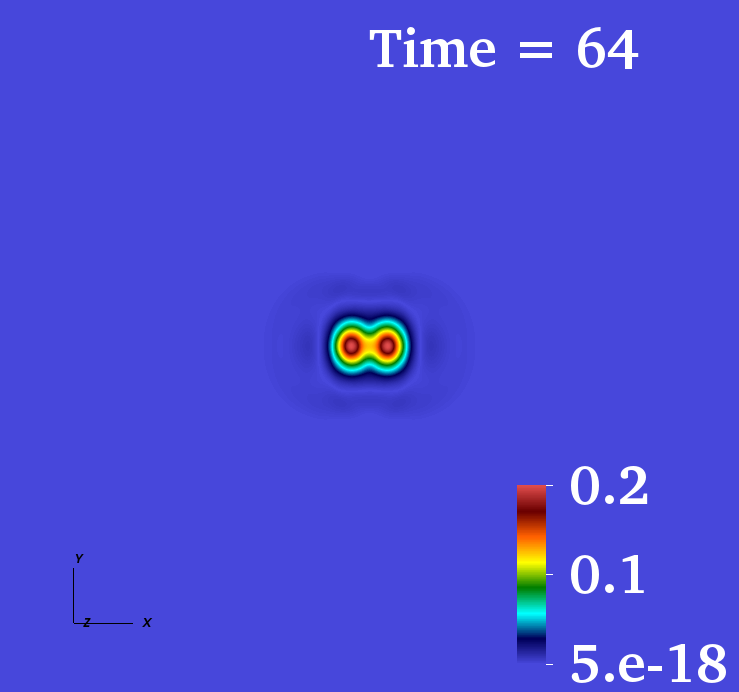}\\
\includegraphics[width=0.3\linewidth]{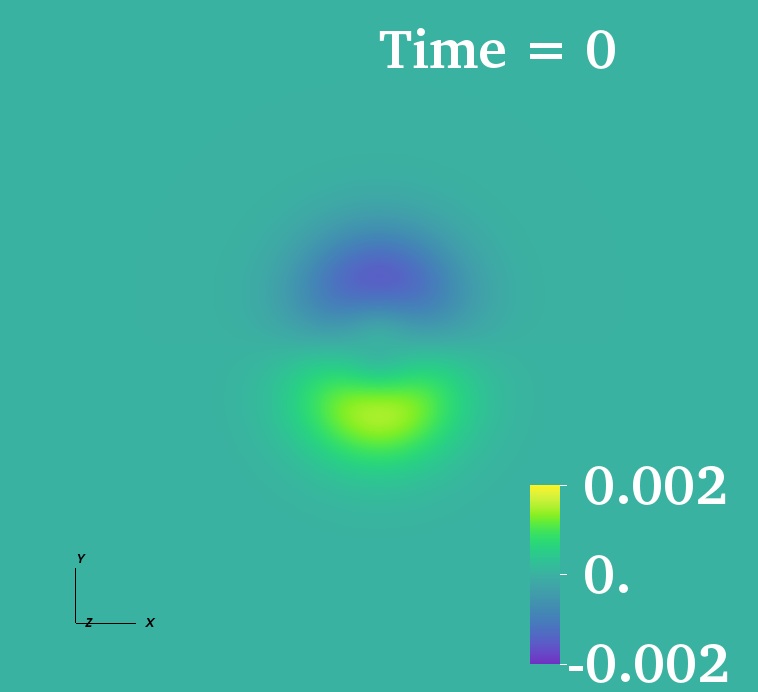}\hspace{-0.005\linewidth}
\includegraphics[width=0.3\linewidth]{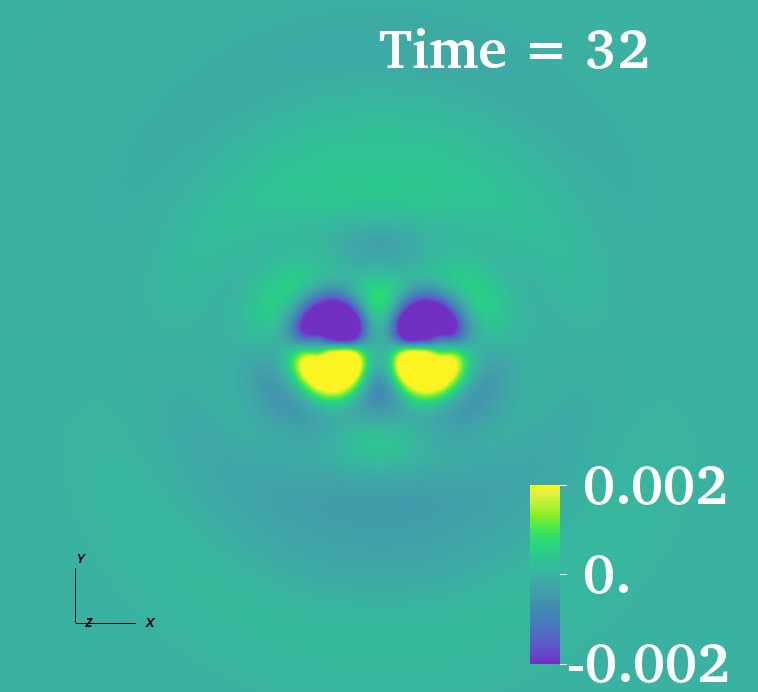}\hspace{-0.005\linewidth}
\includegraphics[width=0.3\linewidth]{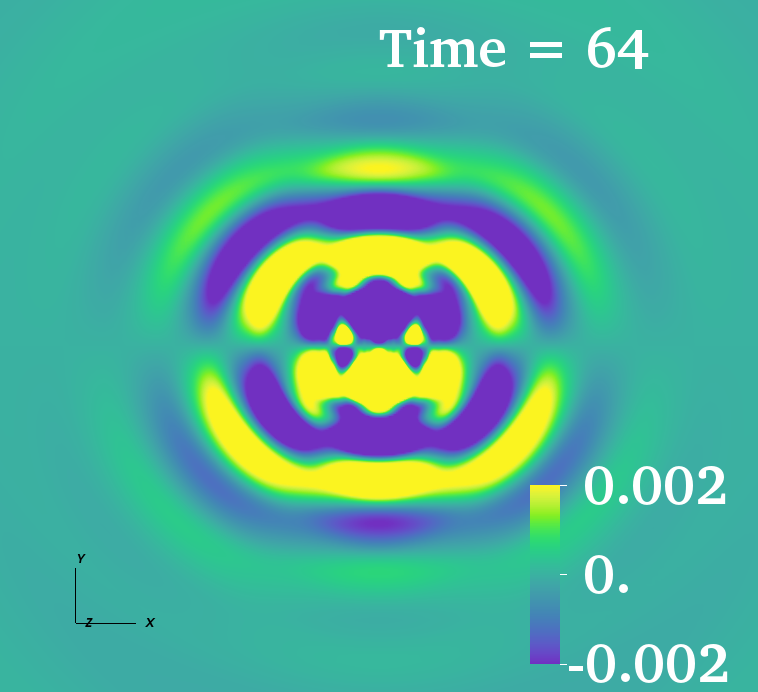}\\
\includegraphics[width=0.3\linewidth]{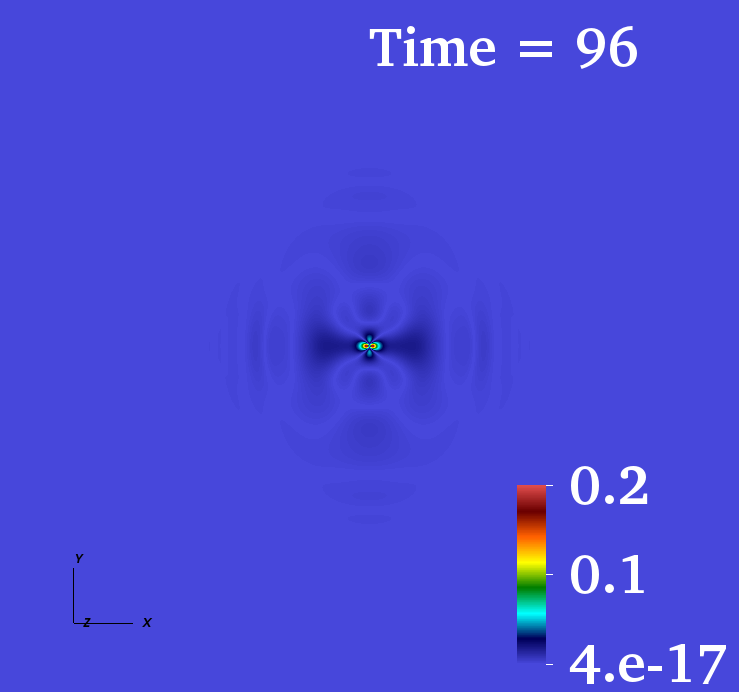}\hspace{-0.005\linewidth}
\includegraphics[width=0.3\linewidth]{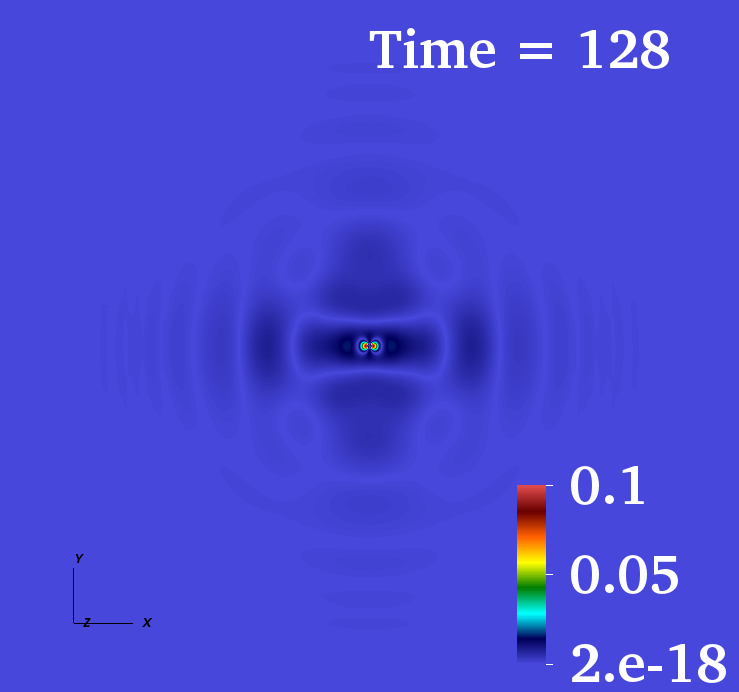}\hspace{-0.005\linewidth}
\includegraphics[width=0.3\linewidth]{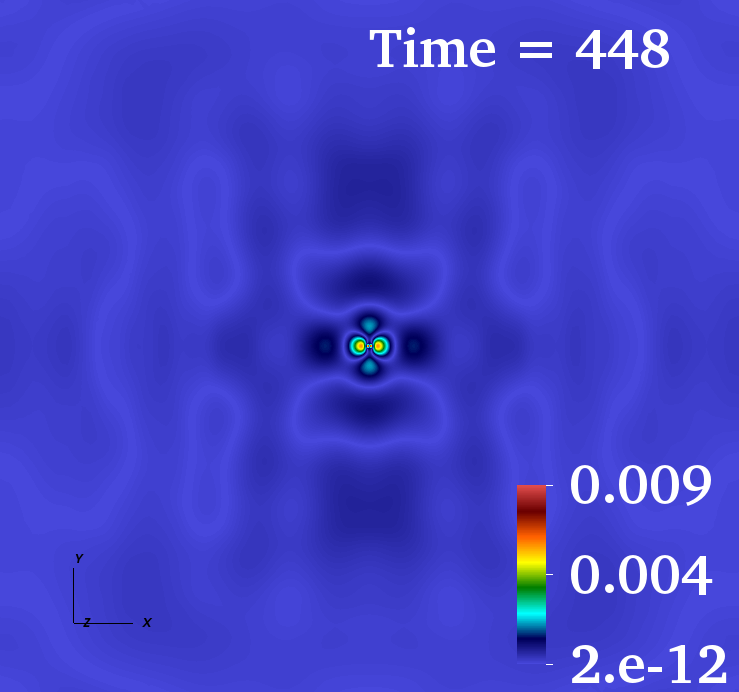}\\
\includegraphics[width=0.3\linewidth]{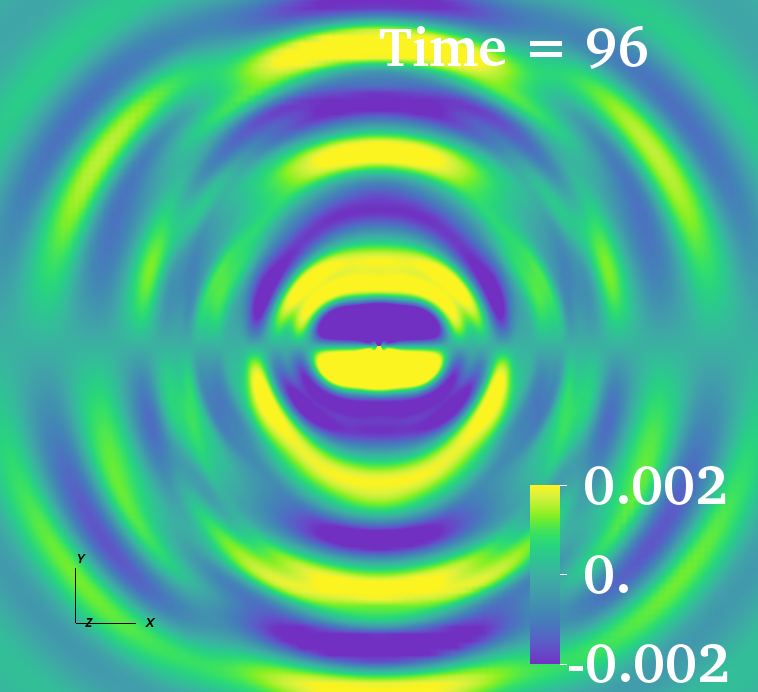}\hspace{-0.005\linewidth}
\includegraphics[width=0.3\linewidth]{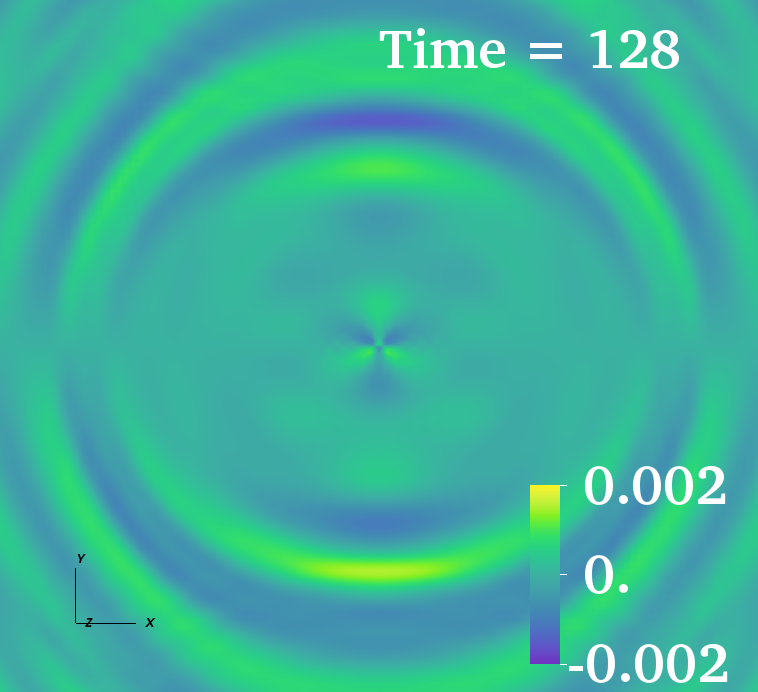}\hspace{-0.005\linewidth}
\includegraphics[width=0.3\linewidth]{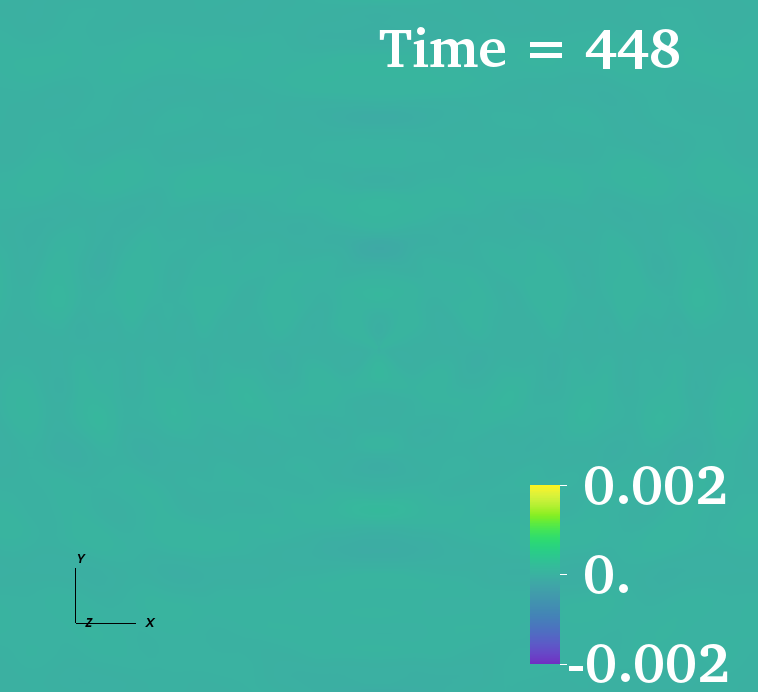}
\caption{Snapshots of the time evolution in the equatorial slice ($z=0$) of $|\Re(\Phi)|$ during the head-on collision of two BSA (cf. Table~\ref{tab:table1}). The $x$ component of the electric field is plotted on top of the energy density. The axionic coupling is $k_{\rm{axion}}=0$ (top panels) and $k_{\rm{axion}}=0.43$ (bottom panels). Time is given in code units.\label{fig:kk0} }
\end{center}
\end{figure}

After considering the evolution of isolated stars and their critical threshold, we have investigated how such BSs behave dynamically. In particular, we studied fully nonlinear head-on collisions, including GW emission. We have performed several such collisions of equal-mass BSs, corresponding to BSA and BSD in Table~\ref{tab:table1}, varying the value of the coupling to study how the collision dynamics, the outcome of the merger, and the GW emission are affected. Our results are summarized in Figs.~\ref{fig3}--\ref{fig10}.

The previous section showed that there is a critical value for the coupling above which the BS becomes unstable and decays to a dilute (less compact) solution. It is then possible to avoid the formation of a BH (BSA) or trigger the axionic instability of two stable BSs (BSD) {\it after} the merger. Interestingly, as we will observe, for collisions with large coupling values, the GW emission comes mainly from the EM field due to the energy transfer from the scalar field to the EM field.

\subsubsection{BSA case: the collision of two compact stars}
We first consider the head-on collision of two BSA stars that in the zero coupling case leads to BH formation. In the left panel of Fig.~\ref{fig3} we plot the time evolution of the minimum value of the lapse function $\alpha$ and the amplitude of the real part of scalar field for several values of $k_{\rm{axion}}$.
When the minimum value of $\alpha$ drops below a certain value it typically signals horizon formation, which happens for $k_{\rm{axion}} = \lbrace{0.03, 0.28, 0.43}\rbrace$, indicating the BS collapse to a BH after the collision. However, for $k_{\rm{axion}}\geq0.71$, the collapse is prevented and the outcome is a dilute BS. A large EM emission is triggered even before the collision. In this case, the final BS configuration obtained after the merger seems to be the same (exhibiting almost the same oscillation frequency of the field and minimum value of the lapse) regardless of the value of the coupling $k_{\rm axion}$ chosen (see middle and right panels of Fig.~\ref{fig3}).

\begin{figure}[tp]
\begin{center}
\begin{tabular}{ p{0.32\linewidth}  }
\centering   $k_{\rm{axion}}=0.71$
\end{tabular}\\
\includegraphics[width=0.295\linewidth]{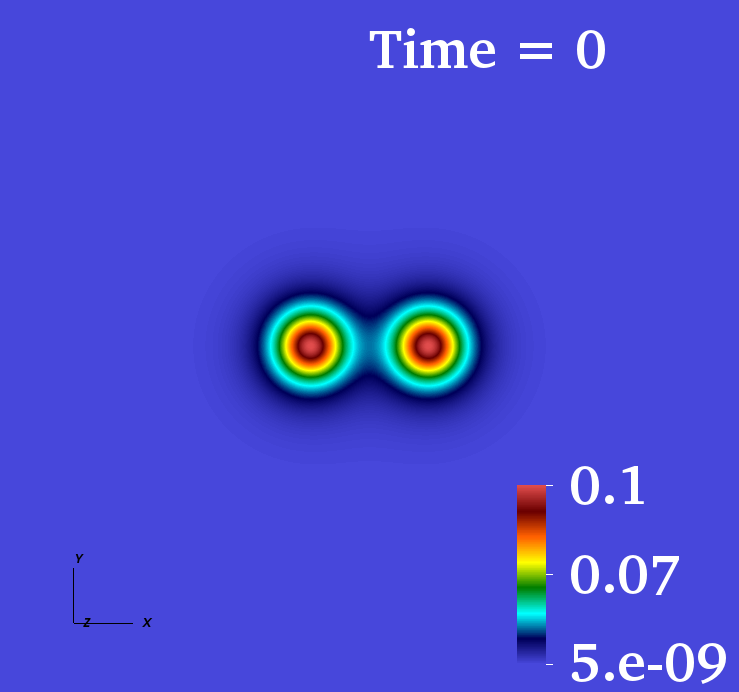}\hspace{-0.005\linewidth}
\includegraphics[width=0.295\linewidth]{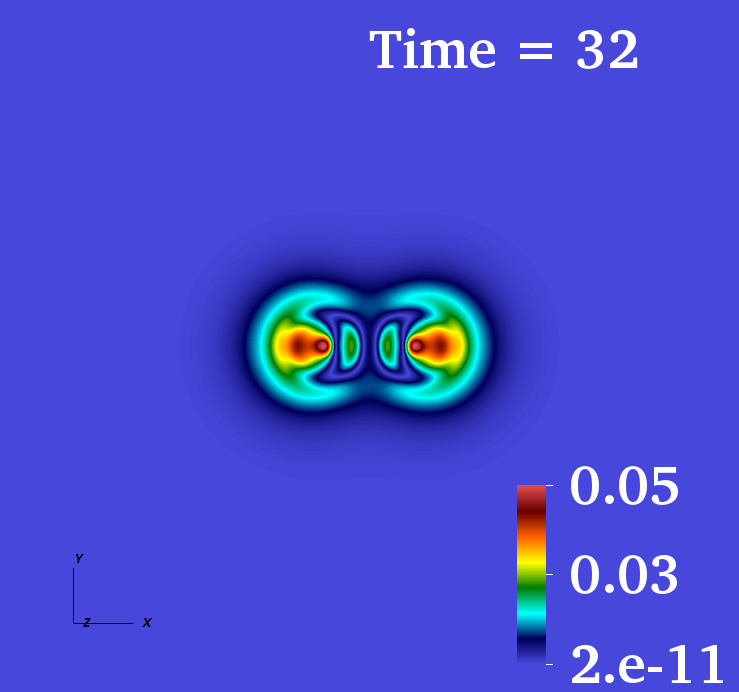}\hspace{-0.005\linewidth}
\includegraphics[width=0.295\linewidth]{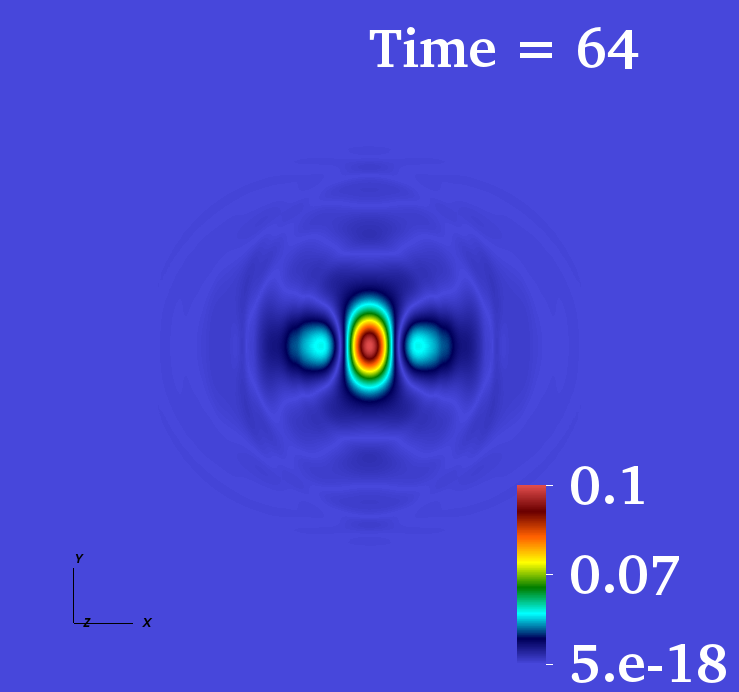}\\
\includegraphics[width=0.295\linewidth]{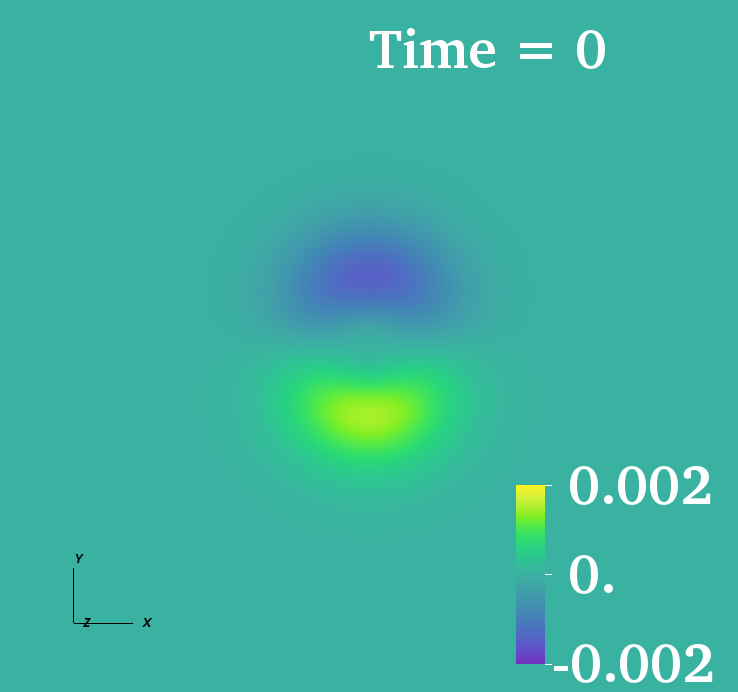}\hspace{-0.005\linewidth}
\includegraphics[width=0.295\linewidth]{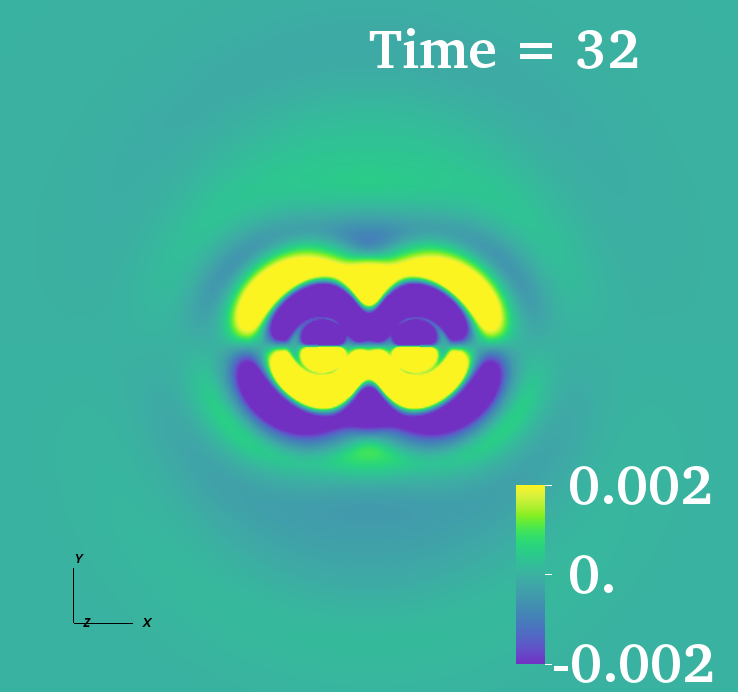}\hspace{-0.005\linewidth}
\includegraphics[width=0.295\linewidth]{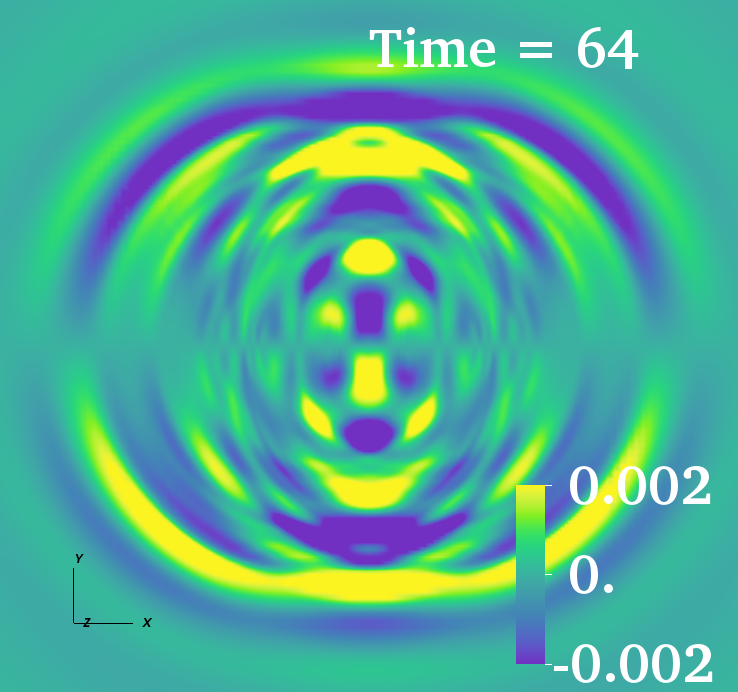}\\
\includegraphics[width=0.295\linewidth]{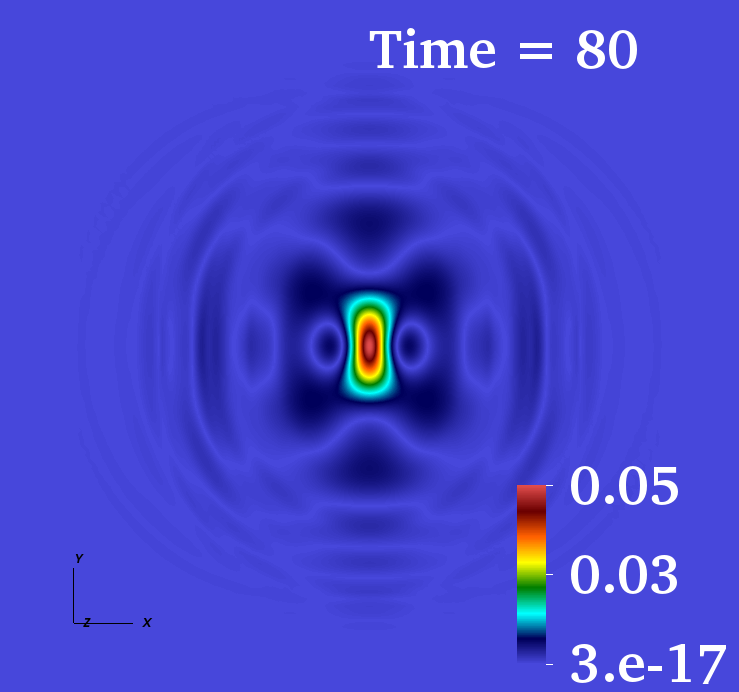}\hspace{-0.005\linewidth}
\includegraphics[width=0.295\linewidth]{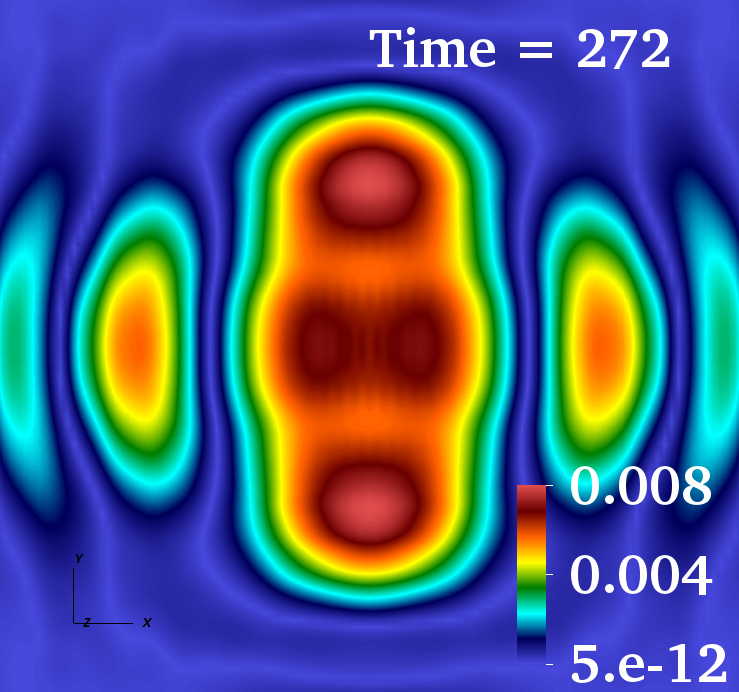}\hspace{-0.005\linewidth}
\includegraphics[width=0.295\linewidth]{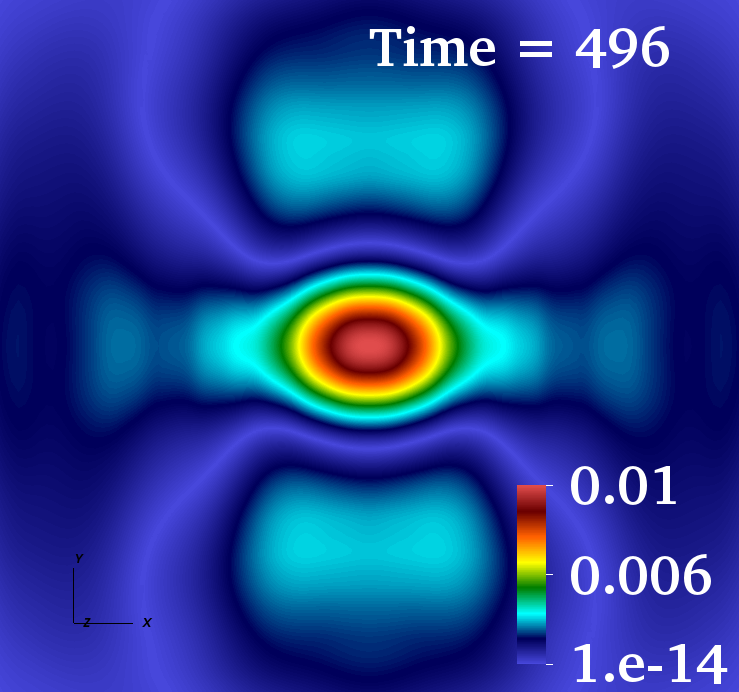}\\
\includegraphics[width=0.295\linewidth]{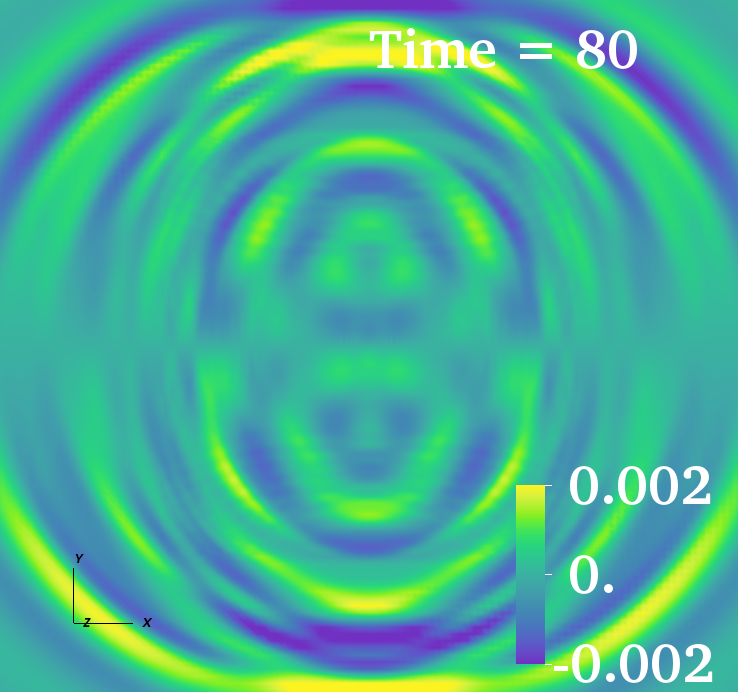}\hspace{-0.005\linewidth}
\includegraphics[width=0.295\linewidth]{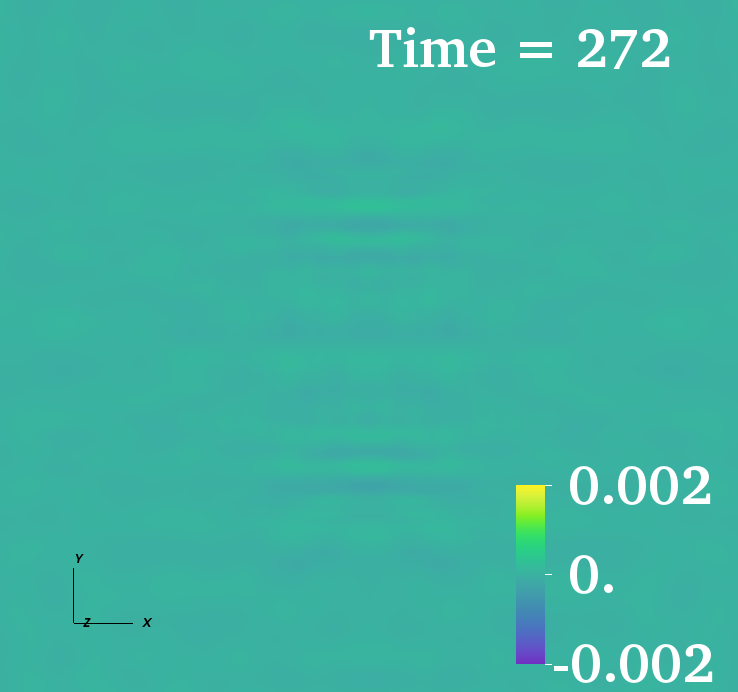}\hspace{-0.005\linewidth}
\includegraphics[width=0.295\linewidth]{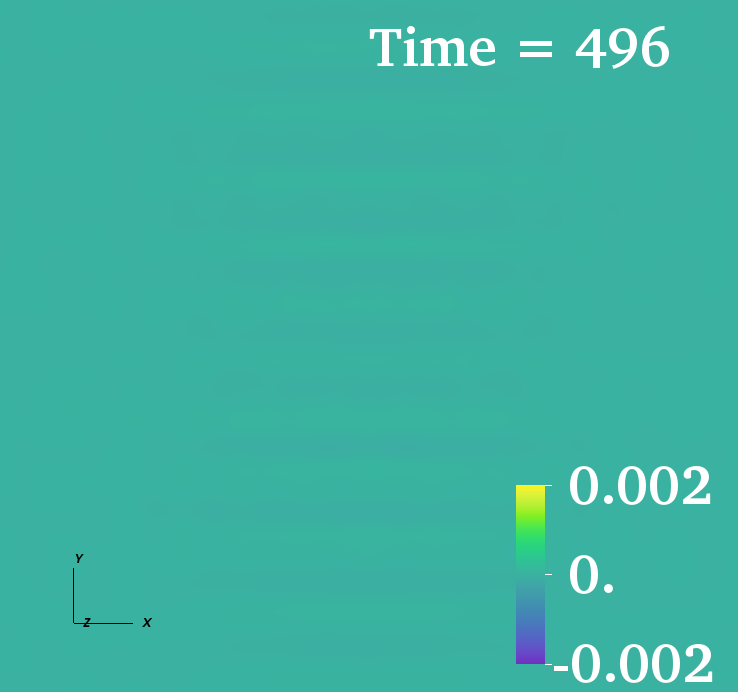}\\
\smallskip
\begin{tabular}{ p{0.32\linewidth}  }
\centering   $k_{\rm{axion}}=1.13$
\end{tabular}\\
\includegraphics[width=0.295\linewidth]{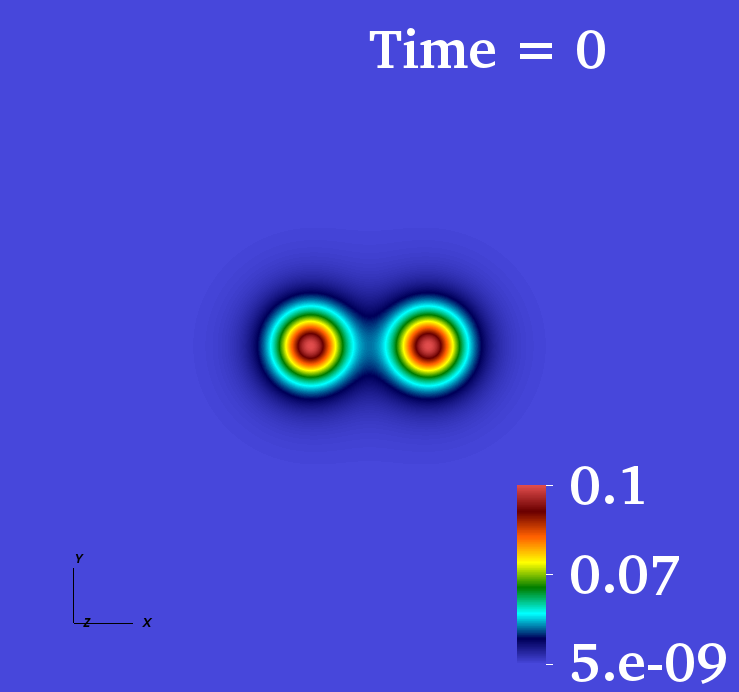}\hspace{-0.005\linewidth}
\includegraphics[width=0.295\linewidth]{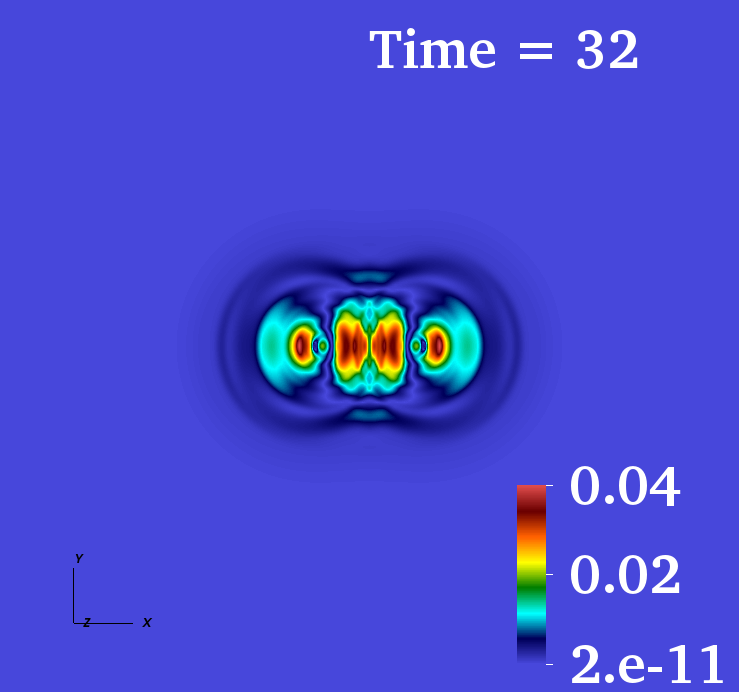}\hspace{-0.005\linewidth}
\includegraphics[width=0.295\linewidth]{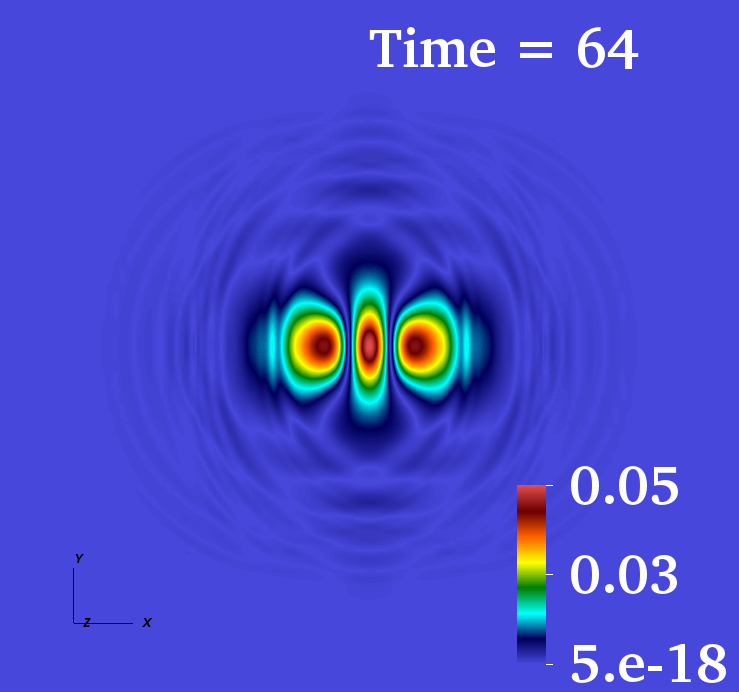}\\
\includegraphics[width=0.295\linewidth]{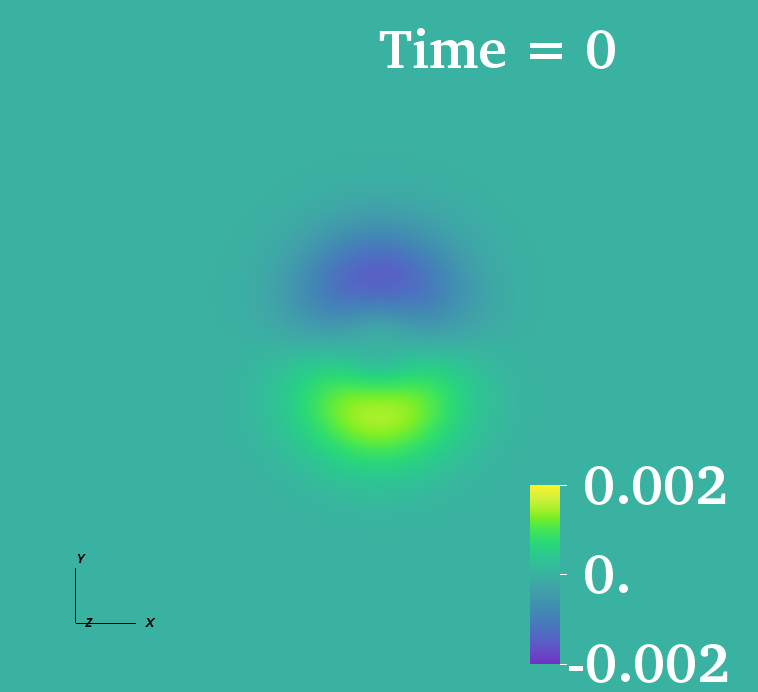}\hspace{-0.005\linewidth}
\includegraphics[width=0.295\linewidth]{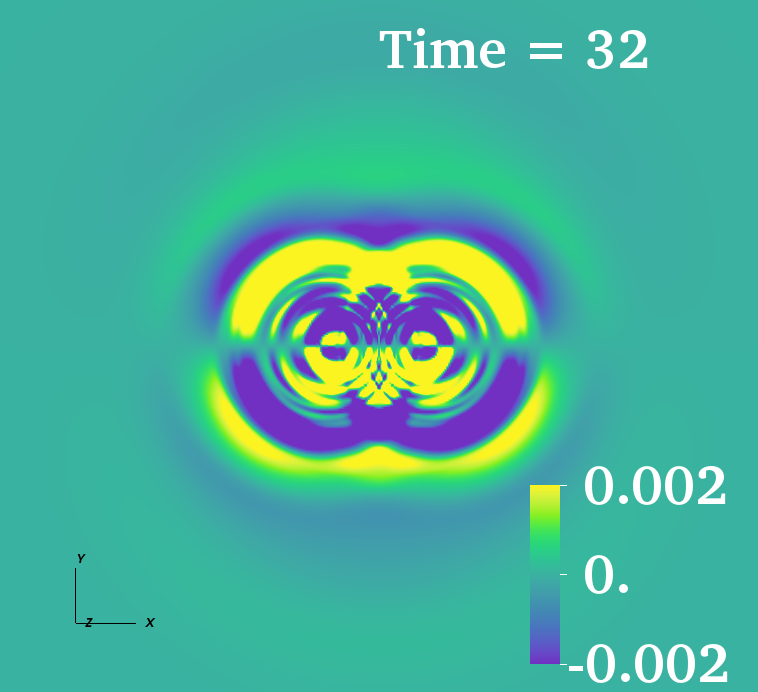}\hspace{-0.005\linewidth}
\includegraphics[width=0.295\linewidth]{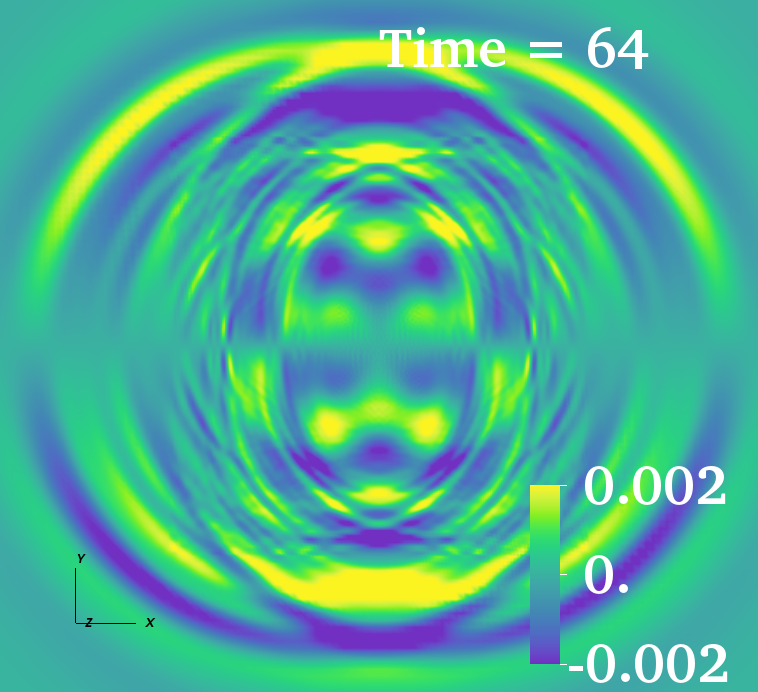}\\
\includegraphics[width=0.295\linewidth]{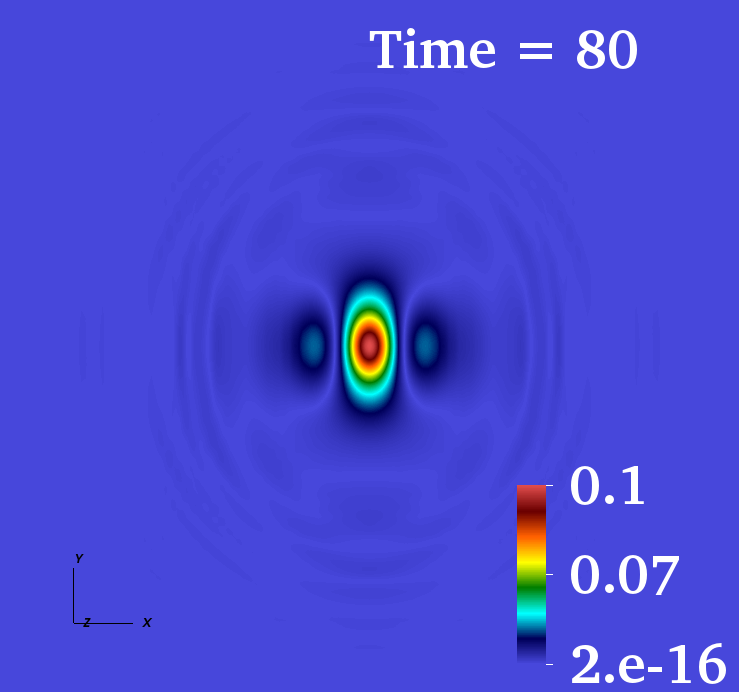}\hspace{-0.005\linewidth}
\includegraphics[width=0.295\linewidth]{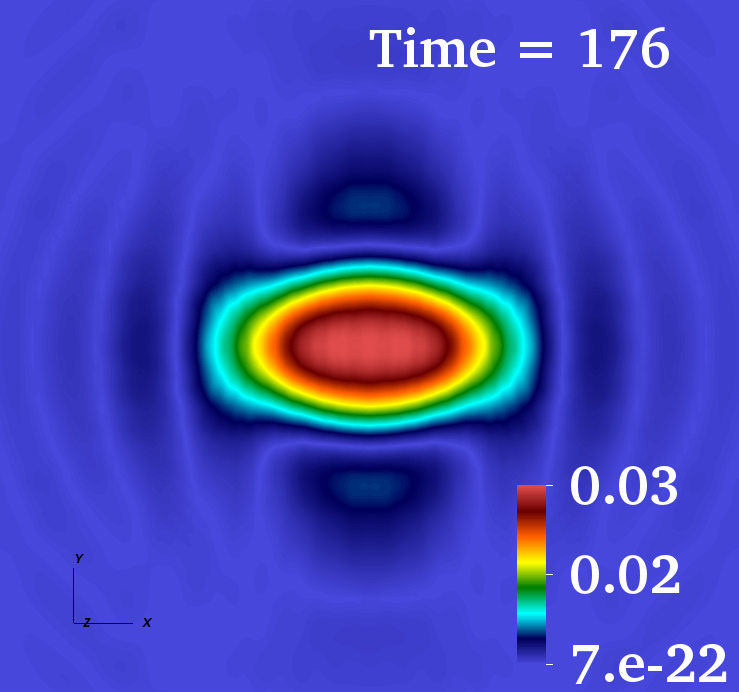}\hspace{-0.005\linewidth}
\includegraphics[width=0.295\linewidth]{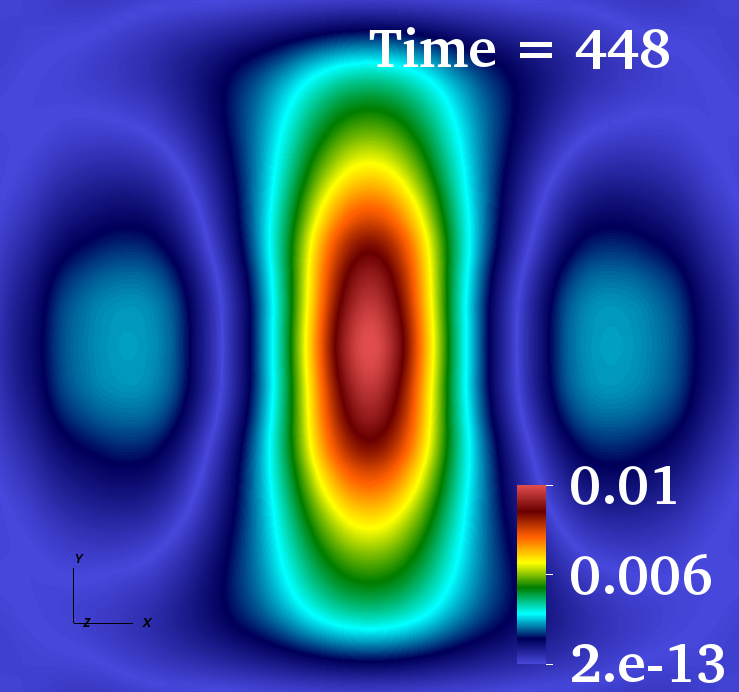}\\
\includegraphics[width=0.295\linewidth]{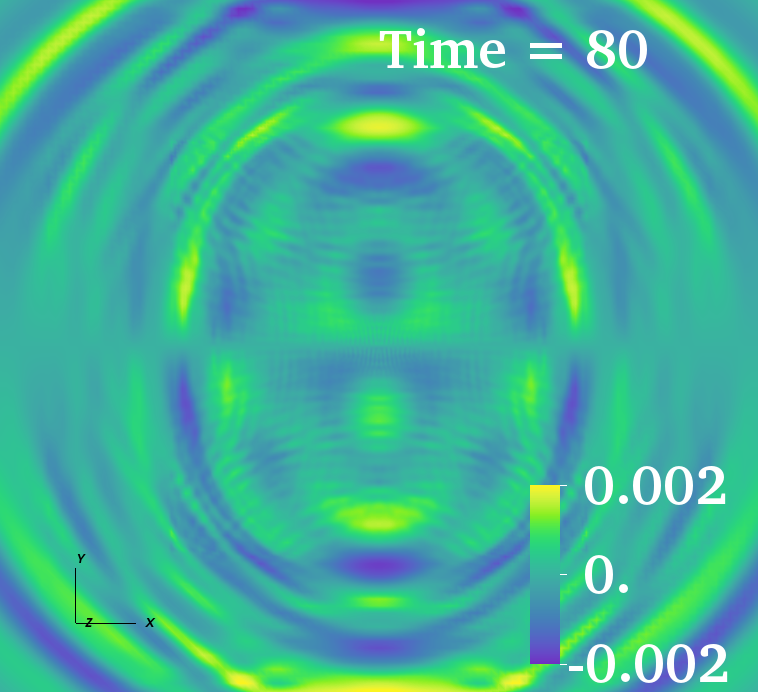}\hspace{-0.005\linewidth}
\includegraphics[width=0.295\linewidth]{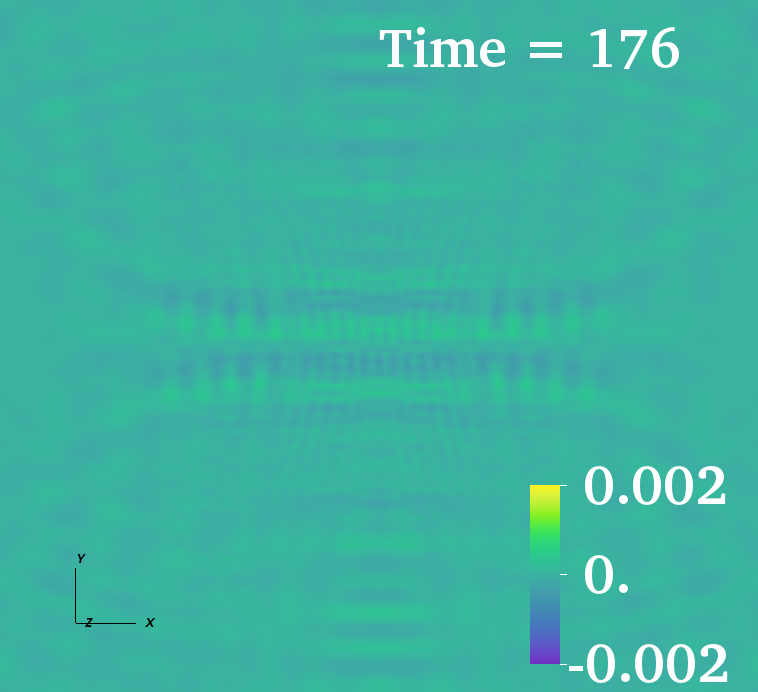}\hspace{-0.005\linewidth}
\includegraphics[width=0.295\linewidth]{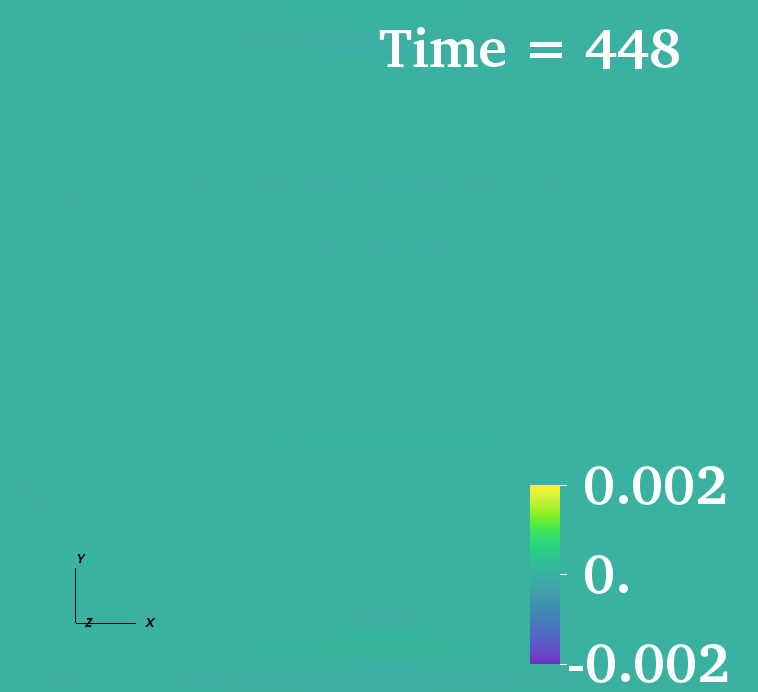}
\caption{Snapshots of the time evolution in the equatorial slice ($z=0$) of $|\Re(\Phi)|$ during the head-on collision of two BSA. The $x$ component of the electric field is plotted on top of the energy density. The axionic coupling is $k_{\rm{axion}}=0.71$ (top panels) and $k_{\rm{axion}}=1.13$ (bottom panels). Time is given in code units.\label{fig:kk2p5} }
\end{center}
\end{figure}
We plot in Figs.~\ref{fig:kk0}-\ref{fig:kk2p5} the snapshots of the time evolution of the
absolute value of the amplitude real part of the scalar field $|\Re(\Phi)|$ and of the $x$-component of the electric field, $E^{x}$, in the equatorial plane ($z=0$). We will discuss three illustrative cases. For $k_{\rm{axion}}=0$
there is no EM emission (only the residual one due the initial EM field distribution) as expected since the scalar and the EM fields are not coupled. After the collision, the final star collapses to a BH around $t\mu\sim80$. On the other hand, for $k_{\rm{axion}}=0.43$, the BSs interact with the EM field even before the collision, when both stars are separated. In this case, large amplitude EM waves are emitted, but a BH is still formed. In the isolated BS case, this value of the coupling causes the decay to a dilute BS solution that, in principle, should not be compact enough to lead to BH formation. However, here the timescale of the collision is shorter than the energy transfer process and the BSs collapse at the time of the merger.

For large values of the coupling
(in particular, $k_{\rm{axion}}=0.71$ and $k_{\rm{axion}}=1.13$), the energy transfer from the scalar field to the
EM field starts even earlier and is significant enough to prevent the collapse and BH formation. Therefore, the final object is not a BH but a less compact BS and there is an important emission of EM waves.

\begin{figure*}[thpb]
  \includegraphics[width=0.32\textwidth]{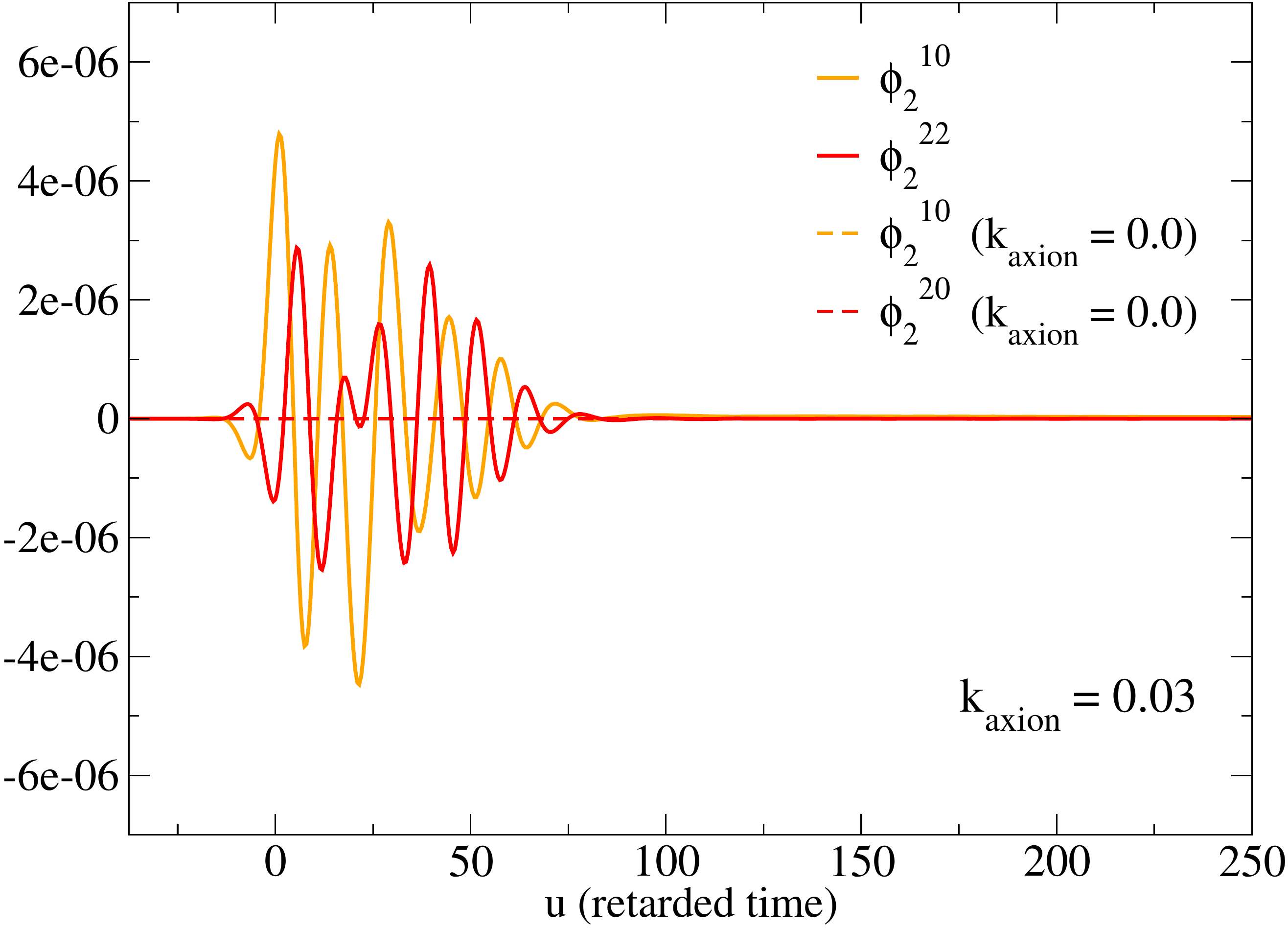} 
  \includegraphics[width=0.32\textwidth]{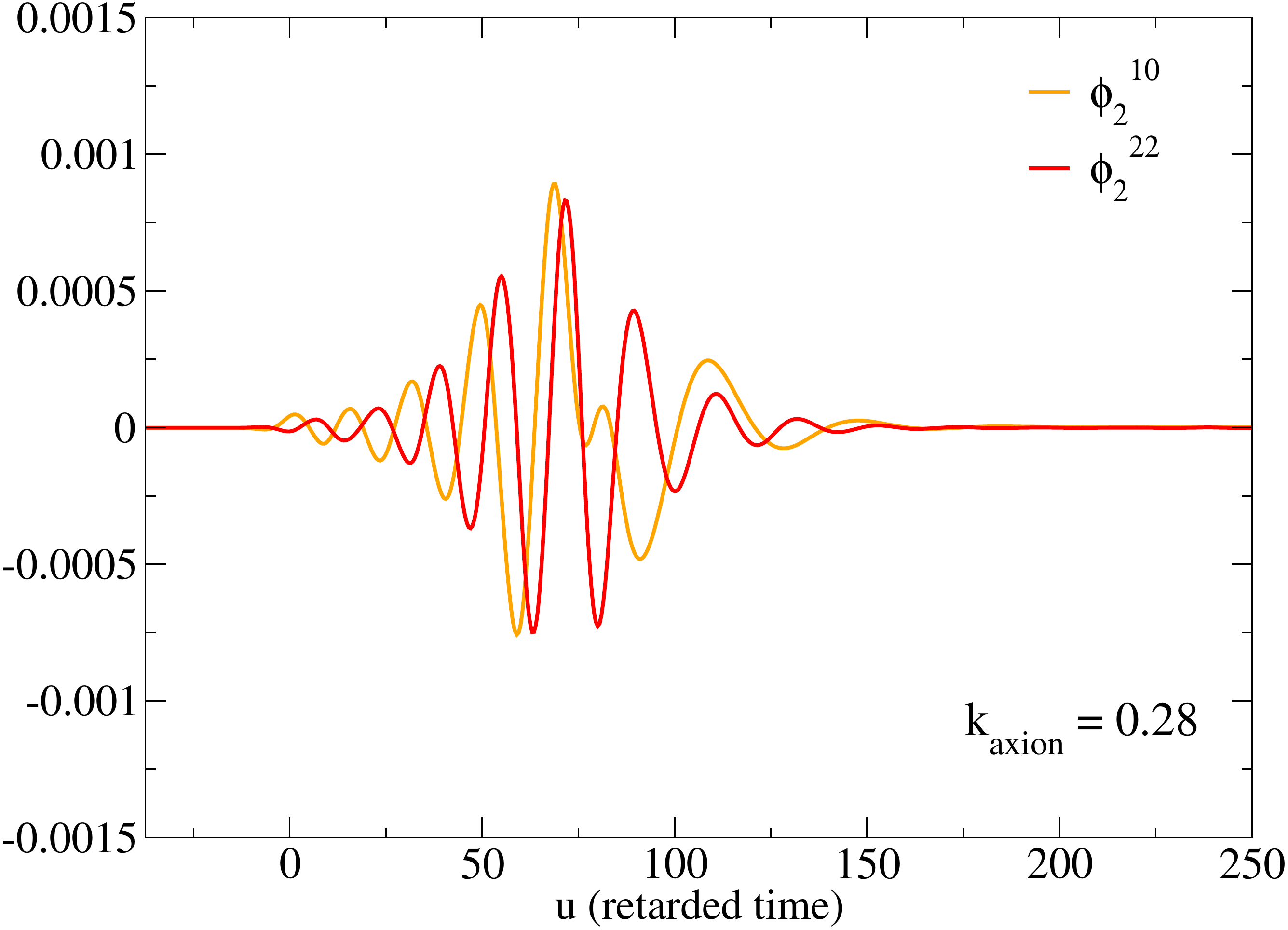} 
  \includegraphics[width=0.32\textwidth]{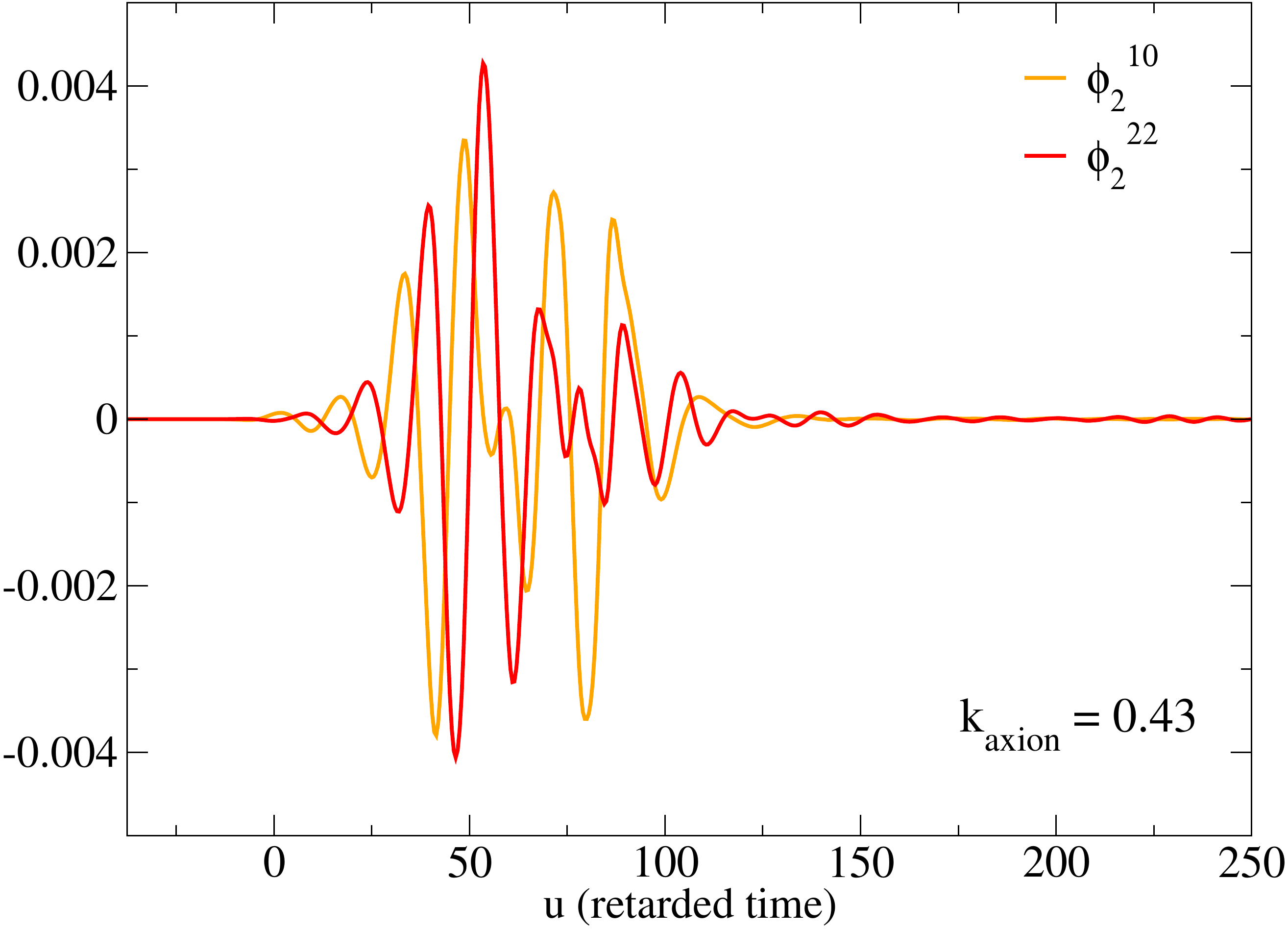}  \\
  \includegraphics[width=0.32\textwidth]{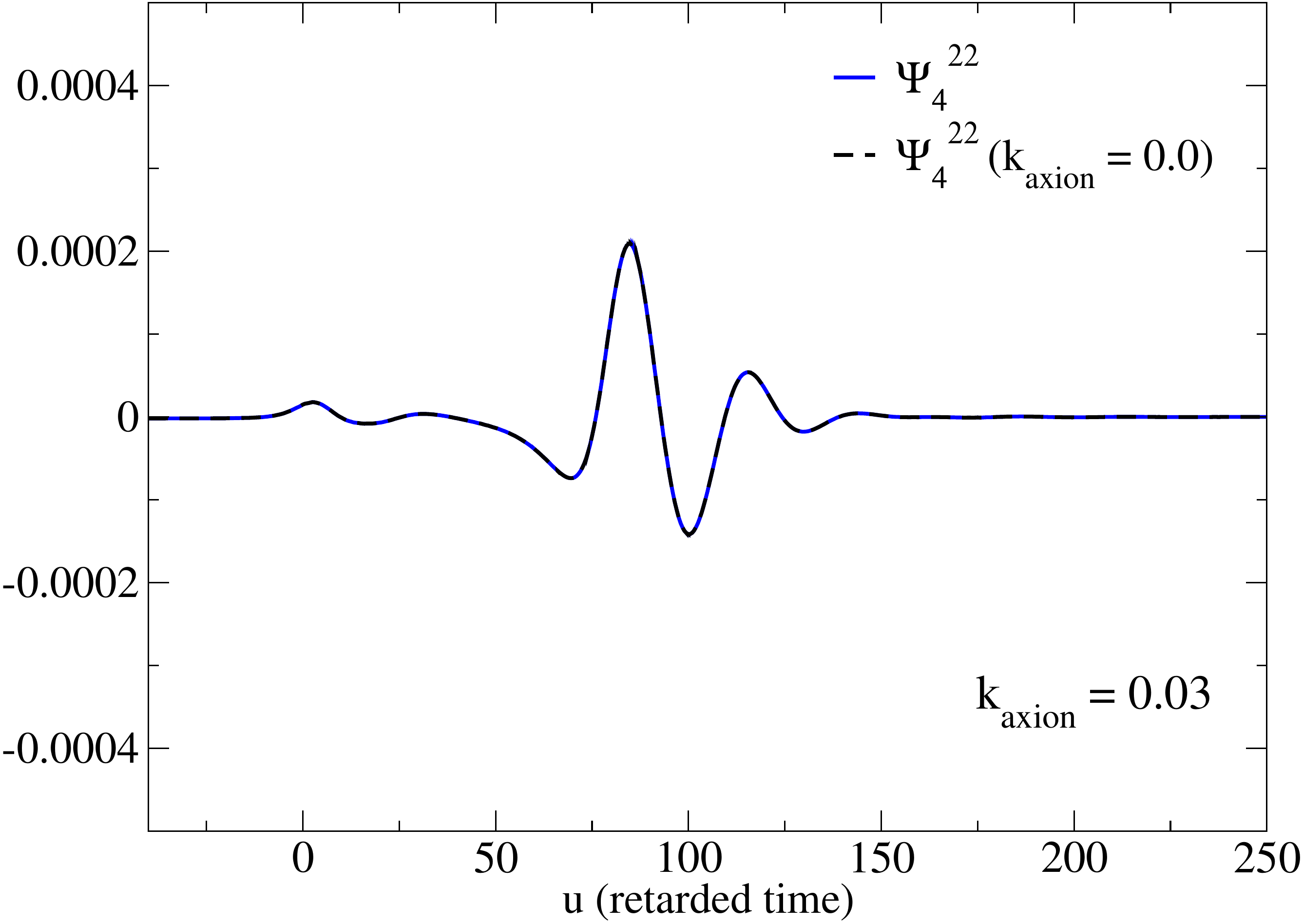} 
  \includegraphics[width=0.32\textwidth]{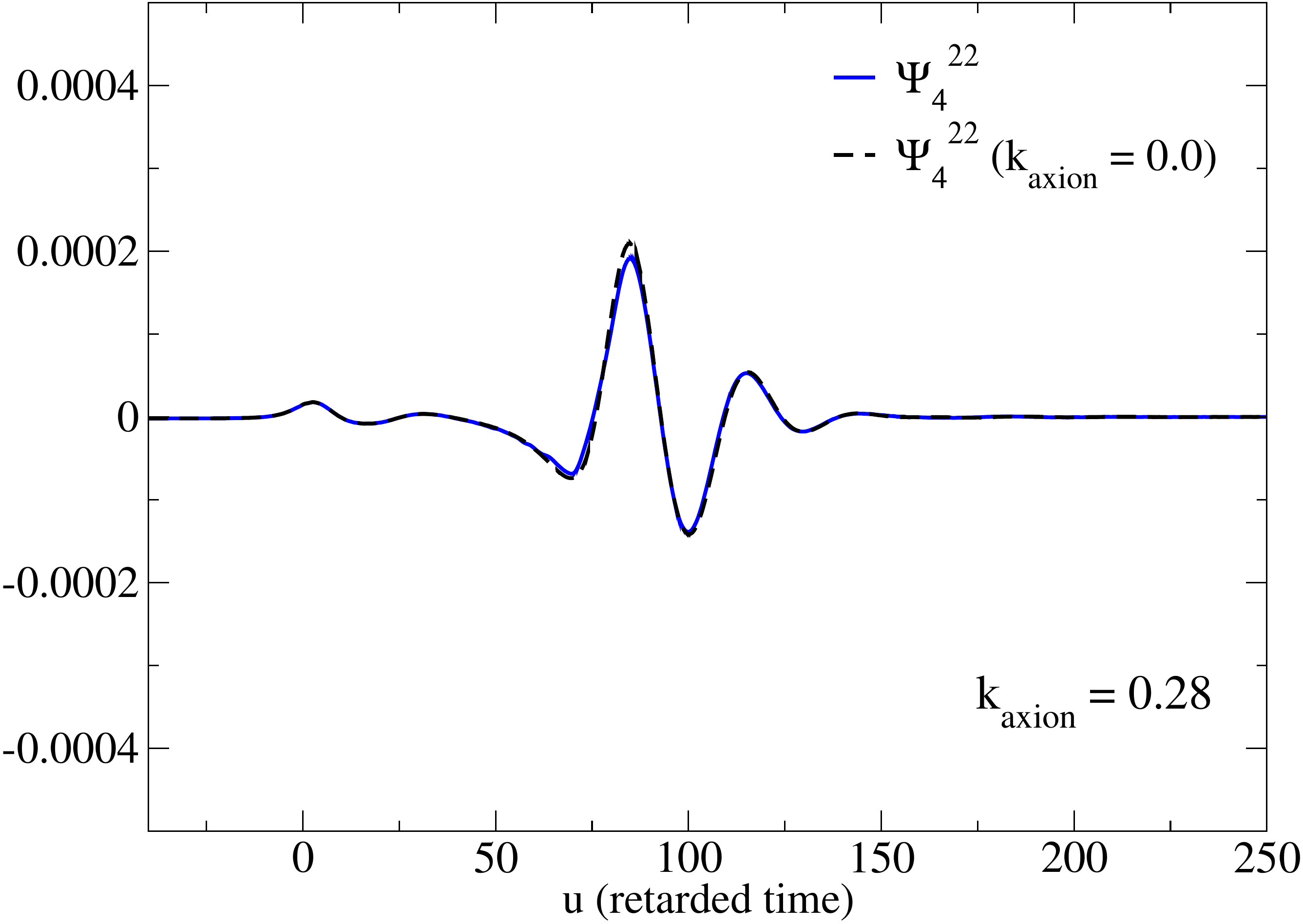}  
  \includegraphics[width=0.32\textwidth]{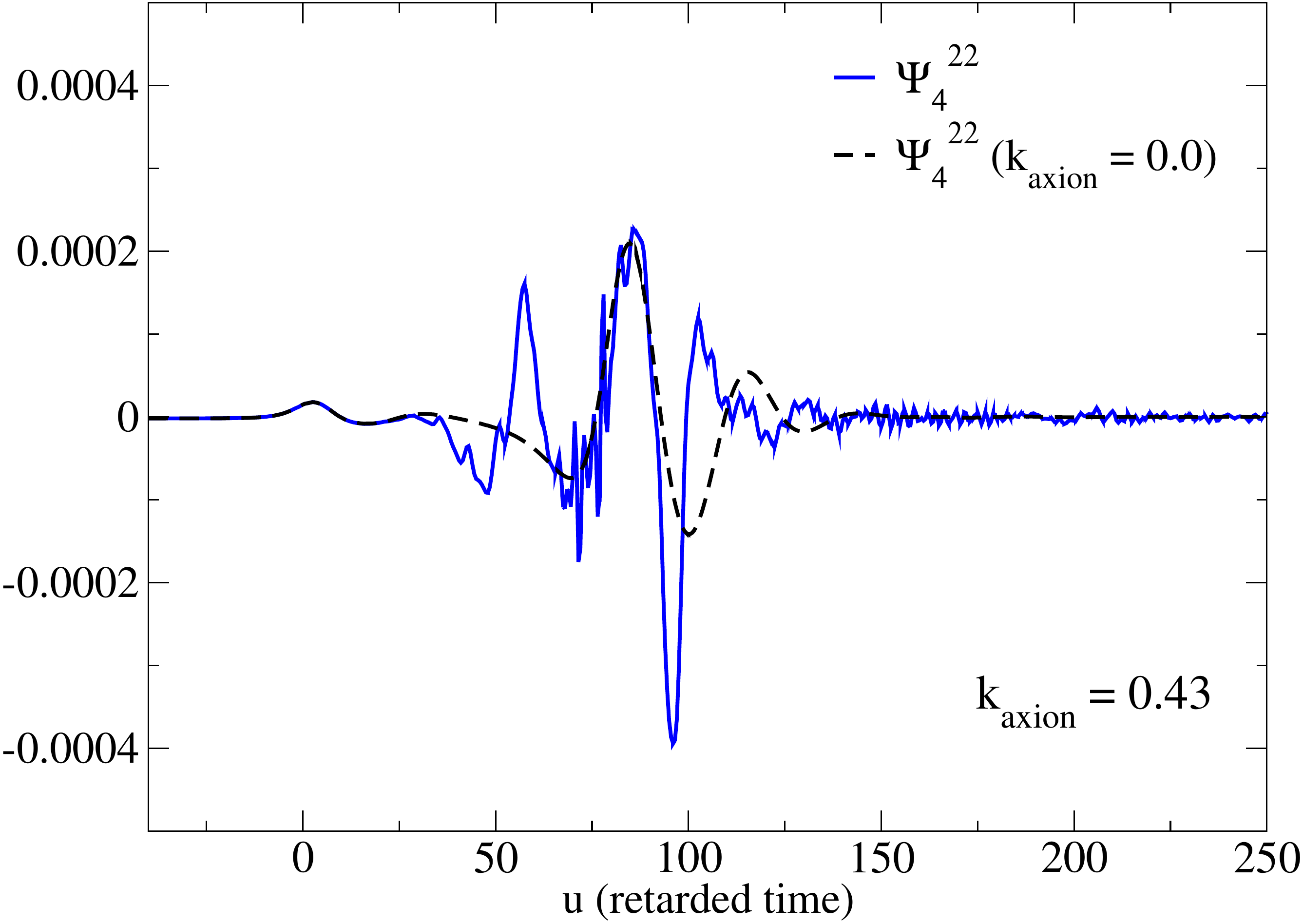} 
  \caption{NP scalars $\phi_{2}^{lm}$ and $\Psi_{4}^{22}$ during collisions of BSA stars, as a function of the retarded time for different values of coupling $k_{\rm{axion}}=\lbrace0.03,0.28,0.43\rbrace$.\label{fig6}}
\end{figure*}

\begin{figure*}[thpb]
  \includegraphics[width=0.32\textwidth]{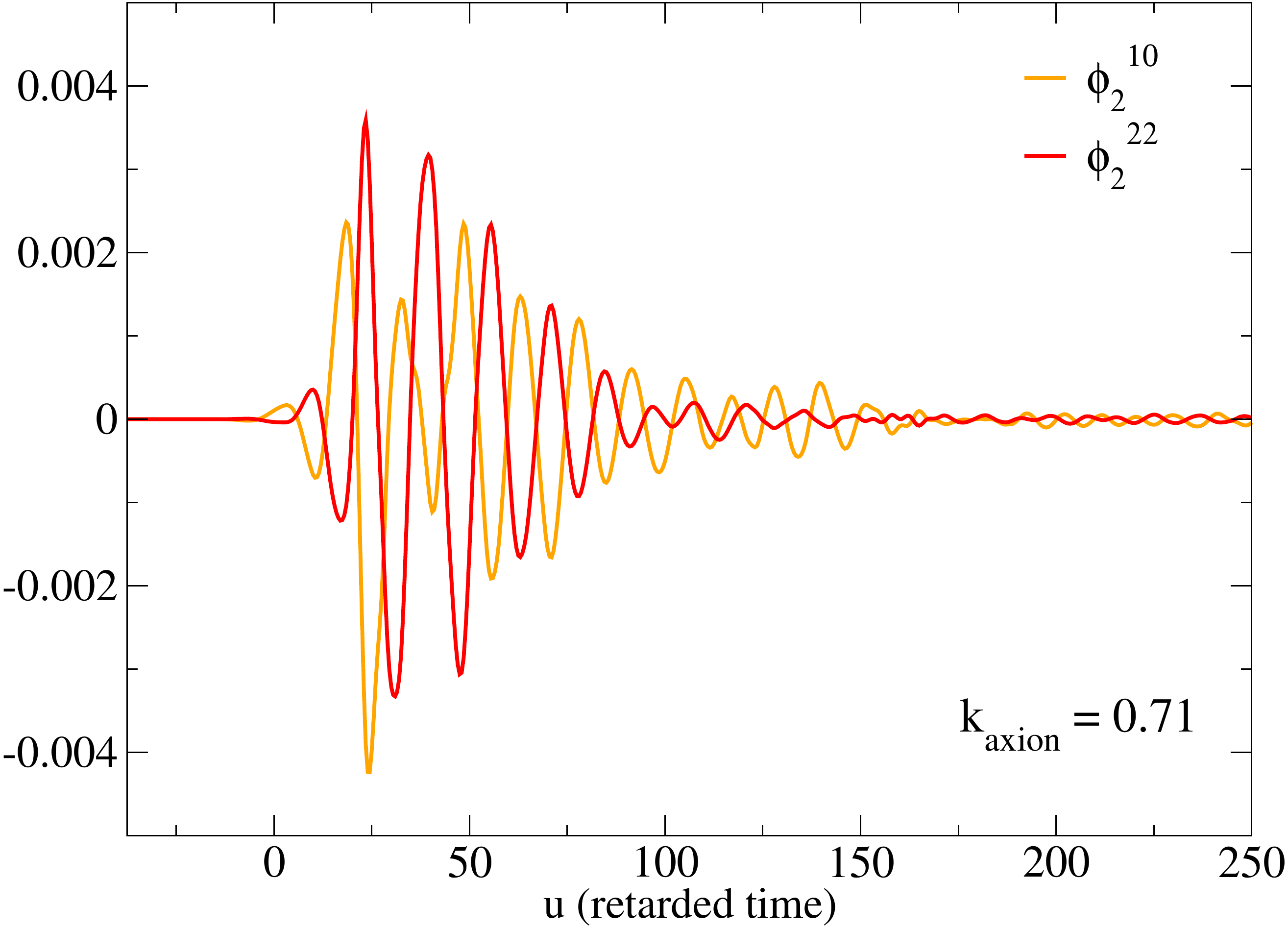}
  \includegraphics[width=0.32\textwidth]{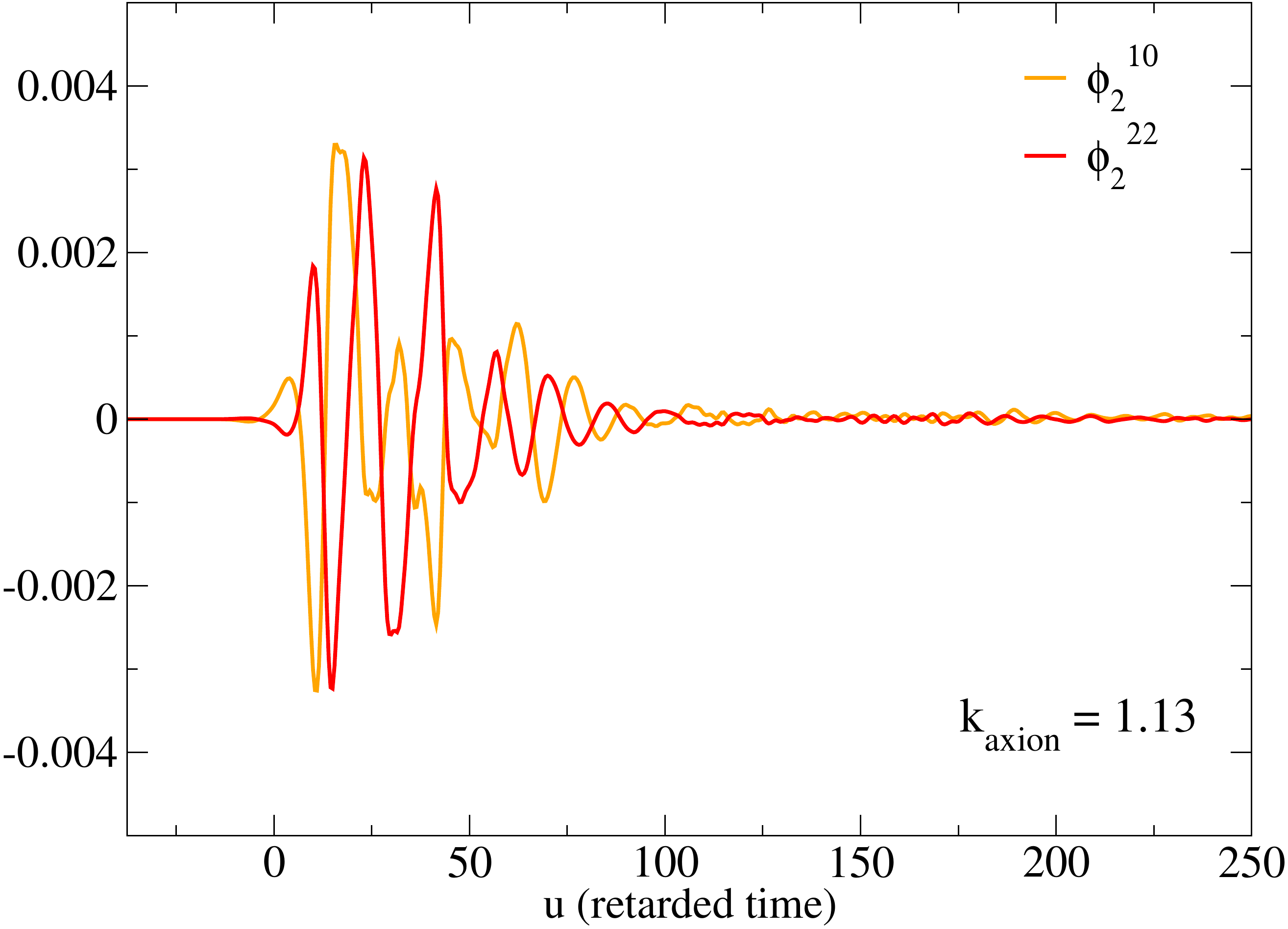} 
  \includegraphics[width=0.32\textwidth]{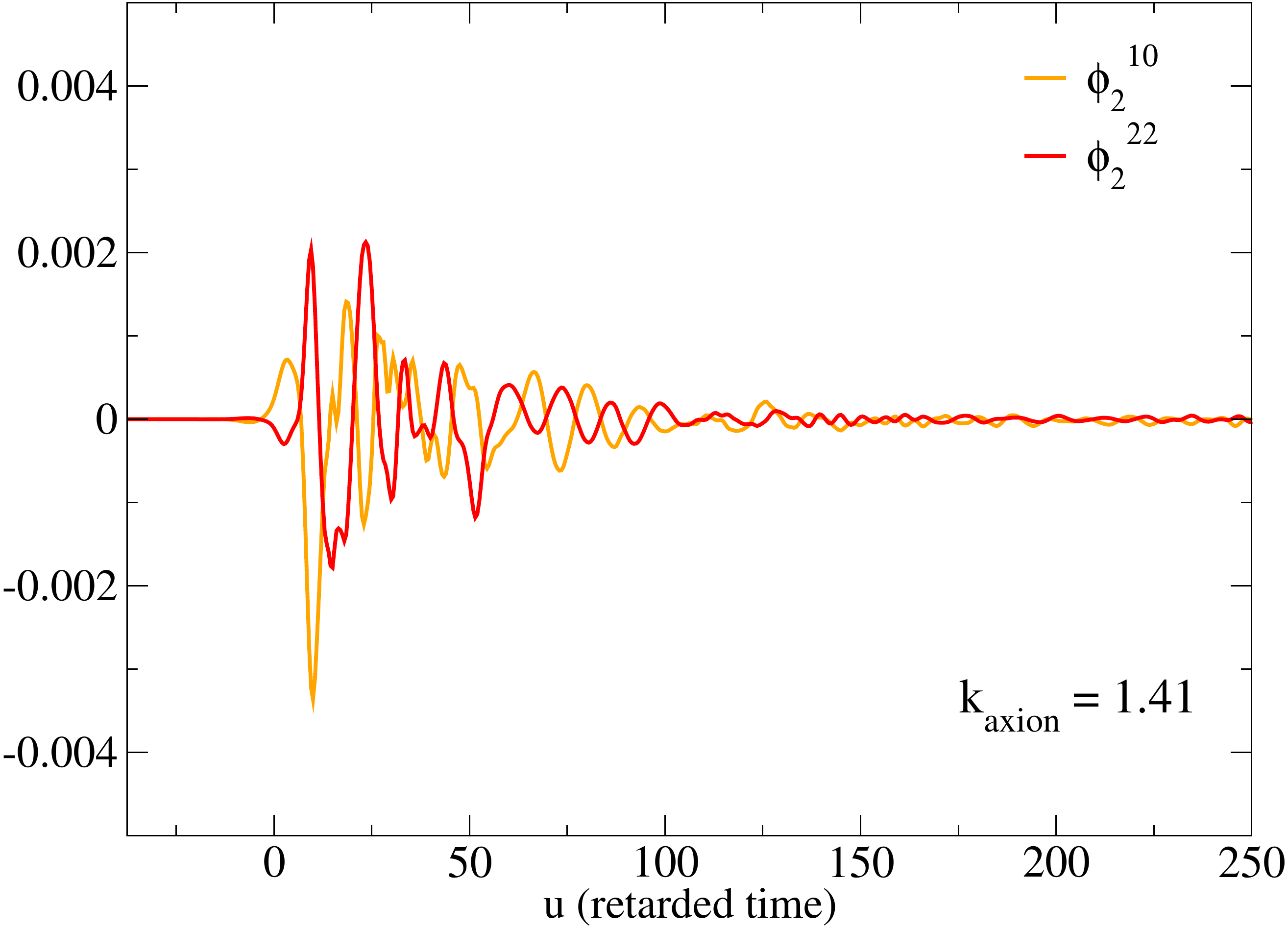} \\
  \includegraphics[width=0.32\textwidth]{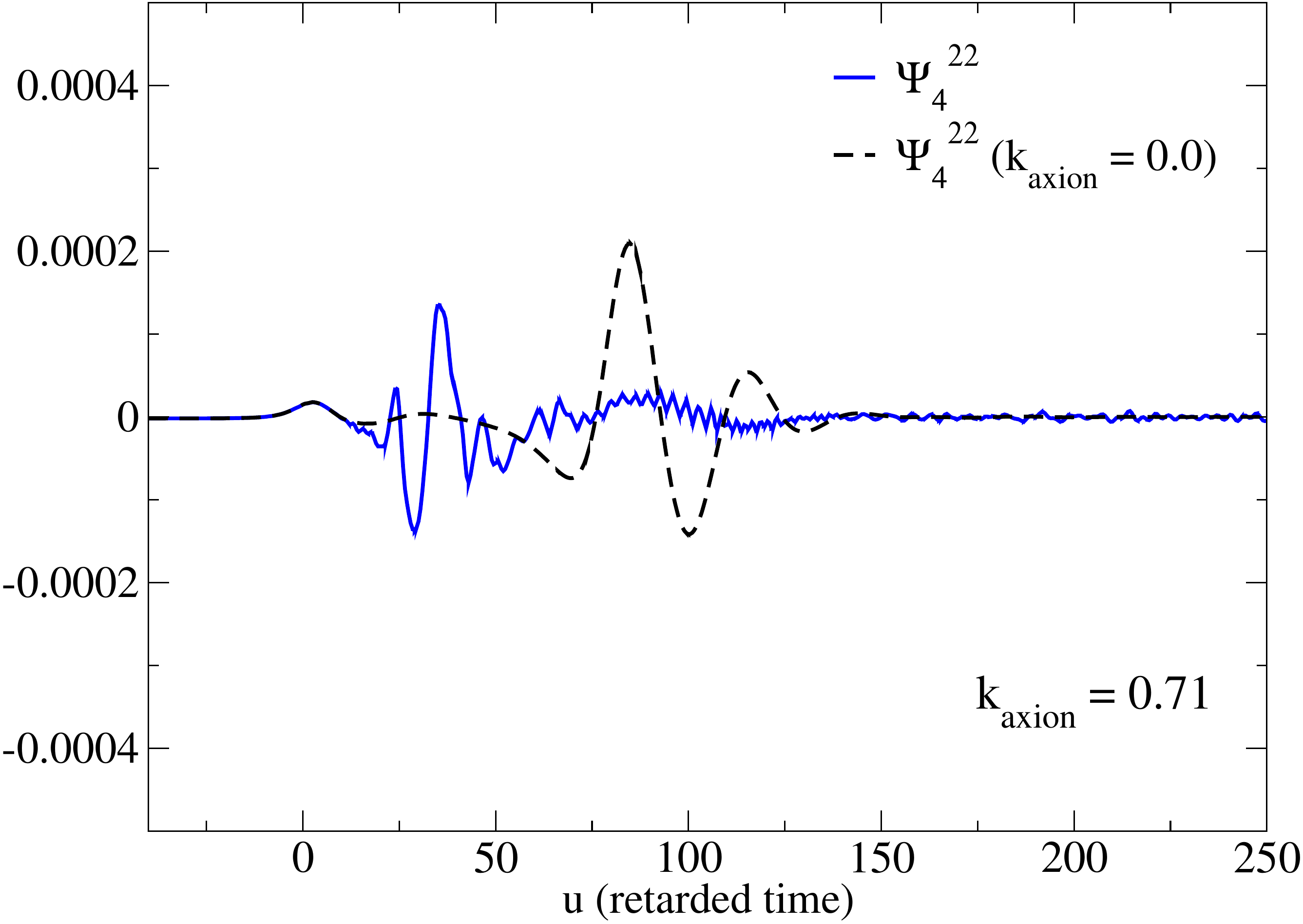} 
  \includegraphics[width=0.32\textwidth]{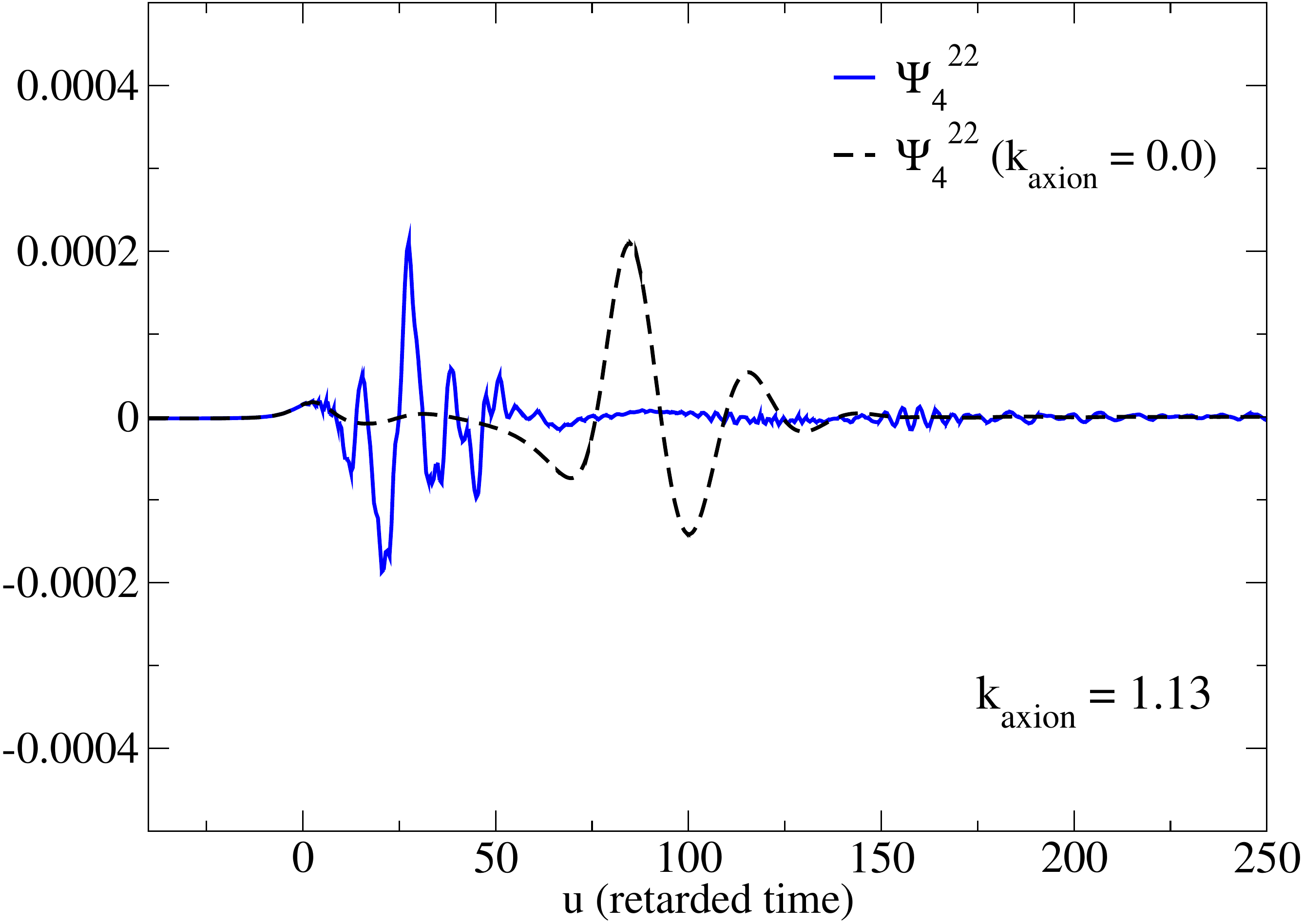}
  \includegraphics[width=0.32\textwidth]{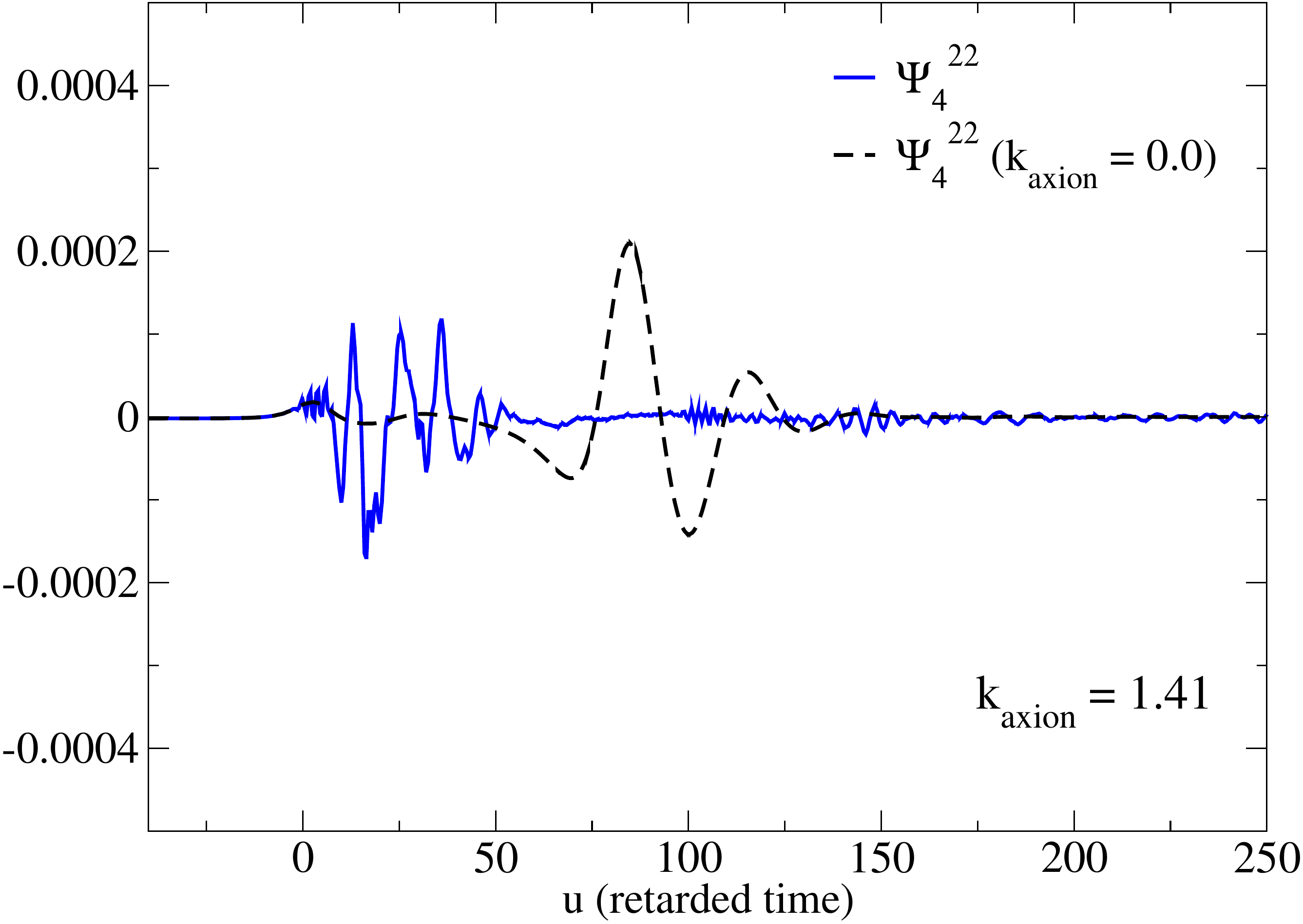}
  \caption{NP scalars $\phi_{2}^{lm}$ and $\Psi_{4}^{22}$ during collisions of BSA stars, as a function of the retarded time for different values of coupling $k_{\rm{axion}}\lbrace0.71,1.13,1.41\rbrace$. \label{fig7}}
\end{figure*}

\begin{figure*}[thpb]
\centering
\includegraphics[scale = 0.217]{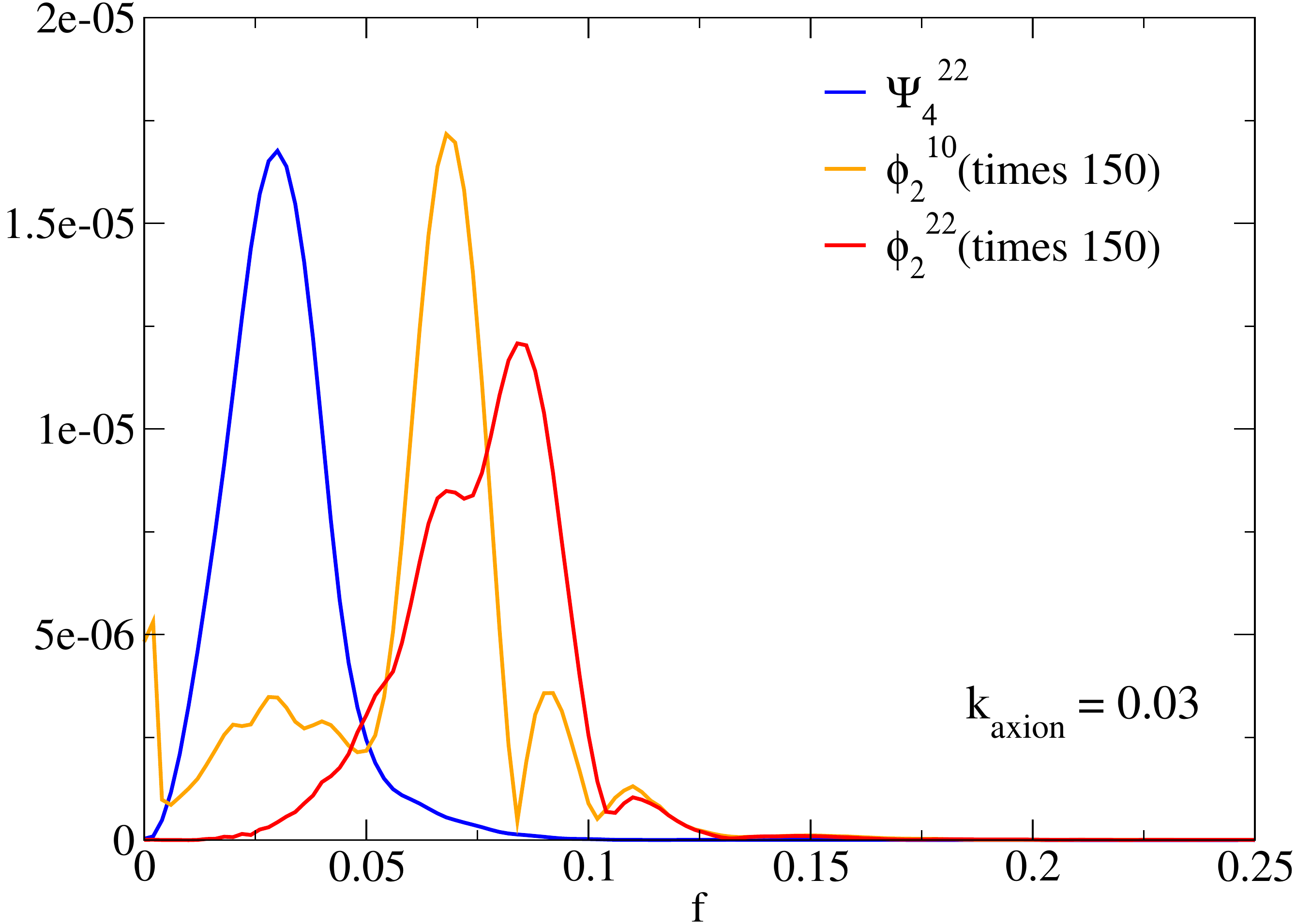}
\includegraphics[scale = 0.217]{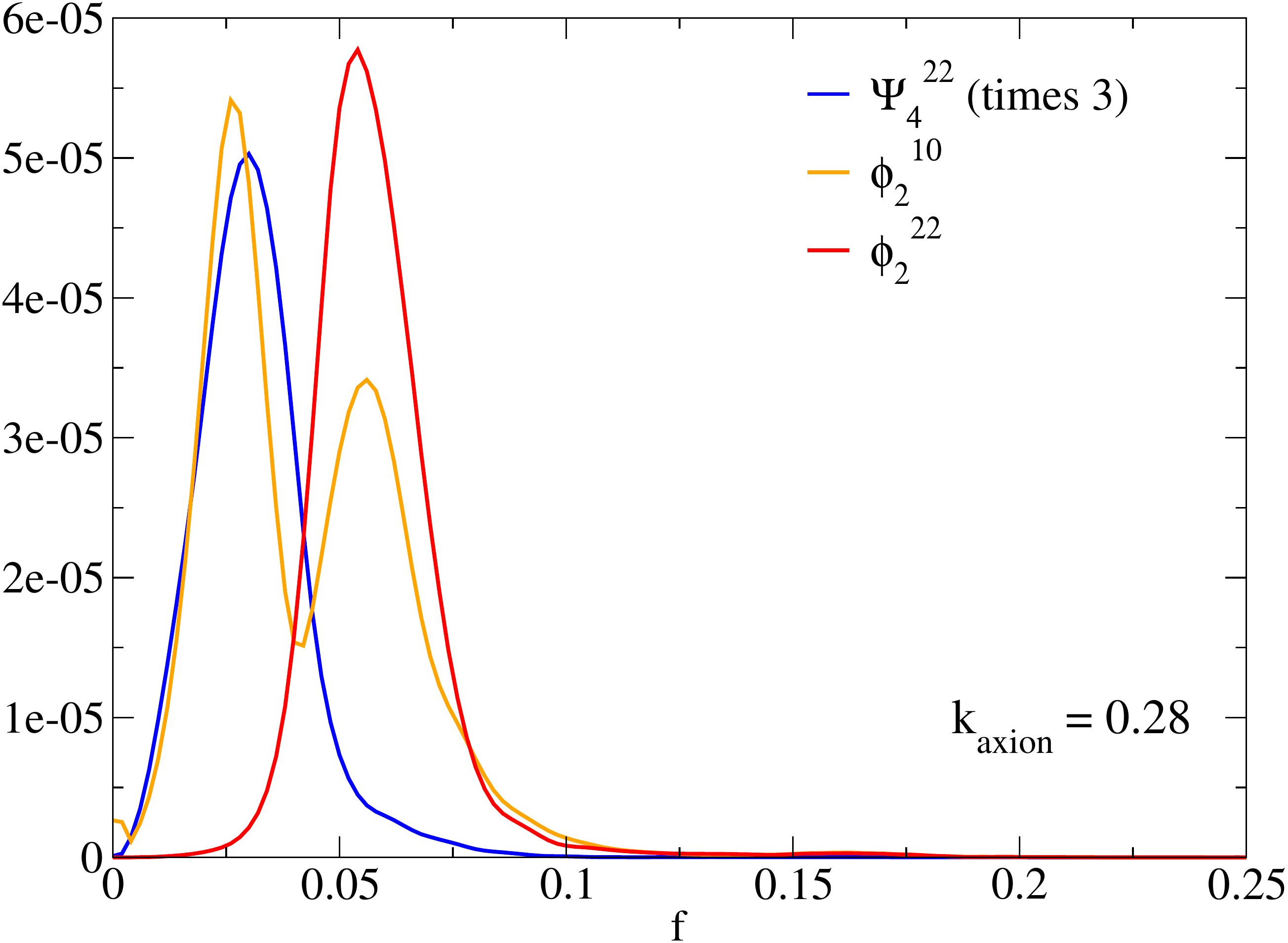}
\includegraphics[scale = 0.217]{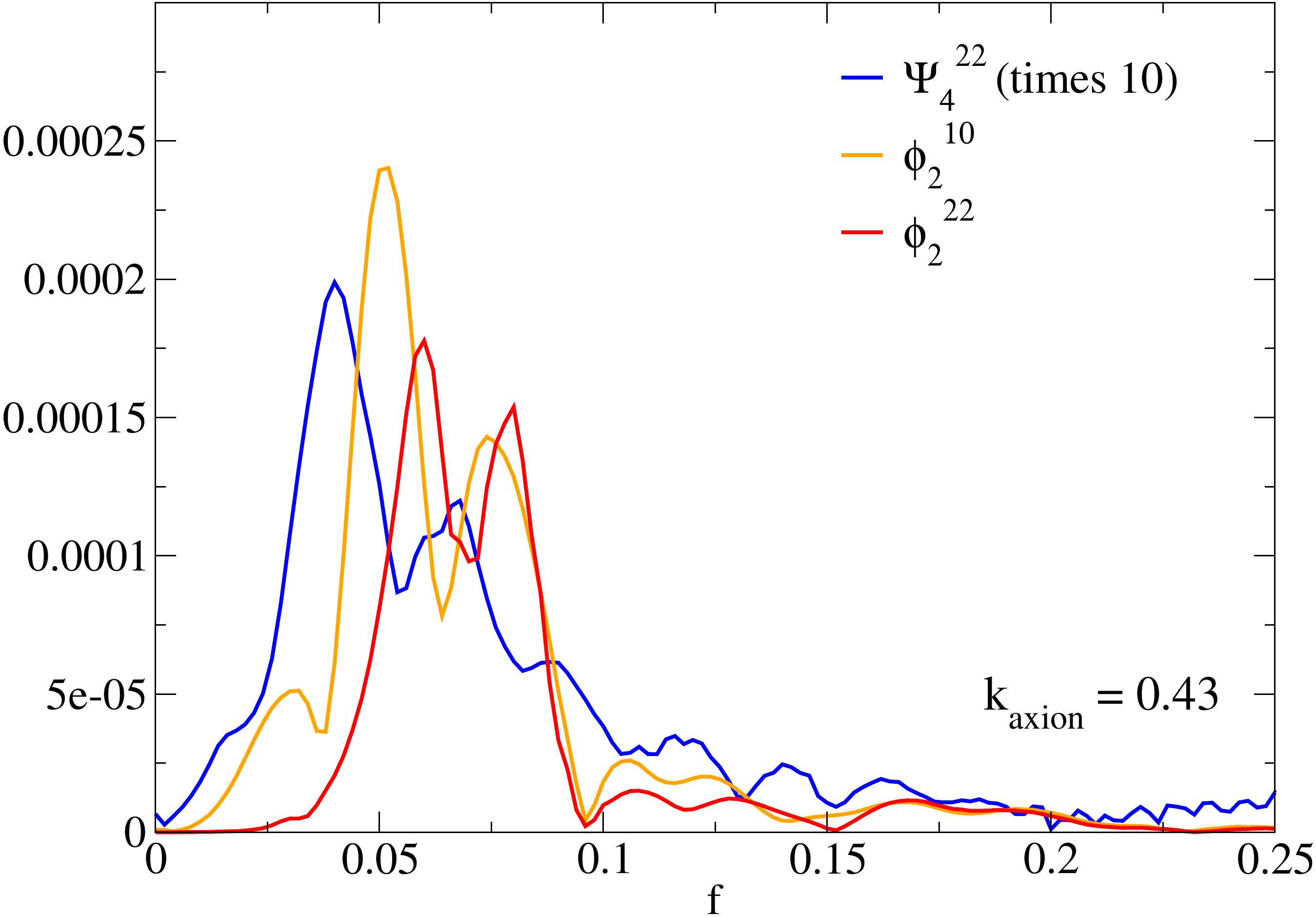} \\
\includegraphics[scale = 0.217]{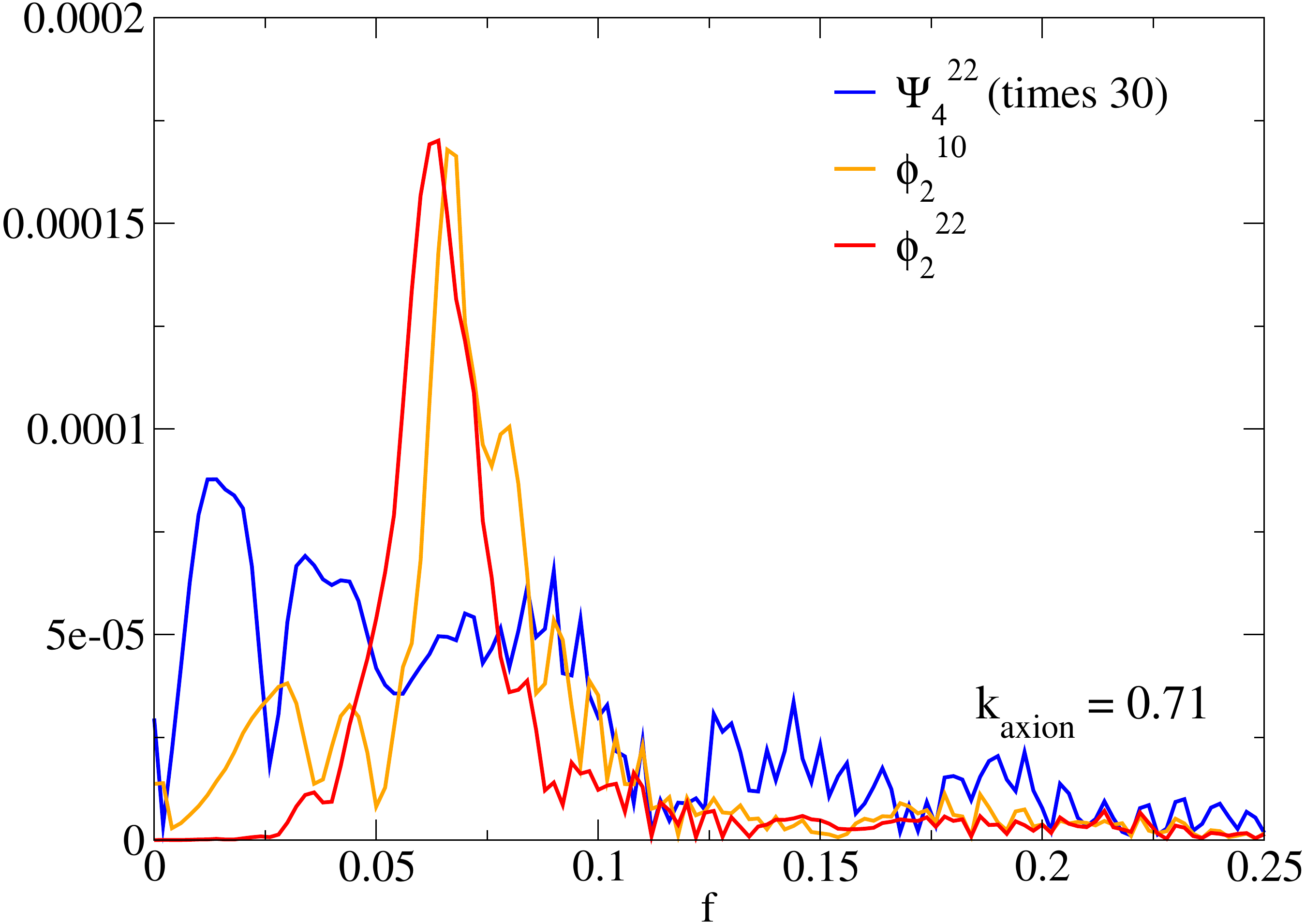}
\includegraphics[scale = 0.217]{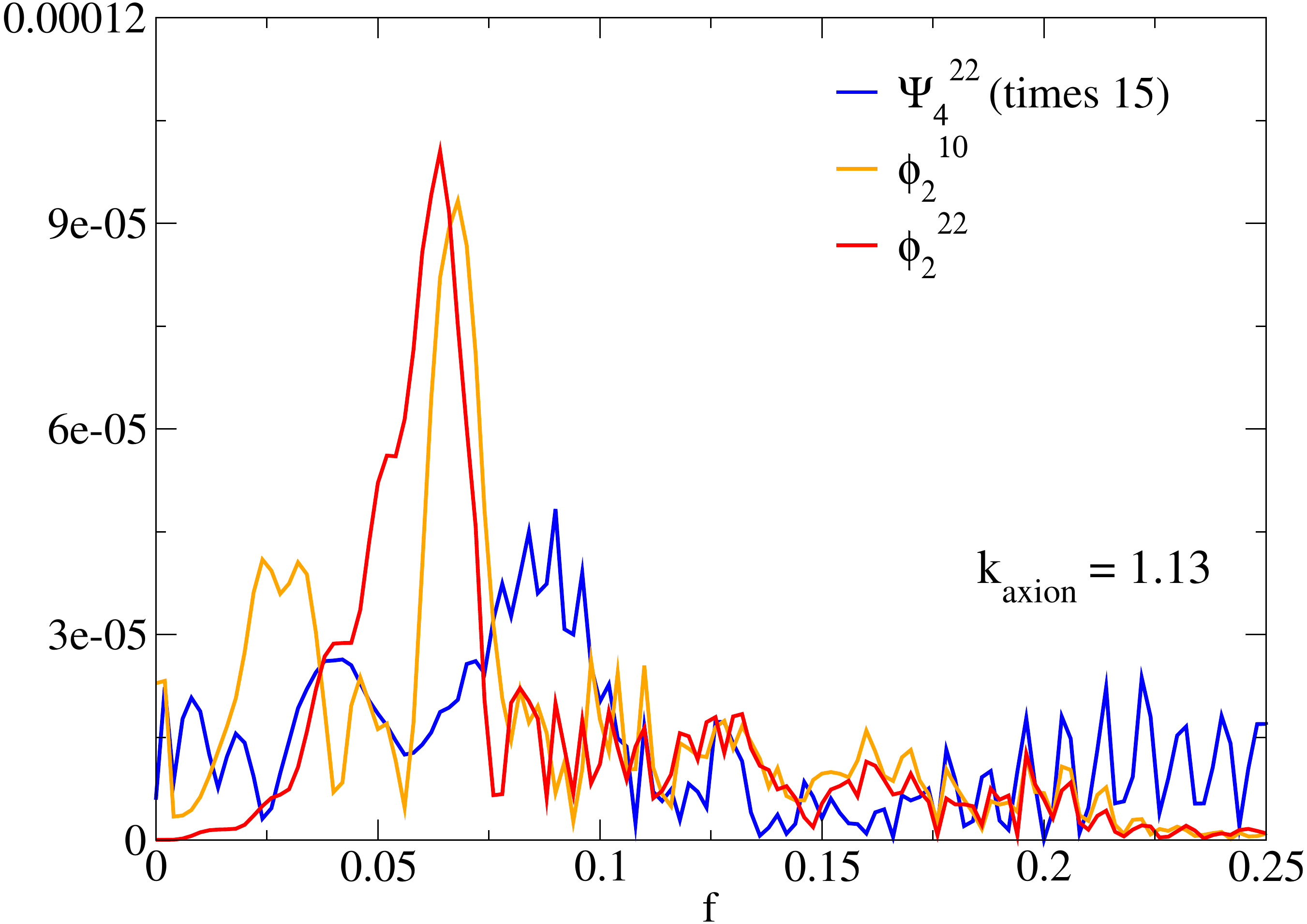}
\includegraphics[scale = 0.217]{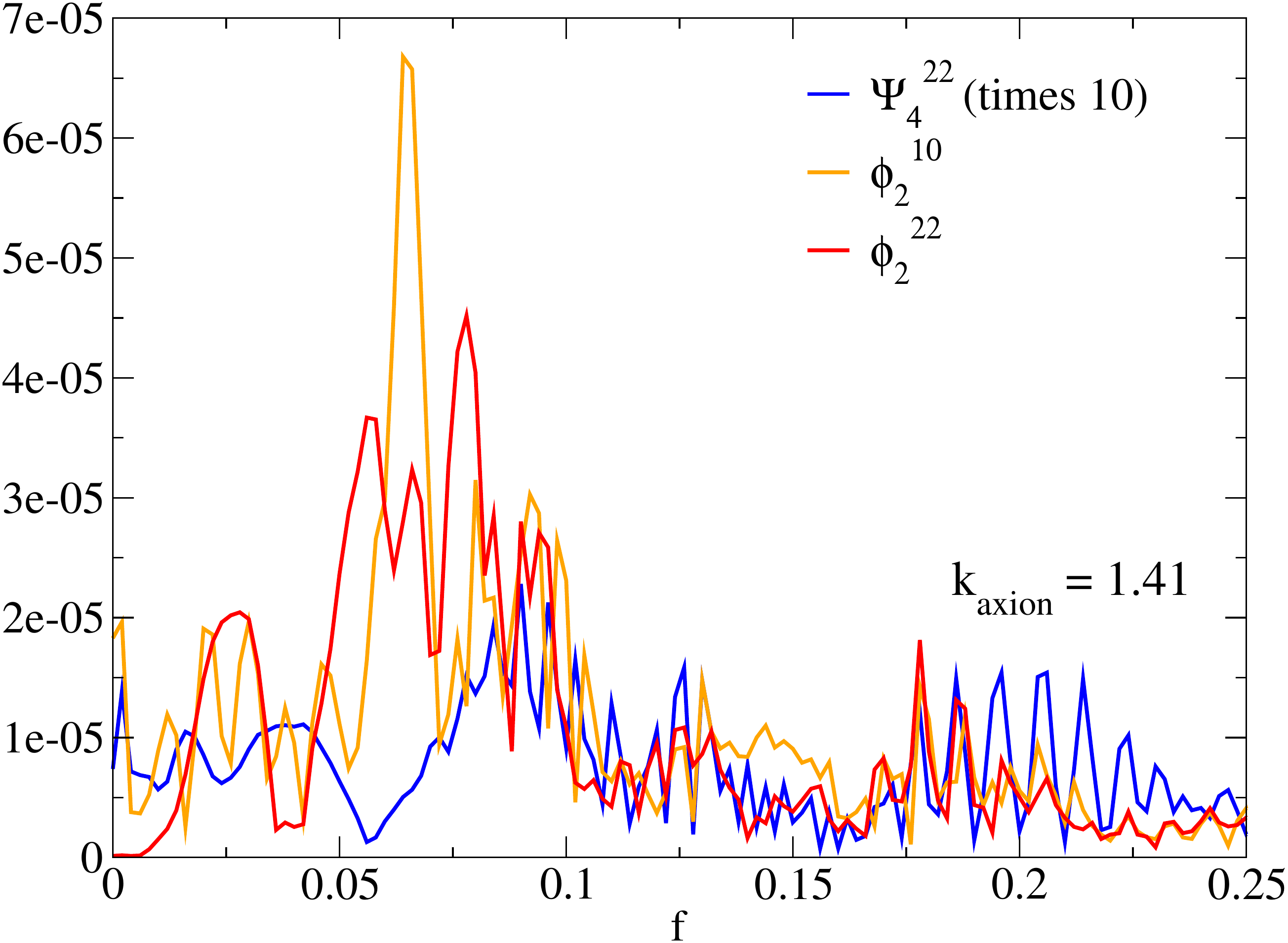}
\caption{Discrete Fourier transform of the EM and GW emissions during collisions of BSA stars, for different values of the coupling $k_{\rm{axion}}=\lbrace0.03,0.28,0.43,0.71,1.13,1.41\rbrace$.\label{fig8}}
\end{figure*}

In Figs.~\ref{fig6} and~\ref{fig7}, we display the GW signal and the EM counterpart emitted during the BS head-on collision as a function of the retarded time, $u=t-R_{\rm{ext}}$ where $R_{\rm{ext}}$ is the extraction radius. We compute the NP scalars defined in Eqs.~(\ref{eq:multipole_Psi4}) and~(\ref{eq:multipole_Phi2}) for both emissions: we show the $l=2, m=2$ mode of $\Psi_4$, and the $l=1$, $m=0$ and $l=m=2$ modes of $\phi_2$, which are the dominant modes. The dashed black line in the bottom panels correspond to the gravitational waveform in the zero coupling case for reference. The same previous trend is observed: the amplitude of the EM wave increases with $k_{\rm{axion}}$ as in the isolated BS case. In the cases shown in Fig.~\ref{fig7}, the value of the coupling is large enough to let the EM field interact with the BSs even before the collision. The GW emission starts almost at beginning of the evolution when the initial EM pulse reacts to the BSs through the coupling, which happens way before the merger.
We have studied the region of the parameter space of the coupling in which the EM emission has a large amplitude and becomes non-linear, greatly perturbing the stars. Figures~\ref{fig6} and \ref{fig7} show the GW departing from the case in which the stars are insensitive to the EM field, with $k_{\rm{axion}}=0.0$ (compare the blue solid and the black dashed lines for $\Psi_{4}$) to the regime where the EM is non-linear and induces its own GW emission.
The amplitude of the EM emission increases three orders of magnitude when going from $k_{\rm{axion}}=0.03$ to $k_{\rm{axion}}=0.28$. However, when the coupling is sufficiently large, the energy transfer is less efficient and the peak amplitude decreases (see bottom left panel of Fig.~\ref{fig7}). For large values of the coupling, the EM emission has an imprint in the GW emission and there is almost no gravitational signal coming at the time of the collision, due to the stars quickly becoming less compact. In all cases, the GW and EM emissions are burst-like and quickly decay when the system leads to the new configuration (BH or dilute BS).
The frequency analysis of the GW and EM waves reveals a correlation that depends on the coupling (see Fig.~\ref{fig8}). Small couplings, below the critical value, do not show such interaction, but for $k_{\rm{axion}}=0.28$, the dominant frequency of the $l=1$, $m=0$ mode of the EM emission approximately coincides with the frequency of the GW. For large couplings, it is the other way around: its the GW that approaches the frequency of the EM wave, exhibiting the energy transfer from the scalar field to the EM field. The EM wave becomes then self-gravitating and emits its own gravitational radiation. This is seen in the bottom panels of Fig.~\ref{fig8}, where it becomes clear that the GW frequency increases. The larger the coupling (see Fig.~\ref{fig8}), the higher the frequency of the GW emission.

\subsubsection{BSD case: triggering the instability after the collision}
We now turn to the case where the two stars are not compact and massive enough to collapse after merger. We perform the head-on collision of two BSD star models. Moreover, we choose the coupling to be initially subcritical for the isolated BSD star with $k_{\rm{axion}}=0.37$, becoming supercritical soon after the collision when the new BS is forming\footnote{As we remarked, we expect this to be an attractor solution of evolving BSs, which by accretion would get close to the critical point, getting stuck there by the parametric instability.}.

\begin{figure*}[thpb]
\centering
\includegraphics[scale = 0.3]{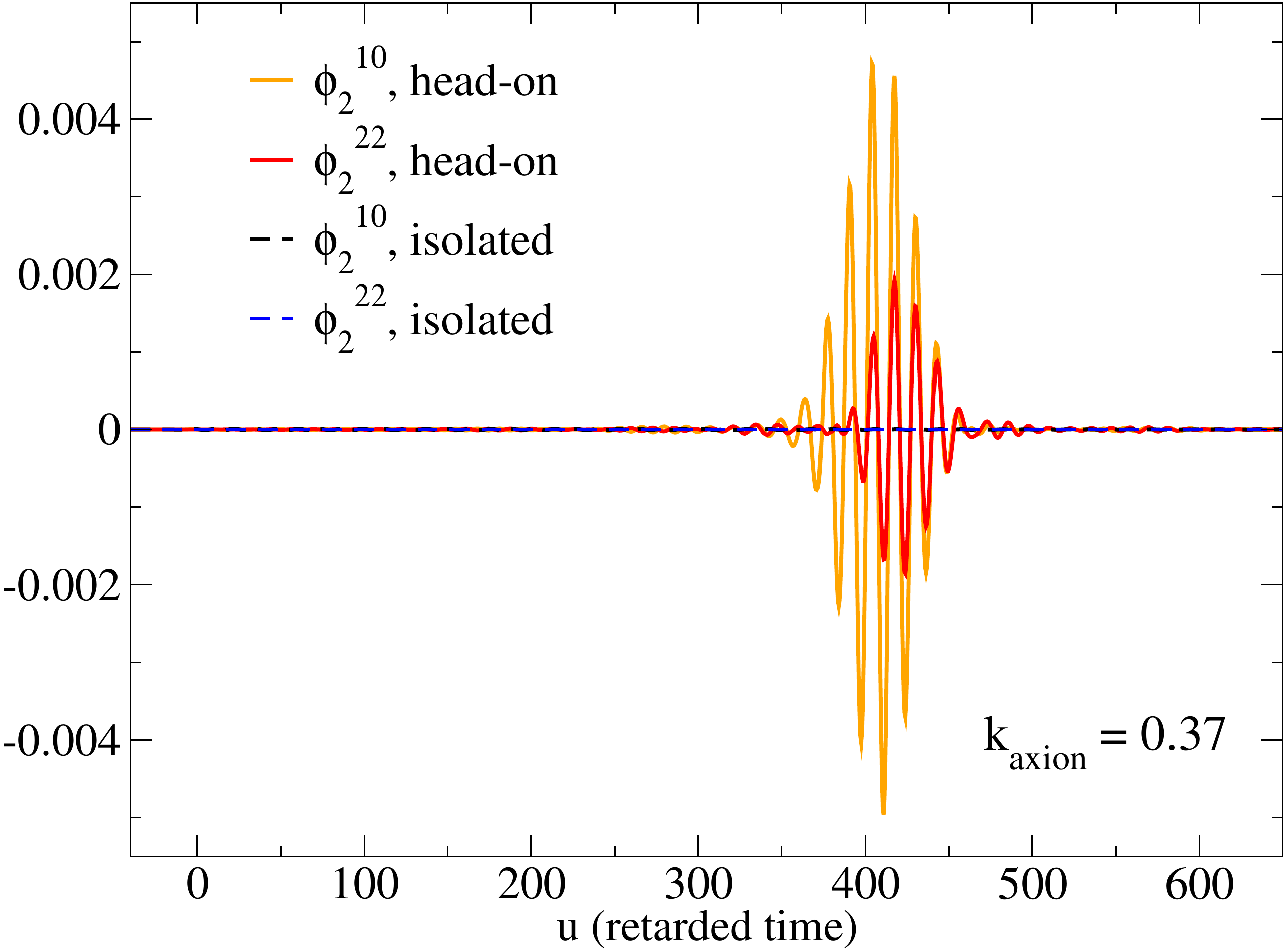}
\includegraphics[scale = 0.3]{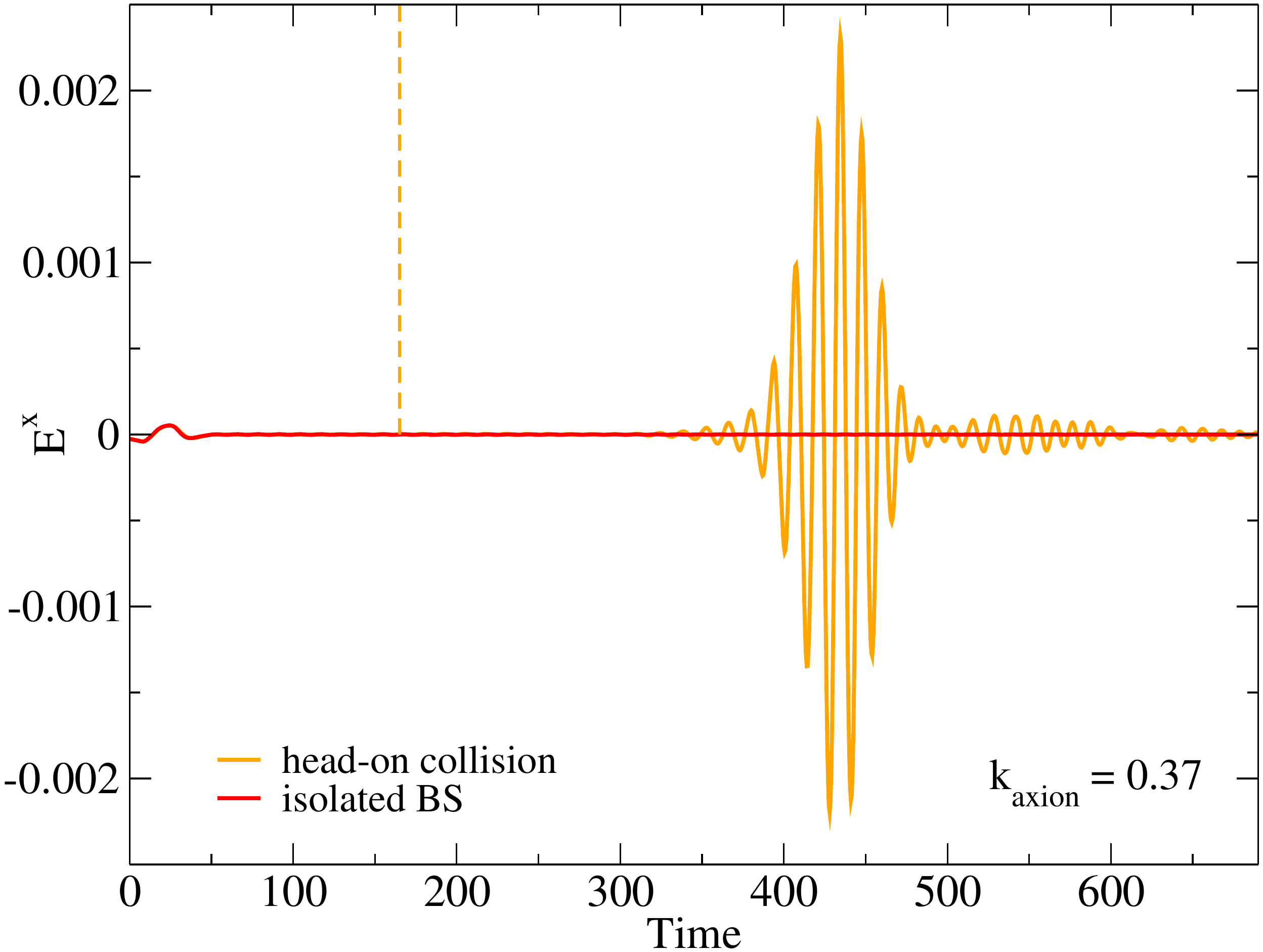}
\caption{{\bf Left panel:} NP scalars $\phi_{2}^{lm}$ as a function of the retarded time for the
evolution of the {\it isolated} BSD and the head-on collision (cf. Table~\ref{tab:table1}), for $k_{\rm{axion}}=0.37$ extracted at $r=40$.  {\bf Right panel:} Time evolution of the amplitude of $x$-component of the electric field $E^{x}$ extracted at $(x, y, z) = (12, 12, 12)$ for BSD. The vertical dashed orange line signals the approximate time at which the two stars collide. \label{fig10}}
\end{figure*}

The final object has an approximate central value of the scalar field $\Phi_0\sim0.09$ and mass between $M\sim0.5$ and $0.6$, similar to configuration BSB. To illustrate this different scenario we plot in Fig.~\ref{fig10} the $x$-component of the electric field $E^{x}$ and the NP scalars $\phi_{2}^{lm}$ as a function of time for the isolated case and the head-on collision. It shows that while in the isolated case there is no growth of the EM field, in the latter the coupling triggers the emission of EM waves.
%

\section{Final remarks}\label{sec:final}
We have investigated the dynamics of BSs described by a complex scalar field non-minimally coupled to the EM field by choosing different values of the coupling $k_{\rm axion}$ in a highly-dynamical scenario. First we studied isolated BSs that are perturbed with a coupling above their corresponding critical value. In this case, a large EM emission can be triggered leading to the decay of the star to a less compact configuration. In a binary system, the head-on collision of two massive BSs produces the emission of a short burst of GWs due to the merger and BH formation. However, for large values of the coupling there is a large energy transfer from the scalar field to the EM field triggering the emission of EM waves that can prevent the formation of compact BSs or the collapse to a BH in the case of BS mergers. If the value of $k_{\rm axion}$ is large enough, the EM waves become non-linear and can even have a substantial impact on the gravitational waveform.

This work is interesting in the context of multi-messenger events, where both gravitational and EM radiation would be emitted. In fact, we argued that accretion-induced growth of BSs will cluster them close to the critical coupling threshold. Hence, mergers of similar mass BSs should always lead to electromagnetic counterparts.
Here, we have considered a complex bosonic field coupled to the EM field through an axionic coupling in BS binary head-on collisions. The system studied here is a toy model of a simple scenario potentially emitting both GW and EM waves, and the values of the coupling chosen are not compatible with observations. However, a natural extension would be to perform quasi-circular mergers: while large couplings would produce a decay of the BSs resulting in a system emitting almost no detectable GWs and likely in a frequency range outside LIGO-Virgo-KAGRA sensitivity, small couplings would probably leave almost unchanged the BSs and lead to a EM burst during the late inspiral and the collision that could be detected as a multi-messenger event.

\begin{acknowledgments}
M.Z.\ acknowledges financial support provided by FCT/Portugal through the IF programme, grant IF/00729/2015, and
by the Center for Research and Development in Mathematics and Applications (CIDMA) through the Portuguese Foundation for Science and Technology (FCT -- Funda\c{c}\~ao para a Ci\^encia e a Tecnologia), references UIDB/04106/2020, UIDP/04106/2020 and the projects PTDC/FIS-AST/3041/2020 and CERN/FIS-PAR/0024/2021.
V.C.\ is a Villum Investigator and a DNRF Chair, supported by VILLUM FONDEN (grant no.~37766) and by the Danish Research Foundation. V.C.\ acknowledges financial support provided under the European
Union's H2020 ERC Advanced Grant ``Black holes: gravitational engines of discovery'' grant agreement
no.\ Gravitas–101052587.
N.S.G. acknowledges financial support by the Spanish Ministerio de Universidades, reference UP2021-044, within the European Union-Next Generation EU.
This work has further been supported by  the  European  Union's  Horizon  2020  research  and  innovation  (RISE) programme H2020-MSCA-RISE-2017 Grant No.~FunFiCO-777740.
This project has received funding from the European Union's Horizon 2020 research and innovation programme under the Marie Sklodowska-Curie grant agreement No 101007855.
We thank FCT for financial support through Project~No.~UIDB/00099/2020.
We acknowledge financial support provided by FCT/Portugal through grants PTDC/MAT-APL/30043/2017 and PTDC/FIS-AST/7002/2020.
The results of this research have been achieved using the DECI resource Snellius based in The Netherlands at SURF with support from the PRACE aisbl,
the Navigator cluster, operated by LCA-UCoimbra, through project~2021.09676.CPCA, the Servei d'Inform\`atica de la Universitat
de Val\`encia and the Argus and Blafis cluster at the U. Aveiro.

\end{acknowledgments}

\bibliography{References}

\end{document}